\documentclass[a4paper,11pt]{article}
\pdfoutput=1 

\usepackage{jheppub} 

\usepackage{bm,amsmath,amssymb,slashed,graphicx,%
            enumerate,alltt,xspace,multirow,xcolor,mathrsfs}
\usepackage{fancyvrb}
\usepackage{booktabs}
\usepackage{graphicx}
\usepackage{subcaption}
\usepackage{xspace}
\usepackage[utf8]{inputenc}
\usepackage{bm}
\usepackage{amssymb}
\usepackage{lineno}
\usepackage{placeins} 

\makeatletter
\g@addto@macro\bfseries{\boldmath}
\makeatother

\definecolor{labelkey}{rgb}{0,0.5,0.0}

\usepackage{listings}
\definecolor{darkgreen}{rgb}{0,0.4,0}
\definecolor{darkblue}{rgb}{0,0,0.5}
\definecolor{grey}{rgb}{0.5,0.5,0.5}

\newcommand{\pythiaeight}{\texttt{Pythia8.3}\xspace}

\newcommand{\as}{\alpha_s}

\newcommand{\lep}{e e}
\newcommand{\had}{pp}
\newcommand{\dis}{e p}

\newcommand{\Mb}[1]{M_{b}^{(#1)}}
\newcommand{\Sb}[1]{S_{b}^{(#1)}}
\newcommand{\MbSD}{M_{b,\,{\rm SD}}}
\newcommand{\SbSD}{S_{b,\,{\rm SD}}}
\newcommand{\bo}{\ensuremath{b}}
\newcommand{\bqcd}[1]{\ensuremath{\beta_{#1}}}
\newcommand{\muf}{\ensuremath{\mu_{\scriptscriptstyle F}}}
\newcommand{\mur}{\ensuremath{\mu_{\scriptscriptstyle R}}}
\newcommand{\mul}{\ensuremath{\mu_{\scriptscriptstyle L}}}
\newcommand{\mulbar}{\ensuremath{\bar{\mu}_{\scriptscriptstyle L}}}
\newcommand{\mum}{\ensuremath{\mu_{\scriptscriptstyle M}}}
\newcommand{\vm}{\ensuremath{v_{\scriptscriptstyle M}}}
\newcommand{\muc}{\mu_{\scriptscriptstyle {\rm ISR}}}
\newcommand{\Q}{Q}
\newcommand{\lnxmuR}{\ensuremath{\ln\frac{\mur}{\mul}}}
\newcommand{\lnxmuRsq}{\ensuremath{\ln^2\frac{\mur}{\mul}}}

\newcommand{\lnxM}{\ensuremath{\ln\frac{\Q}{\mul}}}
\newcommand{\lnxMsq}{\ensuremath{\ln^2\frac{\Q}{\mul}}}
\newcommand{\PhiB}{\ensuremath{\Phi_{\scriptscriptstyle {\rm B}}}}
\newcommand{\lobs}{LTS\xspace}

\newcommand{\cF}{{\cal F}}

\newcommand{\ee}{e^+e^-}

\newcommand{\FNLL}{\mathcal{F}_{\rm NLL}}
\newcommand{\FNNLL}{\mathcal{F}_{\rm NNLL}}
\newcommand{\RpNLL}{\mathcal{R}'_{\mathrm{NLL}}}

\newcommand{\dFsc}{\ensuremath{\delta \cF_{\rm sc}}}
\newcommand{\dFwa}{\ensuremath{\delta \cF_{\rm wa}}}
\newcommand{\dFrec}{\ensuremath{\delta \cF_{\rm rec}}}
\newcommand{\dFhc}{\ensuremath{\delta \cF_{\rm hc}}}
\newcommand{\dFcorrel}{\ensuremath{\delta \cF_{\rm correl}}}
\newcommand{\dFclust}{\ensuremath{\delta \cF_{\rm clust}}}

\newcommand{\dSdglap}[1]{\ensuremath{ \delta \Sigma^{#1}_{\rm
      \scriptscriptstyle{DGLAP}}}}
\newcommand{\dSframe}[1]{\ensuremath{ \delta \Sigma^{#1}_{\rm \scriptscriptstyle{frame}}}}

\collaborationImg{\includegraphics[width=0.12\textwidth]{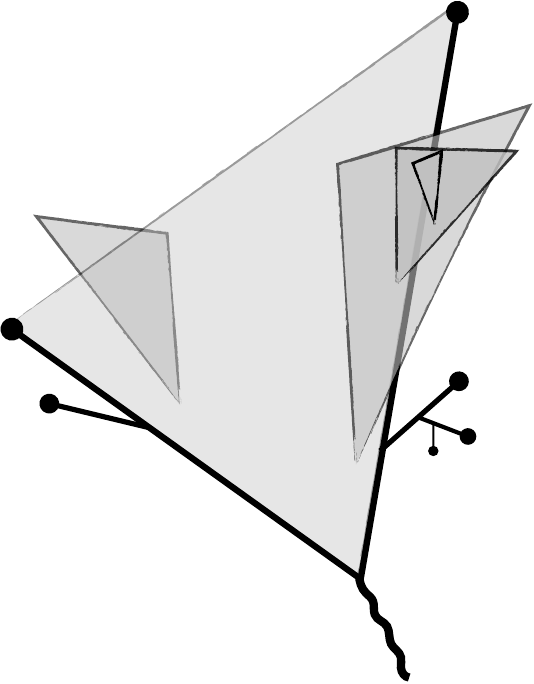}}
\preprint{CERN-TH-2025-213, Nikhef 2025-016}

\title{A new suite of Lund-tree observables to resolve jets}

\author[a]{Melissa van Beekveld,}%
\author[b]{Luca Buonocore,}%
\author[b,c]{Silvia Ferrario Ravasio,}%
\author[b]{Pier Francesco Monni,}%
\author[d]{Alba Soto-Ontoso,}%
\author[e]{Gregory Soyez}%

\affiliation[a]{Nikhef, Theory Group, Science Park 105, 1098 XG, Amsterdam, The Netherlands}
\affiliation[b]{CERN, Theoretical Physics Department, CH-1211 Geneva 23,
Switzerland}
\affiliation[c]{Dipartimento di Fisica, Università di Torino, and INFN, Sezione di Torino,
Via P. Giuria 1, I-10125 Torino, Italy}
\affiliation[d]{Departamento de Física Teórica y del Cosmos, Universidad de Granada, Campus de Fuentenueva, E-18071 Granada, Spain}
\affiliation[e]{IPhT, Universit\'{e} Paris-Saclay, CNRS UMR 3681, CEA Saclay, F-91191 Gif-sur-Yvette, France}

\emailAdd{mbeekvel@nikhef.nl}
\emailAdd{luca.buonocore@cern.ch}
\emailAdd{silvia.ferrarioravasio@unito.it}
\emailAdd{aontoso@ugr.es}
\emailAdd{pier.monni@cern.ch}
\emailAdd{gregory.soyez@ipht.fr}

\abstract{
  We introduce a class of collider observables, named Lund-Tree
  Shapes (\lobs), defined from declustering trees originating
  from the Lund jet plane representation of the QCD radiation pattern
  in multi-jet scattering processes.
  At the differential level, they are continuous global variables akin
  classical event shapes and $n\to n+1$ jet-resolution parameters,
  which probe the geometry and hierarchical structure of the radiation
  in an event. At the integrated, cumulative level, they naturally
  define $n$ jet rates, providing a jet-multiplicity-based
  characterisation of multi-jet final states.
  Their definition applies to scattering processes with any number of
  resolved jets in the final state, as well as to groomed jets. They
  are thus usable as resolution variables in the context of
  higher-order calculations via phase-space slicing, matching
  fixed-order calculations to parton showers, and testing the
  logarithmic accuracy of shower algorithms.
  From a theoretical viewpoint, such observables feature a simple
  all-order structure and are free of non-global logarithmic
  corrections.
  As an initial application, we derive
  next-to-next-to-leading-logarithmic accurate predictions for
  processes with two QCD legs at $e e$, $pp$ and $e p$ colliders, and
  matched predictions to next-to-next-to-leading order for the LHC,
  discussing aspects of collider phenomenology.}


\begin{document}

\maketitle

\section{Introduction}
\label{sec:intro}
Understanding the pattern of QCD radiation in high-energy collisions
is a central goal of the modern collider-physics programme. Hard
scattering events produce jets of collimated particles whose internal
structure encodes fundamental aspects of strong
interactions. Observables such as jet rates, event shapes, and jet
substructure variables have long been used to probe this radiation,
testing the accuracy of perturbative predictions and of parton
showers. Yet, despite their success, many of these observables face
inherent limitations: some are sensitive to non-global logarithms,
others are largely affected by underlying event activity at hadron
colliders, and some do not generalise easily to multi-leg scattering
processes.
To overcome these limitations, in this article we exploit the
Lund-jet-plane~\cite{Andersson:1988gp,Dreyer:2018nbf} representation 
of QCD radiation as a robust way to define new observables that are 
both theoretically tractable and experimentally accessible.

With this technology, we introduce Lund-Tree Shapes (\lobs). For
processes with two emitting legs, these observables were first
introduced (at next-to-leading-logarithmic accuracy, NLL) for the
purpose of testing the logarithmic accuracy of parton showers in
$\lep$~\cite{Dasgupta:2020fwr}, $\had$~\cite{vanBeekveld:2022ukn} and
$\dis$~\cite{vanBeekveld:2023chs} collisions. In this article, we
present a generalisation to processes with any number of resolved
jets, as well as to the case of groomed jets.
These observables serve a dual purpose. At the differential level,
they act as continuous event-shape or jet-resolution-parameter like
variables (cf., e.g.,
Refs.~\cite{Dasgupta:2003iq,Banfi:2010xy,Stagnitto:2025air} and
references therein for an overview), probing the fine-grained geometry
of radiation pattern in multi-jet processes. At the integrated,
cumulative level, the same observables naturally define jet rates,
providing a resolution-based characterisation of multi-jet events and
a natural definition of jet vetoes with improved theoretical
properties.

The advantages of \lobs are manifold. Their definition is general,
applicable to events with any number of resolved jets and across
different collider environments ($\lep$, $\had$, $\dis$).
From a phenomenological viewpoint, they offer a tool to study the
radiation pattern in multi-jet processes, and to characterise
multi-jet events in terms of jet multiplicity when used as jet
vetoes~\cite{Banfi:2012jm,Becher:2013xia,Stewart:2013faa,Becher:2014aya,Banfi:2015pju,Gangal:2014qda,Monni:2019yyr,Gangal:2020qik,Clark:2025riz}.
They can be also used to characterise the composition of underlying
event as well as to study the structure of soft QCD radiation in
multi-jet events.
From a theoretical perspective, they possess simple all-order
perturbative properties, avoiding complications such as non-global
logarithms, and are ideally suited for use as resolution variables in
phase-space
slicing~\cite{Catani:2007vq,Gaunt:2015pea,Boughezal:2015dva,Buonocore:2022mle,Abreu:2022zgo,Buonocore:2023rdw,Fu:2024fgj},
matching fixed-order calculations to parton showers (e.g.,
specifically within the methods of
Refs.~\cite{Hamilton:2012rf,Hamilton:2013fea,Alioli:2013hqa,Alioli:2015toa,Alioli:2021qbf,Monni:2019whf,Monni:2020nks,Mazzitelli:2020jio,Gavardi:2023aco,vanBeekveld:2025lpz})
and testing the logarithmic accuracy of event
generators~\cite{Dasgupta:2018nvj,Dasgupta:2020fwr,Hamilton:2020rcu,Karlberg:2021kwr,Hamilton:2021dyz,vanBeekveld:2022zhl,vanBeekveld:2022ukn,vanBeekveld:2023chs,Forshaw:2020wrq,Nagy:2020dvz,Nagy:2020rmk,Herren:2022jej,Assi:2023rbu,Preuss:2024vyu,Hoche:2024dee,vanBeekveld:2024qxs,vanBeekveld:2024wws,Hoche:2025gsb}.
As an application, we present new calculations at
next-to-next-to-leading logarithmic (NNLL) order for 2-leg processes
in $\lep$, $\had$ and $\dis$ (i.e.~deep inelastic scattering, DIS)
collisions, and provide matched predictions to next-to-next-to-leading
order (NNLO) for the LHC. We also study their sensitivity to
hadronisation and underlying event when used for collider
phenomenology.

The paper is organised as follows. In Sec.~\ref{sec:def} we define the
\lobs observables for $\lep$ (Sec.~\ref{sec:ee}), $\had$
(Sec.~\ref{sec:pp}), $\dis$ (Sec.~\ref{sec:dis}) collisions, and for
groomed jets (Sec.~\ref{sec:groom}).
Their resummation up to NNLL accuracy is detailed in
Sec.~\ref{sec:nnll}, where we also discuss the validation of our
results.
In Sec.~\ref{sec:non-pert} we consider the particular case of
$pp\to Z$ production, match our NNLL predictions to NNLO and inspect
the impact of hadronisation and underlying events on the \lobs.
We conclude in Sec.~\ref{sec:conclusions}.
We also provide several additional details about the resummation,
matching and fixed-order checks of our results in the appendix.

\section{\lobs definition}
\label{sec:def}

Our starting point to define the \lobs is the definition
of the Lund jet plane~\cite{Andersson:1988gp,Dreyer:2018nbf}. We start
with the case of final-state radiation, for instance produced in the
decay of a colour singlet in lepton-lepton collisions, and
then discuss the extension to hadron-hadron and lepton-hadron
collisions.

\subsection{Lepton-lepton collisions}
\label{sec:ee}
We start our discussion by considering the $2$-jet
case, and then generalise the definition to an
arbitrary jet multiplicity $N$ in the final state.
We consider the decay of a colour singlet of invariant mass $M_F\equiv\Q$ and
work in its rest frame. The \lobs are defined as
follows.
We cluster the whole event with the
Cambridge~\cite{Dokshitzer:1997in,Wobisch:1998wt} algorithm that uses
the distance measure (between two proto-jets $i$,$j$)
\begin{align}
\label{eq:distanceCA}
d_{ij} = 2\,(1 - \cos\theta_{ij})\,,
\end{align}
until we are left with two exclusive jets $\{j_L,j_R\}$. 
For each of the two exclusive jets, we apply the following declustering procedure.
\begin{enumerate}
\item We start from the last clustering $\widetilde{ij}\to i, j$,
  where $i$ denotes the most energetic (harder) subjet, i.e.
  $E_i > E_j$. The splitting is characterised by a pair of Lund-plane
  coordinates. These can be chosen to be
  $(\eta^{(\widetilde{ij})}, k_t^{(\widetilde{ij})})$, defined
  as~\footnote{An alternative pair of Lund-plane coordinates that can
    considered is the transverse mass $m_t^{(\widetilde{ij})}$ and
    rapidity $y^{(\widetilde{ij})}$ defined as
  \begin{equation}
    m_t^{(\widetilde{ij})}=E_j \sqrt{1-\beta^2_i\beta^2_j\cos^2\theta_{ij}}\,,\quad
    y^{(\widetilde{ij})}=\frac{1}{2}\ln\frac{1+\beta_i\beta_j
    \cos\theta_{ij}}{1-\beta_i\beta_j \cos\theta_{ij}} \,,
  \end{equation}
  where $\beta_k=|\vec{p}_k|/E_k$. One could also consider defining
  the clusterings using the WTA recombination
  scheme~\cite{Bertolini:2013iqa,Larkoski:2014uqa} in defining the
  above coordinates, or using the $d_{ij}$ variable of
  Ref.~\cite{Alioli:2010xd} as a transverse-momentum measure. For the
  sake of concreteness, we will use the pair defined in
  Eq.~\eqref{eq:kt-eta-ee} in the following. All these variants can be
  calculated at NNLL using the methods discussed in this article.}
  \begin{equation}\label{eq:kt-eta-ee}
    \eta^{(\widetilde{ij})}=-\ln\tan\frac{\theta_{ij}}{2}\,, \quad
     k_t^{(\widetilde{ij})}=E_j |\sin\theta_{ij}|\,.
  \end{equation}

\item Using the above definition~\eqref{eq:kt-eta-ee}, we calculate
  the quantity 
  \begin{equation}
    \label{eq:vbee}
    v_b^{(\widetilde{ij})}=\frac{k_t^{(\widetilde{ij})}}{\Q}
    e^{-b |\eta^{(\widetilde{ij})}|}\,,
  \end{equation}
  where $b> -1$ as required for infrared safety. We note that
  additional Lund-tree variables, such as the azimuthal $\psi$
  angle~\cite{Dreyer:2018nbf,Dasgupta:2020fwr,Karlberg:2021kwr,Hamilton:2021dyz}
  could be also used to define additional variants of \lobs.
  
\item We then go back to step 1 of this procedure following the harder subjet $i$,
  i.e.~we only consider the set of \textit{primary} declusterings
  $d_p^{(\widetilde{ij})}$ for each of the two jets $\{j_L,j_R\}$.
  These declusterings are defined by the set of the Lund coordinates
  $d_{p,L/R}^{(\widetilde{ij})}\equiv \{(\eta^{(\widetilde{ij})},
  k_t^{(\widetilde{ij})})\}_{L/R}$, measured using all primary declusterings.
\end{enumerate}
We then define the set of all declusterings in the event as
\begin{equation}
  {\cal I}_{N=2}\equiv d_{p,L}^{(\widetilde{ij})}\cup
  d_{p,R}^{(\widetilde{ij})}\,,
\end{equation}
where $N=2$ denotes the number of resolved jets.
We can now define the two \lobs $\Mb{N=2}$ and $\Sb{N=2}$
as
\begin{equation}\label{eq:LES-ee-2}
\Mb{2} = \max_{j\in {\cal I}_2} v_b^{j}\,,\quad \Sb{2} =
\sum_{j\in {\cal I}_2} v_b^{j}\,.
\end{equation}
The extension to $N=3$ jets can be done by first identifying the
element $j_k$ of ${\cal I}_2$ with the largest
$v_b^{(\widetilde{ij})}$.\footnote{One could also consider using other
  metrics in defining the \textit{leading} third jet in the event,
  like for instance its transverse momentum with respect to a suitably chosen
  axis.}
We then remove this from the initial list ${\cal I}_2$, and instead
add back all the primary declusterings $d_p^{(k)}$ of $j_k$, hence
defining the new input list ${\cal I}_3$
\begin{equation}
{\cal I}_3\equiv\{{\cal I}_2\setminus j_k\}\cup d_p^{(k)}\,.
\end{equation}
This procedure can be recursively extended for an arbitrary jet
multiplicity $N$. 
That is, to build $\mathcal{I}_{N} $ from
$\mathcal{I}_{N-1}$, we identify the $\widetilde{ij}\to i,j$ clustering in
$\mathcal{I}_{N-1}$ (with $E_i>E_j$) corresponding to the largest
$v_b^{(\widetilde{ij})}$, remove it from the list, and add instead all the
primary declusterings from jet $j$. 
With this procedure, we can define the $N$-jet generalisation of the $2$-jet \lobs of Eq.~\eqref{eq:LES-ee-2} as
\begin{equation}\label{eq:LES-ee-N}
\Mb{N} = \max_{j\in {\cal I}_N} v_b^{j}\,,\quad \Sb{N} =
\sum_{j\in {\cal I}_N}v_b^{j}\,.
\end{equation}
We stress that the collinear safety of the $\Mb{N}$ observable is
guaranteed by the use of a clustering procedure.

In the $N=2$ case, these observables were first introduced at NLL in
Ref.~\cite{Dasgupta:2020fwr} to test the logarithmic accuracy of
parton showers in $\lep$ collisions.
The $\Mb{2}$ observable with $b=0$ is close in spirit (up to a
wide-angle soft correction starting at NNLL) to the $y_{23}$
resolution in the Cambridge algorithm~\cite{Dokshitzer:1997in}, which
is known up to NNLL~\cite{Banfi:2001bz,Banfi:2016zlc}.
Similarly, the $\Sb{2}$ observables are related to
thrust~\cite{Brandt:1964sa,Farhi:1977sg,Becher:2008cf} ($b=1$) and
$FC_x$~\cite{Banfi:2004yd,Banfi:2018mcq} and
angularities~\cite{Berger:2003iw,Bell:2018gce,Banfi:2018mcq} ($b=1-x$)
up to the NLL order, while they differ from these observables starting
at NNLL.

\subsection{Hadron-hadron collisions}
\label{sec:pp}
\begin{figure}[tb]
 	\centering
 	\includegraphics[trim={10cm 0cm 10cm 5cm},width=0.8\textwidth,page=1]{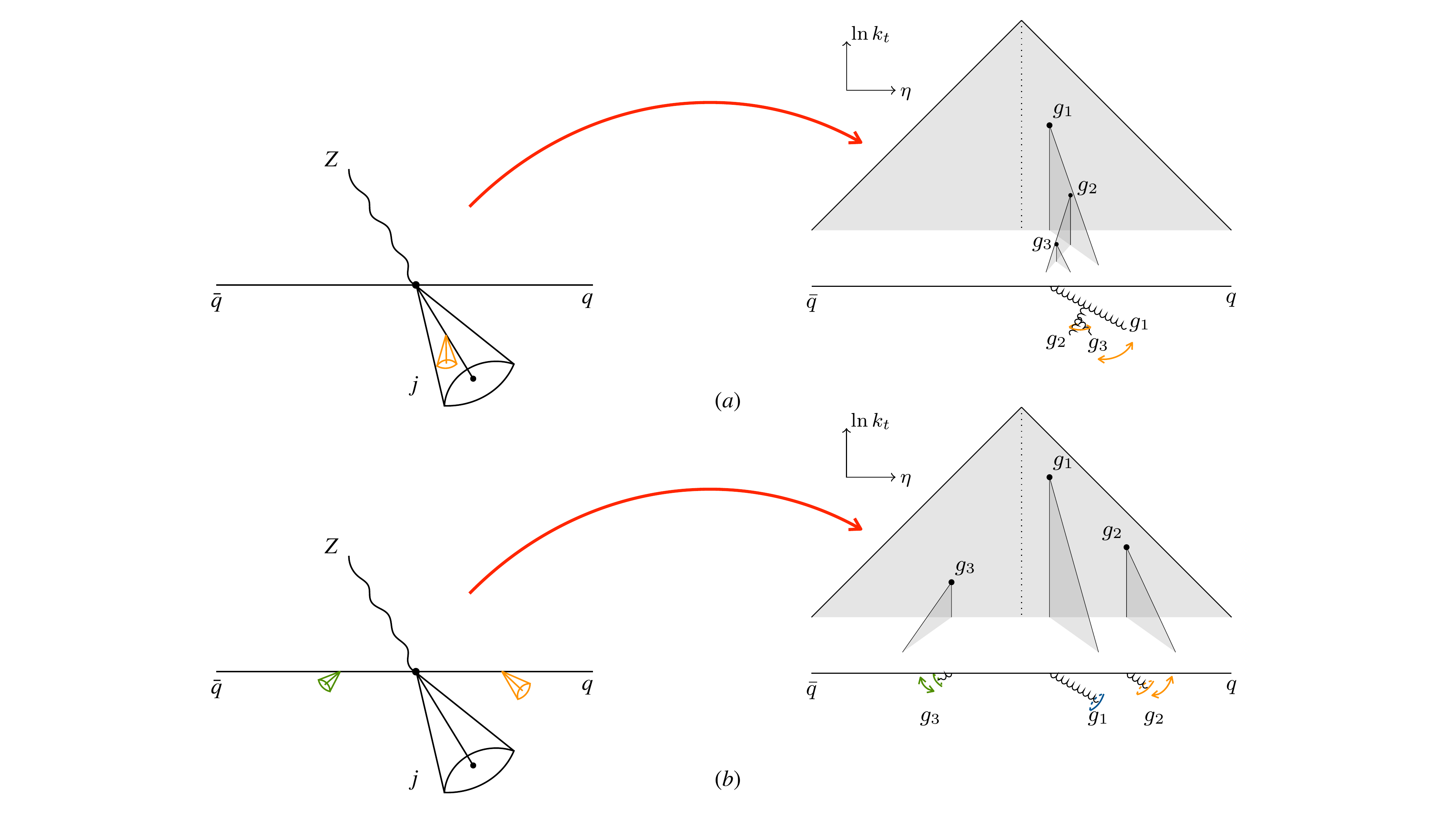}
 	\caption{An illustration of \lobs for two $Z$+jet events: $(a)$
          In this event there are no primary declusterings from the
          beam, while the jet contains a primary declustering
          resulting from the recombination of emissions $g_2$ and
          $g_3$; $(b)$ In this event the jet contains no primary
          declusterings, while there are two primary declusterings
          from the beam corresponding to the (primary) emissions $g_2$
          and $g_3$. The two-sided arrows indicate the Lund measures
          that enter the definition of the \lobs.}
 	\label{fig:correl-zj}
 \end{figure}

We now extend the above definition to the hadron-collider case. We
start by considering the process $pp\to F$, where $F$ denotes a colour-singlet
system of invariant mass $\Q$.
We initially work in the laboratory frame, and then discuss
alternative reference frames in Sec.~\ref{sec:beam-frame}.
We start by considering the case of $N=0$ resolved jets in the final
state~\cite{vanBeekveld:2022ukn}. That is obtained by clustering the
full event with the Cambridge/Aachen (C/A) algorithm~\cite{Dokshitzer:1997in} with radius
$R$. These C/A jets define the input set for the definition of the
\lobs. We denote this by
\begin{equation}
{\cal I}_{N=0} \equiv \{j_1,j_2,\dots,j_m\}\,.
\end{equation}
A natural pair of Lund coordinates for each jet, and the one we will use,
corresponds to the rapidity and transverse momentum of the jet with respect to the beam,
i.e.~$(y^{i},p_{t,i})$. 
Analogously to the lepton colliders case,
using these coordinates, we calculate the quantity
\begin{equation}
    v_b^{j_i}=p_{t,i}
    e^{-b |y^{i}|}\,.
\end{equation}
We then define the \lobs in $pp\to F$ with $N=0$ as
\begin{equation}\label{eq:LES-pp-0}
\Mb{0} = \max_{j\in {\cal I}_0} v_b^{j}\,,\quad \Sb{0} =
\sum_{ j \in {\cal I}_0} v_b^{j}\,.
\end{equation}
This procedure can be extended to the $N=1$ case (e.g.~measured in
$pp\to F +$jet) by removing from the input list ${\cal I}_0$ the jet
with the largest $v_b^{j_i}$, that we denote by $j_k$.
We then add the list of primary declusterings $d_p^{(k)}$
of $j_k$, namely
\begin{equation}
{\cal I}_1\equiv\{{\cal I}_0\setminus j_k\}\cup d_p^{(k)}\,.
\end{equation}
The list of declusterings $d_p^{(k)}$ of the removed jet $j_k$ from ${\cal I}_0$ is constructed as follows.
\begin{enumerate}
\item Starting from the last C/A clustering step
  $\widetilde{ij}\to i, j$ in $j_k$, we denote with $i$ the most energetic
  subjet, i.e.\ $p_{t,i}>p_{t,j}$. 
  We associate to this splitting a pair  of Lund-plane coordinates
  $(y^{(\widetilde{ij})}, k_t^{(\widetilde{ij})})$ on the Lund-plane
  leaf created by the jet $j_k$. 
  These are a measure of the transverse momentum and rapidity of the
  softer subjet $j$ with respect to the harder subjet $i$. They can be
  defined as~\footnote{An alternative definition involves the rapidity
    of $i$ w.r.t.~$j$ given by
    $y^{(\widetilde{ij})}=\frac{1}{2}\ln\frac{E_j+\vec{p}_j\cdot
      \hat{n}_i}{E_j-\vec{p}_j\cdot \hat{n}_i}$ where
    $\hat{n}_i\equiv \vec{p}_i/|\vec{p}_i|$. One could also define the
    Lund transverse momentum as
    $k_t^{(\widetilde{ij})}=p_{t,j}\,\Delta R_{ij}$, as done in the
    original Lund-jet-plane paper~\cite{Dreyer:2018nbf}.}
  \begin{equation}
   y^{(\widetilde{ij})}=|y^k|-\ln \frac{\Delta R_{ij}}{R}\,, \quad
    k_t^{(\widetilde{ij})}=p_{t,j}\frac{\Delta R_{ij}}{R}\,,\quad
     \Delta R_{ij}\equiv\sqrt{\Delta
      y_{ij}^2+\Delta\phi_{ij}^2}\,,
  \end{equation}
  where $y_k$ is the rapidity of the jet $j_k$. The $|y^k|$ term
  ensures that the rapidity of a clustering close to the jet boundary
  approaches that of the jet itself. 
  This aspect is ultimately important to guarantee the absence of
  non-global logarithms in the \lobs (see
  Section~\ref{app:NGL}).
  Note that  $ |y^{(\widetilde{ij})}|\gg 1$ when the two subjets are collinear.

\item On this splitting we calculate
  \begin{equation}
    v_b^{(\widetilde{ij})}=k_t^{(\widetilde{ij})}
    e^{-b |y^{(\widetilde{ij})}|}\,.
  \end{equation}

  \item Go back to step 1 of this procedure following the harder subjet $i$. 
\end{enumerate}
The resulting list of primary declusterings defines $d_p^{(k)}$. Two
example configurations are illustrated in
Fig.~\ref{fig:correl-zj}.

This procedure can be iterated to the case with $N$ jets. For
instance, for $N=2$, the input list ${\cal I}_2$ is obtained by
removing the element of ${\cal I}_1$ with the largest $v_b$ (either
one of the $v_b^{(\widetilde{ij})}$ or one of the $v_b^{j_i}$) and
adding its list of primary declusterings.
The general \lobs are then defined by
\begin{equation}\label{eq:LES-pp-N}
\Mb{N} = \max_{j\in {\cal I}_N} v_b^{j}\,,\quad \Sb{N} =
\sum_{ j \in {\cal I}_N} v_b^{j}\,.
\end{equation}

So far, we have defined the \lobs in the laboratory frame. In some
cases it can be interesting to consider the definition in a different
reference frame. An example is the process $pp\to F$ ($N=0$), where
one may want to work in a frame where the colour singlet $F$ has zero
rapidity, $y_F=0$, that is the partonic centre-of-mass (c.o.m.) frame. The
corresponding definition of the \lobs~\eqref{eq:LES-pp-0} is then
expressed in terms of the quantity
\begin{equation}\label{eq:partonicFrameV}
    v_b^{j_i} = p_{t,i}
    e^{-b |y^{i}-y_F|}\,.
\end{equation}
This will affect the resummation at NNLL. We will come back to this in Sec.~\ref{sec:beam-frame}.

In the $N=0$ case, these observables were first introduced at NLL in
Ref.~\cite{vanBeekveld:2022ukn} to test the logarithmic accuracy of
parton showers in $\had$ collisions.
For $N=0$, observables of the kind written in Eq.~\eqref{eq:LES-pp-N} have been considered before in the
literature.
Observables related to $\Mb{0}$ have been previously studied in the
context of jet vetoes in
Refs.~\cite{Banfi:2012jm,Becher:2013xia,Stewart:2013faa,Becher:2014aya,Banfi:2015pju}
($b=0$) and Refs.~\cite{Gangal:2014qda,Gangal:2020qik,Clark:2025riz}
($b=1$). These types of jet vetoes agree with $\Mb{0}$ for $b=0$ up
to NNLL, but they differ beyond this order due to the details of the
jet algorithm (anti-$k_t$ v. C/A) as well as at fixed order starting
from NNLO by subleading-power corrections. Instead, for $b=1$, they
differ starting at NNLL from the rapidity-dependent jet vetoes
computed and studied in Refs.~\cite{Gangal:2020qik,Clark:2025riz}, due
to their use of the transverse mass instead of transverse momentum in
the observable definition. The latter would correspond to a different
variant of \lobs, not considered here.
At $\had$ colliders, the \lobs constitute a \textit{global}
generalisation of jet vetoes to the multi-jet case.
The $\Sb{0}$ observables are also related to beam
thrust~\cite{Stewart:2010tn} ($b=1$) and
$E_T$~\cite{Halzen:1982cb,Davies:1983di,Altarelli:1986zk,Papaefstathiou:2010bw,Grazzini:2014uha}
($b=0$) up to NLL, but they differ from these at higher orders.

We note that, in contrast with the $e^+e^-$ case where the total
energy is known, in $pp$ we defined the \lobs to have mass
dimension one. While in the $pp\to F$ case one could normalise the
transverse momentum by $\Q$, one has some freedom in defining the hard
scale in the $N$-jet case, where multiple scales in the scattering
process are present. 
For the sake of simplicity, in the $N=0$ resummed calculations given
in Sec.~\ref{sec:nnll} we will normalise $\Mb{0}$ and $\Sb{0}$ by $\Q$
so that they are dimensionless. 
That is, for the rest of this article, we define the dimensionless
counterpart of the observables in the
$\had$ case as
\begin{equation}
	v_b^{j_i} = \frac{p_{t,i}}{Q}
	e^{-b |y^{i}|}\,, \quad \Mb{0} = \max_{j\in {\cal I}_0} v_b^{j}\,,\quad \Sb{0} =
	\sum_{ j \in {\cal I}_0} v_b^{j}\,,
\end{equation}
with $\Q$ being the invariant mass of the colour singlet. An analogous
definition holds for the partonic c.o.m.~frame variant.

\subsection{Lepton-hadron collisions}
\label{sec:dis}
For DIS, it is convenient to start by considering a single
final-state jet and then generalise the definition to arbitrary jet
multiplicities.

We work in the Breit frame~\cite{Webber:1993bm}, defined such that the
incoming parton has energy exactly equal to the total energy of the
final-state partons, and the momentum transfer
\begin{equation}
q^\mu_{\rm dis} = q^\mu_{\rm out}-q^\mu_{\rm in}
\end{equation}
is space-like. In this frame, the incoming parton is conventionally chosen to be anti-aligned with the $z$ axis.

The event is then clustered using the Cambridge algorithm, in analogy with
the lepton-collider case (see Sec.~\ref{sec:ee}). Alongside the usual
pairwise distance
\begin{equation}
d_{ij} = 2\,(1-\cos\theta_{ij})
\end{equation}
between proto-jets, we also consider the distance to the beam,
\begin{equation}
d_{i{\rm B}} = 2\,(1-\cos\theta_{i{\rm B}}),
\end{equation}
and proceed with clustering the event into jets in the standard manner.

All resulting jets correspond to primary emissions from the initial
state, except for the one originating from the fragmentation of the
original final-state parton, which we denote as $j_k$. This jet is
identified as the one carrying the largest light-cone component
anti-parallel to the incoming parton in the Breit
frame. Operationally, it is the jet whose momentum maximises the
scalar product with the proton momentum.

We then decluster this final-state jet following the same procedure as
in the lepton-collider case, denoting the list of its primary
declusterings by $d_p^{(k)}$. If ${\cal I}_0$ denotes the list of
primary declusterings from initial-state radiation, the complete list
of primary declusterings for DIS is
\begin{equation}
\mathcal{I}_1 = \left\{ {\cal I}_0 \setminus j_k \right\} \cup d_p^{(k)}.
\end{equation}

As in the lepton- and hadron-collider cases, there is some freedom in
defining the Lund coordinates that enter the definition of the
\lobs. In the following, we adopt the coordinates in
Eq.~\eqref{eq:kt-eta-ee}, replacing the angle $\theta_{ij}$ with
$\theta_{i{\rm B}}$ for and initial-state declusterings.

The \lobs are then defined as in Eq.~\eqref{eq:vbee} with
\begin{equation}
\Q = \sqrt{-q_{\rm dis}^2}.
\end{equation}
The $N=2$ definition is obtained from the $N=1$ case by replacing, in
${\cal I}_1$, the declustering with the largest $v_b$ (either an
initial- or final-state declustering) with its own primary
declusterings. Higher jet multiplicities are obtained recursively
following the same procedure.

In the $N=1$ case, these observables were first introduced at NLL in
Ref.~\cite{vanBeekveld:2023chs} to test the logarithmic accuracy of
parton showers in $\dis$ collisions.
For $N=1$, observables of the $S^{(1)}_{b}$ type with $b=1$ are
similar to $1$-jettiness or one of the variants of DIS
thrust~\cite{Antonelli:1999kx,Kang:2013nha,Kang:2012zr,Ee:2025scz,Cao:2024ota}
up to NLL, but not beyond.

\subsection{Groomed jets}
\label{sec:groom}
The \lobs variables also serve as jet substructure observables when,
for instance, measured on jets groomed with the
Soft-Drop~\cite{Larkoski:2014wba} or mMDT~\cite{Dasgupta:2013ihk}
grooming procedure.
Their definition now applies to the Lund declusterings within an
energetic jet that passes the grooming condition. This variant of the
\lobs features a milder sensitivity to hadronisation and underlying
event at hadron colliders, enabling the study of the structure of QCD
jets.
To define them, we consider a jet $J$ (this can be, for instance, an
anti-$k_t$ jet at $\had$ colliders or a hemisphere defined with
respect to the thrust/WTA axis at $\lep$ colliders), and groom its
content with the Soft-Drop/mMDT grooming
procedure,
with grooming parameter $z_{\rm cut}$ (and $\beta$ in the Soft-Drop
case). We now apply the Lund-jet declustering~\cite{Dreyer:2018nbf} to
the resulting groomed jet and denote the list of its \textit{primary}
Lund declusterings by
\begin{equation}
{\cal I}_{J_{\rm SD}}\equiv d_{p,\,{\rm SD}}^{(J)}\,,
\end{equation}
constructed as outlined in the previous sections depending on the type
of collider environment considered. Each declustering is characterised
by the usual set of Lund-plane coordinates
$(k_t^{(\widetilde{ij})},\,y^{(\widetilde{ij})})$. For each of these
declusterings we calculate
$v_{b,\,{\rm SD}}^{(\widetilde{ij})}=k_t^{(\widetilde{ij})} e^{-b
  |y^{(\widetilde{ij})}|}$.
The groomed \lobs (g\lobs) are then defined by
\begin{equation}\label{eq:LES-SD}
\MbSD = \max_{j\in {\cal I}_{J_{\rm SD}}} v_{b,\,{\rm SD}}^{j}\,,\quad \SbSD =
\sum_{ j \in {\cal I}_{J_{\rm SD}}} v_{b,\,{\rm SD}}^{j}\,.
\end{equation}
They can be normalised to hard scale of the jet, i.e.~its pre-grooming
transverse momentum ($\had$ colliders) or energy ($\lep$ colliders).
The resummation structure of these observables for $\beta=0$, in the limit
$v\ll z_{\rm cut}$, can be studied, for instance, with the methods of
Refs.~\cite{Frye:2016aiz,Marzani:2017kqd,Marzani:2017mva,Kang:2018vgn,Kardos:2020gty,Anderle:2020mxj,Dasgupta:2021hbh,Dasgupta:2022fim,vanBeekveld:2023lsa}.
For $\beta>0$ additional complications related to non-global
logarithms may arise.
An alternative definition of the groomed \lobs would directly use the
Soft-Drop/mMDT grooming procedure at the level of the declusterings
that enter the \lobs, that is similar in spirit to using a recursive
soft-drop procedure~\cite{Dreyer:2018tjj}.
We do not pursue the groomed versions further here, and leave their detailed
study to future work.

\section{Resummation to NNLL for the two-legs case}
\label{sec:nnll}

In this section we discuss the resummation properties of the \lobs,
and perform explicit calculations up to the NNLL order in the case of
processes with $2$ radiating legs at the lowest order. We will start
with a discussion of the lepton-collider case, and then illustrate the
hadron-collider and DIS cases.

\subsection{$N=2$ case at lepton colliders}

The \lobs at lepton colliders have a very simple resummation
structure. We adopt the resummation formalism of
Refs.~\cite{Banfi:2001bz,Banfi:2014sua,Banfi:2016zlc,Banfi:2018mcq},
in which one expresses the (non-normalised) cumulative cross section for
a given observable $v\in \{\Mb{2},\Sb{2}\}$ as follows
\begin{equation}\label{eq:master}
\Sigma^{\lep}(v)\equiv \int_0^vd v' \frac{d\sigma}{d v'} =
\Sigma^{\lep}_{S}(v) {\cal F}^{\lep}(v)\,.
\end{equation}
The explicit expressions of the ingredients of the above equation
depend on whether we are considering a Born final state consisting of
two quark jets ($Z$ decay) or two gluon jets ($H$ decay). To keep the
notation light, we suppress the dependence on the flavour of the final
state in this section. However we provide results for both quark and
gluon jets in Appendix~\ref{app:ingredients}.

The factor $\Sigma^{\lep}_S(v)$ denotes the cumulative distribution
for a simplified version of the observable. Its definition proceeds by
classifying the multi-parton squared matrix elements into correlated
clusters of emissions~\cite{Banfi:2018mcq}, namely the building blocks
of the multi-parton squared amplitudes that vanish in the limit where
emissions are widely separated in rapidity.
In practice, this classification can be implemented recursively. For
concreteness, let us consider the soft limit. In this regime, matrix
elements factorise into independent eikonal factors whenever
emissions, or subsets of emissions, are widely separated in rapidity.
With two emissions, one can thus identify an uncorrelated contribution
obtained in the limit where the two emissions are widely
separated. The correlated contribution is then defined as the
remainder, which by construction only has support in the region where
the two emissions have comparable rapidities.
This recursive definition extends naturally to higher
multiplicities~\cite{Banfi:2018mcq}.\footnote{For example, for three
  emissions, the procedure generalises as follows. One first subtracts
  the fully uncorrelated contribution, corresponding to three widely
  separated emissions, which factorises into a product of three
  independent eikonal factors. Next, one removes the three
  configurations in which a pair of emissions is widely separated from
  the third, which factorise into a single emission and a correlated
  pair. The remaining term then represents the contribution from three
  correlated emissions.}
An analogous organisation can be carried out in the collinear limit.
One then calculates the quantity
\begin{equation}\label{eq:vsimple}
v_b(k)=\frac{k_t}{\Q}
    e^{-b |y|}\,,
\end{equation}
for each of the correlated clusters, and defines the simplified
observable to be the largest of these.

This definition leads to a remarkably simple resummation structure that, at
NNLL order, can be cast as
\begin{equation}\label{eq:sigmaS_ee}
\Sigma^{\lep}_S(v) =\sigma^{(0)} \left(1+\frac{\alpha_s(\mur)}{2\pi} H^{(1)} +
  \sum_{\ell=1}^2 \frac{\alpha_s(\mur)}{2\pi}\frac{1+b}{1+b-2\lambda}C_{{\rm hc},\,\ell}^{(1)}\right) \exp\{-\mathcal{R}_{\rm NNLL}(v) \}\,,
\end{equation}
where $\sigma^{(0)}$ is the Born cross-section and
$\lambda=\alpha_s(\mur)\beta_0L$, with $L$ being the large logarithms
to resum (cf. Eq.~\eqref{eq:defs}).
The constant $H^{(1)}$ ($C^{(1)}_{\rm hc}$) stems from the hard
(hard-collinear) region.
The sums over $\ell$ reflect the presence of two legs in the process.
All constants as a function of the parameter $b$ are given in
Appendix~\ref{app:constants}.
The NNLL Sudakov radiator $\mathcal{R}_{\rm NNLL}(v)$, which is a function of the parameter $b$, can be written as a sum of contributions $R_\ell(v)$ from each emitting leg $\ell$, which can be in turn decomposed into a soft and collinear term
\begin{equation}\label{eq:sudakov}
\mathcal{R}_{\rm NNLL}(v) =\sum_{\ell=1}^2 R_{\ell}(v)= \sum_{\ell=1}^2 \left[ R^{\ell}_{\text{s}}(v) +  R^{\ell}_{\text{hc}}(v) \right]\,.
\end{equation}
At NNLL accuracy, the soft radiator $R_s^{\ell}$ reads
\begin{align}\label{eq:sudakov-s}
  R_{\rm s}^{\ell}(v)=& -\frac{\lambda}{\as \beta_0} g_1^{\ell}(\lambda) -  g_2^{\ell}(\lambda) -\frac{\as}{\pi} \left(g_3^{\ell}(\lambda) + \delta g_3^{\ell}(\lambda)\right)\,,
\end{align}
and the hard-collinear radiator reads
\begin{align}
  R^\ell_{\rm hc}(v) = - h_2^{\ell}(\lambda) -\frac{\as}{\pi} h_3^{\ell}(\lambda)\,.
\end{align}
The functions $g^\ell_i$, $\delta g_3^\ell$ and $h_i^\ell$ are
provided in Appendix~\ref{app:ingredients}.
An important aspect of the definition of the simplified observable is
that it agrees with the full observable at LL.

Let us now consider the second factor in the r.h.s.\ of
Eq.~\eqref{eq:master}, the \textit{transfer function} ${\cal
  F}^{\lep}(v)$. As the name suggests, this accounts for the difference
between the full observable that is being resummed, in this case
either $\Mb{2}$ or $\Sb{2}$, and the simplified observable encoded in
$\Sigma^{\lep}_S$.
By definition, it is given by
\begin{equation}\label{eq:F-def}
{\cal F}^{\lep}(v) \equiv \frac{\Sigma^{\lep}(v)}{\Sigma^{\lep}_S(v)}\,.
\end{equation}
This object has the following key properties. Firstly, since by
definition $\Sigma_S(v)$ contains all the LL corrections, the transfer
function starts at NLL order. Secondly, the transfer function is
finite in $D=4$ space-time dimensions, which crucially makes it
amenable to a numerical computation. To make this fact manifest, it is
customary to introduce a resolution parameter $\delta$ and recast
Eq.~\eqref{eq:F-def} as
\begin{equation}\label{eq:F-def-delta}
{\cal F}^{\lep}(v) =\lim_{\delta\to 0}
\frac{\Sigma^{\lep}(v)}{\Sigma^{\lep}_S(\delta v)} \frac{\Sigma^{\lep}_S(\delta v)}{\Sigma^{\lep}_S(v)} \,,
\end{equation}
which can be directly used for practical calculations.
The fact that Eq.~\eqref{eq:F-def-delta} is finite in the limit
$\delta\to 0$ stems from the recursive IRC (rIRC)
safety~\cite{Banfi:2004yd} of \lobs: emissions that contribute to
$\Sigma^{\lep}(v)$ with a $v_b \leq \delta v$ can be neglected in the
evaluation of the observable.
Therefore, their contribution to $\Sigma^{\lep}(v)$ factorises from
the one of resolved emissions (i.e.\ with $v_b > \delta v$) and
cancels against the denominator $\Sigma^{\lep}_S(\delta v)$. The
residual dependence on $\delta$ will now balance that of the ratio
$\Sigma^{\lep}_S(\delta v)/\Sigma_S(v)$, leading to a result that is
finite in the $\delta\to 0$ limit.
Following
Refs.~\cite{Banfi:2014sua,Banfi:2016zlc,Banfi:2018mcq,Arpino:2019ozn},
at NNLL these corrections can be parameterised as follows
\begin{subequations}\label{eq:F-ee}
\begin{align}
{\cal F}^{\lep}(v) &\,= {\cal F}^{\lep}_{\rm NLL}(v) +
                     \frac{\alpha_s(\mur)}{\pi}\delta {\cal F}^{\lep}_{\rm NNLL}(v)\,,\\
\delta {\cal F}^{\lep}_{\rm NNLL}(v) &\,=\delta{\cal F}^{\lep}_{\rm sc}(v)+\delta{\cal
    F}^{\lep}_{\rm wa}(v) +\delta{\cal F}^{\lep}_{\rm rec}(v)+\delta{\cal F}^{\lep}_{\rm
    hc}(v)+ \delta{\cal F}^{\lep}_{\rm correl}(v)+\delta{\cal F}^{\lep}_{\rm
    clust}(v)\,.
\end{align}
\end{subequations}

\begin{figure}[t]
	\centering
	\includegraphics[width=0.6\textwidth]{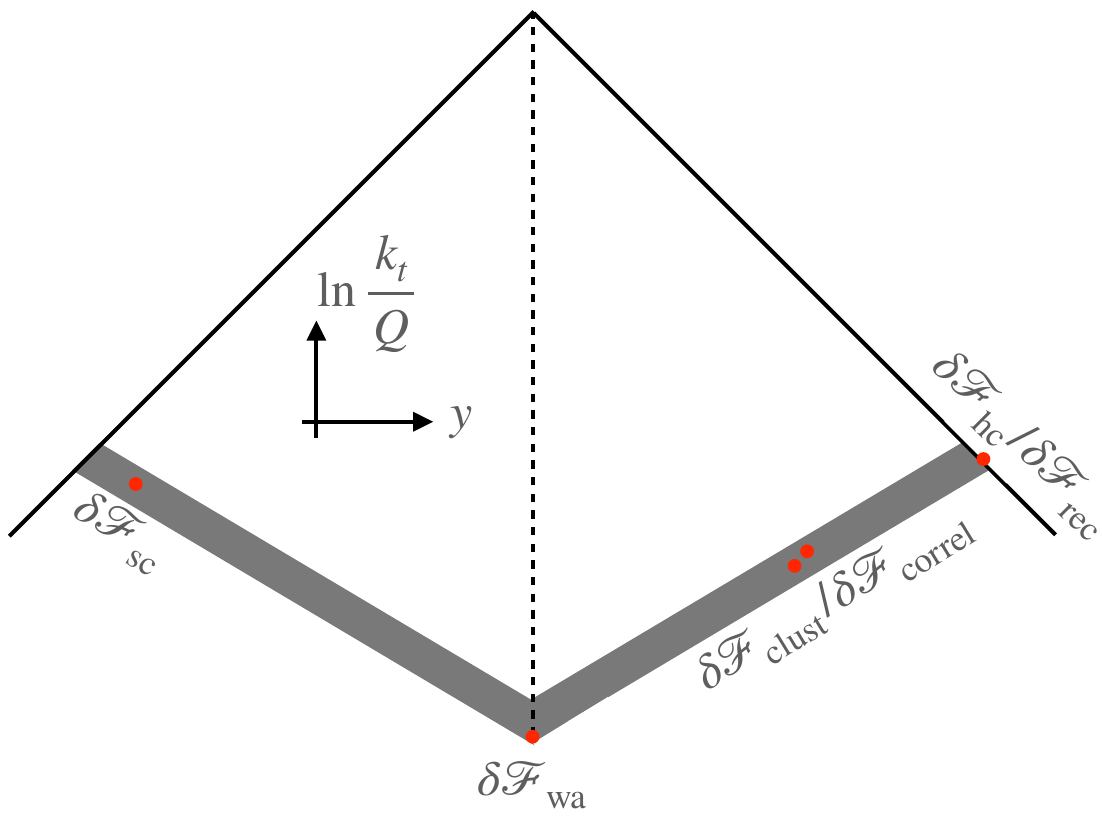}
	\caption{Location of NNLL corrections to the transfer function
          in the Lund plane of the Born $q\bar{q}$ system.}
	\label{fig:LP}
\end{figure}

The structure of Eq.~\eqref{eq:F-ee} can be understood with a simple
power counting argument, that can be explained with the help of the
primary Lund plane in Fig.~\ref{fig:LP}.

At NLL order, ${\cal F}^{\lep}_{\rm NLL}(v)$ is described by an
ensemble of multiple soft \textit{and} collinear emissions widely separated in
rapidity, each having a value $v_i$ of the observable, with
$\delta v\leq v_i \leq v$.
This is represented by the shaded region adjacent to the observable
boundary in Fig.~\ref{fig:LP}, which is populated by such emissions.

Beyond NLL, ${\cal F}^{\lep} (v)$ must encode corrections in regions of the
Lund plane that are parameterically suppressed by ${\cal O}(\alpha_s)$
(counting $L\sim 1/\alpha_s$) with respect to the NLL limit. Each of the
terms in Eq.~\eqref{eq:F-ee} encodes one of these corrections in
different kinematic limits (cf. discussion in
Refs.~\cite{Banfi:2014sua,Banfi:2016zlc,Banfi:2018mcq} for more details).
In short, these are:
\begin{itemize}
\item An extra pair of soft-collinear emissions becoming close in
  rapidity. This correction is encoded into the
  $\delta{\cal F}^{\lep}_{\rm correl}(v)$ (involving the correlated
  part of the double-soft squared amplitude) and
  $\delta{\cal F}^{\lep}_{\rm clust}(v)$ (involving the uncorrelated,
  independent part of the double-soft squared amplitude).
  This contribution starts with terms of order $\alpha_s^2L$.
\item A single extra soft emission at wide angle w.r.t.\ the event
  direction, encoded in the correction $\delta{\cal F}^{\lep}_{\rm
    wa}(v)$. This correction encodes the difference between the
  observable and the simple observable for a soft gluon radiated with
  a wide angle w.r.t.\ its emitter. 
  This contribution starts at order $\alpha_s$.
\item A single extra hard-collinear emission, giving rise to the
  corrections $\delta{\cal F}^{\lep}_{\rm rec}(v)$ and
  $\delta{\cal F}^{\lep}_{\rm hc}(v)$.
  The former, $\delta{\cal F}^{\lep}_{\rm rec}(v)$ is the recoil
  correction which encodes the difference between the observable and
  the simple observable in the presence of a hard-collinear splitting
  kinematics.
  The latter, $\delta{\cal F}^{\lep}_{\rm hc}(v)$, accounts for the
  correction to the matrix element.
\item A single extra soft-collinear emission, with a higher-order
  kernel containing running-coupling terms in the CMW
  scheme~\cite{Catani:1990rr,Banfi:2018mcq,Catani:2019rvy} as well as
  with the correct rapidity bound which was approximated at
  NLL.\footnote{At NLL one expands the rapidity bound for each
    emission about the hard collinear scale, namely
    $|\eta| \leq \ln\frac{\Q}{k_t}\simeq
    \frac{1}{1+b}\ln\frac{1}{v}$.} This correction defines
  $\delta{\cal F}^{\lep}_{\rm sc}(v)$.
\end{itemize}
All the ingredients are provided in Appendix~\ref{app:ingredients}.
In our numerical implementation, we expand out the product
$\Sigma_S \, {\cal F}_{\rm NNLL}$ in Eq.~\eqref{eq:master} so that
$\delta{\cal F}_{\rm NNLL}$ multiplies only the exponential of the
Sudakov radiator $\exp\{-\mathcal{R}_{\rm NNLL}(v)\}$, i.e.\ without
the $\mathcal{O}(\alpha_s)$ constant pre-factor in
Eq.~\eqref{eq:sigmaS_ee}. This throws away subleading N$^3$LL terms.

\subsection{$N=0$ case at hadron colliders}
We will first consider the observable defined in the partonic
c.o.m.~frame (cf. Eq.~\eqref{eq:partonicFrameV} and discussion
thereof) where the colour singlet has zero rapidity. We will then
discuss the laboratory-frame definition of the \lobs in
Sec.~\ref{sec:beam-frame}.
In the hadron-collider case, the structure of the NNLL resummation for
the cumulative distribution $\Sigma ^{\had} (v)$ has to be modified
with respect to the lepton-collider form given in Eq.~\eqref{eq:master}.
A first difference with the lepton-collider case is due to the precise
definition of the observable, and specifically to the different jet
algorithm and the definitions of transverse momentum and rapidity
adopted, as discussed in Sec.~\ref{sec:pp}. This will modify the
correlated and clustering corrections, whose expressions are reported
in Appendix~\ref{app:ingredients}.
A second key difference is due to the presence of initial-state
radiation (ISR) that modifies the structure of the resummation of
collinear logarithms, both in the simple observable $\Sigma_S(v)$, as
well as in the form of the transfer function ${\cal F}$.

To gain intuition, it is instructive to consider the NNLL transfer
function defined in Eq.~\eqref{eq:F-ee} in the final-state radiation
(FSR) case. Among the six NNLL corrections to ${\cal F}^{\lep}$, four
originate from soft dynamics. In the initial-state radiation (ISR)
case, these soft corrections do not modify the longitudinal momentum
fractions of the incoming partons, and hence their functional form
remains identical to that of the FSR case (up to modifications arising
from the differences between observables, as discussed above).

In contrast, the hard-collinear and recoil corrections account for the
effects of hard-collinear splittings in the final state. The
hard-collinear correction describes the modification of the matrix
element due to the emission of hard-collinear radiation, while
treating the corresponding kinematics as soft. The recoil correction,
on the other hand, captures the impact of the recoil induced by the
hard-collinear emission on the observable.

In the ISR case, the corrections associated with an initial-state
hard-collinear splitting differ in several respects. First, since the
beam direction, and hence the observable, is unaffected by transverse
recoil, the corresponding recoil correction is absent. Second, the
hard-collinear correction remains unchanged, as it reflects a
modification of the matrix element rather than of the emission
kinematics, which are treated as soft. Consequently, this correction
does not alter the longitudinal momentum fractions of the incoming
partons.
In addition, a new correction arises, accounting for the longitudinal
recoil induced by the hard-collinear emission. We refer to this as the
DGLAP correction, which will be discussed in detail below. This
correction replaces the recoil correction present in the FSR
case.

To derive the NNLL resummation formula, we start by defining the
contribution of the simple observable, $\Sigma_S ^{\had}(v)$.
The \textit{simple} observable in the hadron-collider case can be
defined analogously to its lepton-collider counterpart given in
Eq.~\eqref{eq:vsimple}, where now the transverse momentum $k_t$ and
rapidity $y$ are taken in the c.o.m.~frame of the partonic scattering.
To formalise the discussion, we consider the production of a colour
singlet system $F$ in a $pp$ collision. For simplicity we assume this
production process to be mediated by a single flavour channel (either
quark or gluon initiated). At the parton level, we will denote the
flavour of the two partons initiating the Born process by $c$, $d$.
For processes that receive a contribution from different flavour
channels one should repeat the discussion below for each of the
relevant channels and add them together.
With a slight abuse of notation, we denote the cumulative cross
section differential in the kinematics of the final state at the Born
level by $\Sigma(v,\PhiB)\equiv d\Sigma/d\PhiB$. The
resummation for the simple observable at NNLL takes the form
\begin{align}\label{eq:sigmaS_pp}
\Sigma^{\had}_S(v,\PhiB) &= \frac{d\hat{\sigma}^{(0)}_{cd}}{d\Phi_B} \exp\{-\mathcal{R}_{\rm NNLL}(v)\}\,\bigg[f_c(x_c,\muc) f_{d}(x_{d},\muc)\left(1+\frac{\alpha_s(\mur)}{2\pi} H^{(1)}_{cd}(\Phi_B)\right)\,\notag\\
&\!\!\! + \frac{\alpha_s(\mur)}{2\pi}\frac{1+b}{1+b-2\lambda} \sum_{i}\left(\big(C^{(1)}_{c i}\otimes f_i\big)(x_c,\muc) \, f_{d}(x_{d},\muc) +\{c\leftrightarrow d\}\right)\bigg]\,,
\end{align}
where $\hat{\sigma}_{cd}^{(0)}$ is the Born partonic cross section for
the $cd$ flavour channel and $C_{ci}^{(1)}$ are one-loop, $b$-dependent coefficient functions (see Appendix~\ref{app:ingredients}). The longitudinal momentum fractions
$x_{c,d}$ are defined as
\begin{equation}
x_{c,d} = \frac{\Q}{\sqrt{s}} e^{\pm y_F}\,, 
\end{equation}
and $\Q$ and $y_F$ denote the invariant mass and the rapidity of the
colour singlet $F$ produced in the hadronic collision of c.o.m. energy
$\sqrt{s}$. The ISR scale $\muc$ is defined as
\begin{align}
	\label{eq:col-scale}
\muc  =	\muf e^{-\frac{L}{1+\bo}}\,.
\end{align}
The expressions of all the ingredients in Eq.~\eqref{eq:sigmaS_pp} are
given in Appendix~\ref{app:ingredients}.
In Eq.~\eqref{eq:sigmaS_pp}, the Sudakov radiator
$\mathcal{R}_{\rm NNLL}(v)$ depends on the flavour configuration of
the initial-state partons $c$ and $d$, but it is flavour
diagonal. Here we suppress the dependence on the flavour indices to
simplify the notation. Conversely, the coefficient functions
$C^{(1)}_{ij}$ modify the flavour of the initial partons flowing into
the hard scattering. The operator $\otimes$ denotes the standard
convolution over the longitudinal momentum fraction given by
\begin{equation}
(f\otimes g)(x) = \int_{x}^1\frac{d z}{z} f(z) g\left(\frac{x}{z}\right)\,.
\end{equation}
We now discuss the transfer function. Starting at NNLL, the emission
of hard-collinear radiation modifies the longitudinal momentum
fraction of the initial-state partons. Therefore, the cumulative
cross section $\Sigma^{\had}(v,\PhiB)$ cannot be expressed as
a simple product of $\Sigma^{\had}_S(v,\PhiB)$ and a transfer
function in momentum space. Instead, this factorised, multiplicative
structure holds after taking a double Mellin transform over the
momentum fractions $x_c$ and $x_d$ (see, for instance, the
discussion in Ref.~\cite{Bizon:2017rah}). When expressed in momentum
space, up to NNLL accuracy, the cumulative distribution
$\Sigma^{\had}(v,\PhiB)$ can be conveniently cast as
\begin{equation}\label{eq:sigma_pp}
  \Sigma^{\had}(v,\PhiB) = \Sigma^{\had}_{S}(v,\Phi_{\rm
    B}){\cal F}^{\had}(v) +\dSdglap{\had}(v,\PhiB)\,,
\end{equation}
where
\begin{subequations}\label{eq:F-pp}
\begin{align}
{\cal F}^{\had}(v) &\,= {\cal F}^{\had}_{\rm NLL}(v) +
                     \frac{\alpha_s(\mur)}{\pi}\delta {\cal F}^{\had}_{\rm NNLL}(v)\,,\\
\delta {\cal F}^{\had}_{\rm NNLL}(v) &\,=\delta{\cal F}^{\had}_{\rm sc}(v)+\delta{\cal
    F}^{\had}_{\rm wa}(v) +\delta{\cal F}^{\had}_{\rm
    hc}(v)+ \delta{\cal F}^{\had}_{\rm correl}(v)+\delta{\cal F}^{\had}_{\rm
    clust}(v)\,.\label{eq:F-pp-decomp}
\end{align}
\end{subequations}
Above we have omitted the simple dependence of ${\cal
  F}^{\had}$ on the invariant mass $Q$ of the singlet
system and on the flavour of the initial-state partons.
As anticipated, in Eqs.~\eqref{eq:F-pp-decomp} the recoil correction is
absent, and the new DGLAP correction is included in
Eq.~\eqref{eq:sigma_pp}.
In the above decomposition, the NNLL transfer function
${\cal F}^{\had}(v)$ encodes the effect of emissions that do not
affect the longitudinal momentum fraction of the initial-state
partons. These are either terms of soft origin, that correspond to the
hadron-collider counterpart of the terms ${\cal F}^{\lep}_{\rm NLL}$,
$\delta{\cal F}^{\lep}_{\rm sc}$, $\delta{\cal F}^{\lep}_{\rm wa}$,
$\delta{\cal F}^{\lep}_{\rm correl}$,
$\delta{\cal F}^{\lep}_{\rm clust}$ in Eq.~\eqref{eq:F-ee}, or
corrections to the squared amplitude in the hard-collinear limit,
corresponding to $\delta{\cal F}^{\lep}_{\rm hc}$.
 Their calculation can be carried out exactly as in the lepton-collider case, with the
 only difference being in the functional form of the observable which
 is different in the hadron-collider case.
 Their expressions are reported in Appendix~\ref{app:ingredients}.

 The new term $\dSdglap{\had}$ describes the change in the
 longitudinal momentum fraction (i.e.~longitudinal recoil) of the
 initial-state partons.
 To understand the structure of $\dSdglap{\had}$, it is instructive to
 study the contribution of an initial-state, hard-collinear emission
 to the cross section.
 To be concrete, let us consider the case of a quark initiated process
 (e.g. Drell-Yan) and consider the flavour-diagonal $q\to q$
 transition mediated by the emission of initial-state gluons. The
 corresponding contribution of one hard-collinear emission to the real
 radiation is obtained by integrating, over the emission phase space,
 the emission probability
  \begin{equation}~\label{eq:real-hc}
   \frac{\alpha_s(k_t)}{2\pi}\frac{d\phi}{2\pi}\frac{ d
     k_t}{k_t} \,\frac{d z}{z}\, P^{(0)}_{qq}(z)
     f_c\left(\frac{x_c}{z}, \mu \right) f_{d}(x_d, \mu) +\{c\leftrightarrow d\}\,,
 \end{equation}
 where $ P_{qq}^{(0)}(z)$ denotes the tree-level unregularised
 splitting functions, cf.~Appendix~\ref{app:ingredients}.
 It is convenient to recast Eq.~\eqref{eq:real-hc} into the sum of a
 term that encodes the change in the longitudinal momentum fraction of
 the incoming parton and a term that does not modify it.
  This can be done by recasting Eq.~\eqref{eq:real-hc} as
 \begin{align}~\label{eq:real-hc-2}
\frac{d\phi}{2\pi}\frac{ d
     k_t}{k_t}\,dz  \frac{\alpha_s(k_t)}{2\pi}\Bigg[&\left(\frac{2 C_F}{1-z} + \gamma_q\,\delta(1-z)\right) f_c(x_c,\mu)f_d(x_d,\mu )
     \\&+ \frac{\hat{P}^{(0)}_{qq}(z)}{z}
     f_c\left(\frac{x_c}{z},\mu\right) f_d(x_d,\mu)+\{c\leftrightarrow d\}\Bigg]\,,\nonumber
 \end{align}
 with $\hat{P}$ now denoting the regularised splitting function and
 $\gamma_q=-3/2 \,C_F$. The parameterisation in
 Eq.~\eqref{eq:real-hc-2} conveniently singles out the terms in the
 real radiation that modify the longitudinal momentum fraction of the
 incoming partons (second term in Eq.~\eqref{eq:real-hc-2}), which
 will be encoded in the new contribution to the transfer function
 $\dSdglap{\had}$. Instead, the first term in Eq.~\eqref{eq:real-hc-2}
 will be accounted for in various parts of ${\cal F}^{\had}(v)$ in
 Eq.~\eqref{eq:F-pp}, namely  $\FNLL$,
 $\delta{\cal F}^{\had}_{\rm sc}(v)$ and $\delta{\cal F}^{\had}_{\rm hc}(v)$
 (cf. Refs.~\cite{Banfi:2004yd,Banfi:2014sua,Banfi:2016zlc}).

 The derivation of $\dSdglap{\had}$ can be easily performed in Mellin
 space, where the transform of the cumulative cross section
 $\Sigma^{\had}(v,\PhiB)$ can be expressed as a \textit{product} of
 the transform of $\Sigma^{\had}_{S}(v,\PhiB)$ and a transfer
 function.\footnote{This product is to be understood as a matrix
   product in flavour space.}~In this space, the Mellin transform of
 $\dSdglap{\had}$ arises from considering a single insertion (at NNLL)
 of the Mellin-transform of the second term in
 Eq.~\eqref{eq:real-hc-2} in an ensemble of initial-state,
 soft-collinear emissions widely separated in rapidity. Its inverse
 transform yields the momentum space expression can be written as
  \begin{align}\label{eq:Fdglap-master}
 \dSdglap{\had}(v,&\PhiB) =\lim_{\delta \to 0} \frac{d\hat{\sigma}^{(0)}_{cd}}{d\PhiB}  \,e^{-\mathcal{R}_{\rm NLL}(\delta v)} \Bigg\{\sum_{n=0}^{\infty}\frac{1}{n!}\prod_{i=1}^n \int_{\delta}^1\frac{d\zeta_i}{\zeta_i}
\RpNLL(v)\notag\\
     & \hspace{-1cm}\times 
      \frac{\alpha_s( Q v^{\frac{1}{1+b}})}{\pi (1+b)}\int_{\delta}^{1} \frac{d\zeta_{\rm hc}}{\zeta_{\rm
       hc}}\left(\int_{x_c}^1\frac{ dz}{z}
        \hat{P}_{qq}^{(0)}(z) f_c\left(\frac{x_c}{z}, Q v^{\frac{1}{1+b}}\right) f_d\left(x_d,  Q v^{\frac{1}{1+b}}\right)+\{c\leftrightarrow d\} \right) \notag \\
     & \hspace{-1cm}  \times \left[ \Theta\left(v-V(\{\zeta_i\},\zeta_{\rm hc})\right)-\Theta(1-\zeta_{\rm hc})\Theta\left(v-V(\{\zeta_i\})\right)\right]\Bigg\},
 \end{align}
 where we define $\zeta_i=v_i/v$ and $V(\{\zeta_i\})$ ($V(\{\zeta_i\},\zeta_{\rm hc})$) denotes the
 the observable in the presence of an ensemble of soft-collinear
 emissions $\{\zeta_i\}$ (and one hard-collinear emission
 $\zeta_{\rm hc}$) all widely separated in rapidity in the partonic
 c.o.m.~frame.
 The structure of Eq.~\eqref{eq:Fdglap-master} can be understood as
 follows.
 The $\RpNLL(v)$ kernels (cf. Appendix~\ref{app:ingredients})
 describe the emission of widely-separated, soft-collinear radiation
 from the initial state legs. The regularised kernel $\hat{P}$
 describes instead the insertion of a single hard-collinear emission,
 also widely separated in rapidity from the rest of the radiation. For
 the latter, the scale of the coupling and that of the parton
 densities is expanded, with NNLL accuracy, about the value of the
 transverse momentum of the hard-collinear emission that is set by the
 observable, that is
 $v=v_b(k_{\rm hc}) = \left(k_{t,\,{\rm hc}}/Q\right)^{1+b}$, which
 leads to $k_{t,\,{\rm hc}} = Q v^{\frac{1}{1+b}}$.

 The generalisation to any flavour channel, including the non-diagonal
 ones, simply amounts to the following replacement
\begin{equation}
\hat{P}_{qq}^{(0)}(z) f_c\left(\frac{x_c}{z}, Q v^{\frac{1}{1+b}}\right)\to \sum_i \hat{P}_{ci}^{(0)}(z) f_i\left(\frac{x_c}{z}, Q v^{\frac{1}{1+b}}\right),
\end{equation}
and similarly for $c\leftrightarrow d$.
For the two observables considered in this article, we have
\begin{align}
V(\{\zeta_i\},\zeta_{\rm hc})&=\sum_i \zeta_i+\zeta_{\rm hc}\,,&&
  \text{for } \Sb{0}\,,\\
V(\{\zeta_i\},\zeta_{\rm hc})&=\max\left\{\{\zeta_i\},\zeta_{\rm hc}\right\}\,,&&
  \text{for } \Mb{0}\,,
\end{align}
leading to $\dSdglap{\had}(v) =0$ for $\Mb{0}$, while for $\Sb{0}$ one finds
 \begin{align}
\dSdglap{\had}(v) &=   \frac{d\hat{\sigma}^{(0)}_{cd}}{d\PhiB}
                      \lim_{\delta\to 0}e^{-\mathcal{R}_{\rm NLL}(v)} \delta^{\RpNLL(v)}\sum_{n=0}^{\infty}\frac{1}{n!}\prod_{i=1}^n \int_{\delta}^1\frac{d\zeta_i}{\zeta_i}
\RpNLL(v) \nonumber \\ 
       & \hspace{-1cm}\times  \frac{\alpha_s(Q v^{\frac{1}{1+b}})}{\pi (1+b)}\int_{\delta}^{1}\frac{d\zeta_{\rm hc}}{\zeta_{\rm hc}}\left(\sum_i\int_{x_c}^1\frac{ dz}{z}
 \hat{P}_{ci}^{(0)}(z) f_i\left(\frac{x_c}{z}, Q v^{\frac{1}{1+b}}\right) f_d\left(x_d , Q v^{\frac{1}{1+b}}\right)+\{c\leftrightarrow d\}
         \right)  \nonumber \\ & \hspace{-1cm}\times  \left[\Theta(1-\sum_i \zeta_i
                                 -\zeta_{\rm hc})-\Theta(1-\zeta_{\rm hc})\Theta(1-\sum_i \zeta_i)\right].
 \end{align}
 The above integrals can be evaluated by following the steps in
 Appendix~C.3 of Ref.~\cite{Banfi:2014sua}. Finally, in the scale of
 the coupling and the PDFs we also expand the hard scale $Q$ about the
 renormalisation and factorisation scales, respectively, which will
 allow us to estimate perturbative uncertainties via scale
 variations. We thus obtain, for $\Sb{0}$
\begin{align}\label{eq:Fdglap-Sb}
\dSdglap{\had}(v) =&-\frac{d\hat{\sigma}^{(0)}_{cd}}{d\PhiB}
  e^{-\mathcal{R}_{\rm NLL}(v)} {\cal F}_{\rm \tiny NLL}(v) \frac{\alpha_s(\mur)}{\pi}\left(\psi^{(0)}(1+\RpNLL(v))+\gamma_E\right)
\frac{1}{1+\bo-2\lambda} \nonumber\\
&\times \sum_i \left[ (\hat{P}_{ci}\otimes f _i)(x_{c},\muc) \, f_d(x_d, \muc) + \{ c\leftrightarrow d\}\right]\,,
\end{align}
with $\muc$ given in Eq.~\eqref{eq:col-scale}.
In practice, we evaluate Eq.~\eqref{eq:Fdglap-Sb} with the NNLL
radiator
$e^{-\mathcal{R}_{\rm NLL}(v)}\to e^{-\mathcal{R}_{\rm NNLL}(v)}$,
which only introduces subleading, N$^3$LL corrections in our
predictions for $\Sb{0}$.

\subsubsection{Results for different reference frames}\label{sec:beam-frame}

The formulas we have described so far in this section are valid if the
observable is computed in the frame where the colour singlet has zero
rapidity, which we refer to as the partonic c.o.m.~frame. However, it
is also interesting to calculate such observables by defining the
rapidity in the laboratory frame, that is in the c.o.m.~frame of the
colliding hadrons.  It is easy to see that, at NLL accuracy, the two
calculations are identical.
At NNLL, we have three corrections, two of which are of soft origin.
The first is a modification to the soft Sudakov radiator
\( R_s^{\ell} \), defined in Eq.~\eqref{eq:sudakov}, whose expression
is reported in Appendix~\ref{sec:sudakovrad}.  This arises due to a
change in the definition of the simple observable, which, in the case
of the partonic c.o.m.~frame, reads:
\begin{equation}
  V_{\rm s}^{(\ell)}(k_t^{(\ell)}, y^{(\ell)}) = \frac{k_t^{(\ell)}}{Q}e^{-\bo y^{(\ell)}}.
\end{equation}
If we focus on the case in which $\ell=\ell_1$, i.e.~radiation
collinear to the beam aligned with the positive $z$-axis direction,
the simple observable for the hadronic c.o.m.~frame variant becomes
\begin{equation}
  V_{\rm s,\, c.o.m.
  }^{(\ell_1)}(k_t^{(\ell_1)}, y^{(\ell_1)})  = \lim_{y_{\rm c.o.m.} \to +\infty} \frac{k_t^{(\ell_1)}}{Q} e^{-\bo |y_{\rm c.o.m.}|} 
  = e^{\bo y_F} \frac{k_t^{(\ell_1)}}{Q} e^{-\bo y^{(\ell_1)}},
  \label{eq:Vcms1}
\end{equation}
where $y_{\rm c.o.m.}$ and $y_F$ are the rapidity of the emission and
of the colour singlet in the hadronic c.o.m.~frame, and
$y^{(\ell_1)}=y_{\rm c.o.m.}-y_F$.  Thus, we see that our simple
observable now reads
\begin{equation}
  V_{\rm s,\, c.o.m.}^{(\ell)} (k_t^{(\ell)}, y^{(\ell)})= d_{\ell} V_{\rm s}^{(\ell)}(k_t^{(\ell)}, y^{(\ell)}) , 
  \qquad \mbox{with } d_{\ell} = e^{\pm b y_F},
\end{equation}
where the $+$ ($-$) sign corresponds to $\ell_1$ ($\ell_2$).  
From Eq.~(B.5) of Ref.~\cite{Banfi:2014sua}, we observe that
\( g_2^{(\ell)} \) acquires a correction proportional to
\( \ln d_{\ell} \), which is equal and opposite in sign for both legs
and hence disappears in the sum.  Similarly, from Eq.~(B.6) of the
same reference, \( g_3^{(\ell)} \) acquires a term proportional to
\( \ln d_{\ell} \), which also vanishes in the sum over legs.
However, \( g_3^{(\ell)} \) also contains a quadratic term%
\footnote{Strictly speaking, the \( g_i^{(\ell)} \) of Ref.~\cite{Banfi:2014sua} are equal to \( g_i^{(\ell)} + h_i^{(\ell)} \),  
i.e., they are the sum of the soft and collinear radiator.  
Hence, some of the terms linear in \( \ln d_\ell \) would actually be in \( h_3^{(\ell)} \) (i.e., the collinear radiator).  
The quadratic term, however, is of purely soft origin.}
 that survives and amounts to the replacement $g_3^{(\ell)} \to
 g_3^{(\ell)} + \Delta g_3^{(\ell)}$, with
\begin{align}
\Delta g_3^{(\ell)} = & -\frac{C_{\ell}}{(1-2\lambda)(1+\bo -2\lambda)} \ln^2 d_{\ell} = 
-\frac{2\bo^2\,y_F^2\,C_{\ell}}{(1-2\lambda)(1+\bo -2\lambda)} \nonumber \\
= & -C_{\ell}\, \bo\, y_F^2 \frac{1}{\alpha_s(Q)}\left[ \frac{\alpha_s(Q e^{-L/(1+\bo)})}{1+\bo}-\alpha_s(Q e^{-L})\right].
\label{eq:framesoft1}
\end{align}
The second contribution of soft origin arises from a wide-angle
correction to the transfer function, hence contributing to
\( \delta \cF^{\had}_{\rm wa} \). This correction vanishes in the
partonic c.o.m.~frame (cf. comment in Appendix.~\ref{app:swa}), but it
is non-zero in the hadronic c.o.m.~frame. Indeed, assuming for simplicity that
\( y_F < 0 \), we see that for leg \( \ell_1 \), in the region where
\( 0 < y^{(\ell_1)} < y_F \), we have
\begin{equation}
  V_{\rm wa,\, c.o.m.} (k_t^{(\ell_1)}, y^{(\ell_1)}) = \frac{k_t^{(\ell_1)}}{Q} e^{-\bo |y^{(\ell_1)} - y_F|} =  
  \frac{k_t^{(\ell_1)}}{Q} e^{+\bo(y^{(\ell_1)} - y_F)},
\end{equation}
where the exponent has the opposite argument compared to the simple
observable in Eq.~\eqref{eq:Vcms1}.
Starting from Eq.~(C23) of Ref.~\cite{Banfi:2014sua} (which also applies to the \( \Mb{0} \) observable), we can thus calculate this correction,
obtaining
\begin{align}
\delta \cF^{\had}_{\rm wa} &= - \sum_{\ell} \frac{2C_\ell \alpha_s(Q)}{\alpha_s(Q e^{-L})}\cF_{\rm NLL}\int_0^{\infty} d y^{(\ell)} \ln \frac{V_{\rm wa,\, c.o.m.}}{V_{\rm s,\, c.o.m.}} \nonumber \\
& = - \frac{2C_\ell \alpha_s(Qe^{-L})}{\alpha_s(Q )}\cF_{\rm NLL} \int_0^{ y_F} d y^{(\ell_1)} 2\bo(y^{(\ell_1)} - y_F) \nonumber \\
& = 2C_\ell \,\bo \,y_F^2 \frac{\alpha_s(Qe^{-L})}{\alpha_s(Q)}\cF_{\rm NLL}.
\label{eq:framesoft2}
\end{align}
Combining the terms in Eqs.~\eqref{eq:framesoft1}
and~\eqref{eq:framesoft2}, we obtain that the total correction of soft
origin is
\begin{align}
\exp\left( -\frac{\alpha_s}{\pi} \sum_{\ell=\ell_1}^{\ell_2}\Delta g_3^{(\ell)} \right) \left(1+\frac{\alpha_s}{\pi} \frac{\delta \cF^{\had}_{\rm wa}}{\cF_{\rm NLL}}  \right)
& = 1 +\frac{\alpha_s(Q e^{-\frac{L}{1+\bo}})}{\pi} 2C_{\ell} \bo y_F^2 +\mathcal{O}({\rm N^3 LL}).
\end{align}

\begin{figure}[tb]
  \begin{center}
    \includegraphics[width=0.7\textwidth]{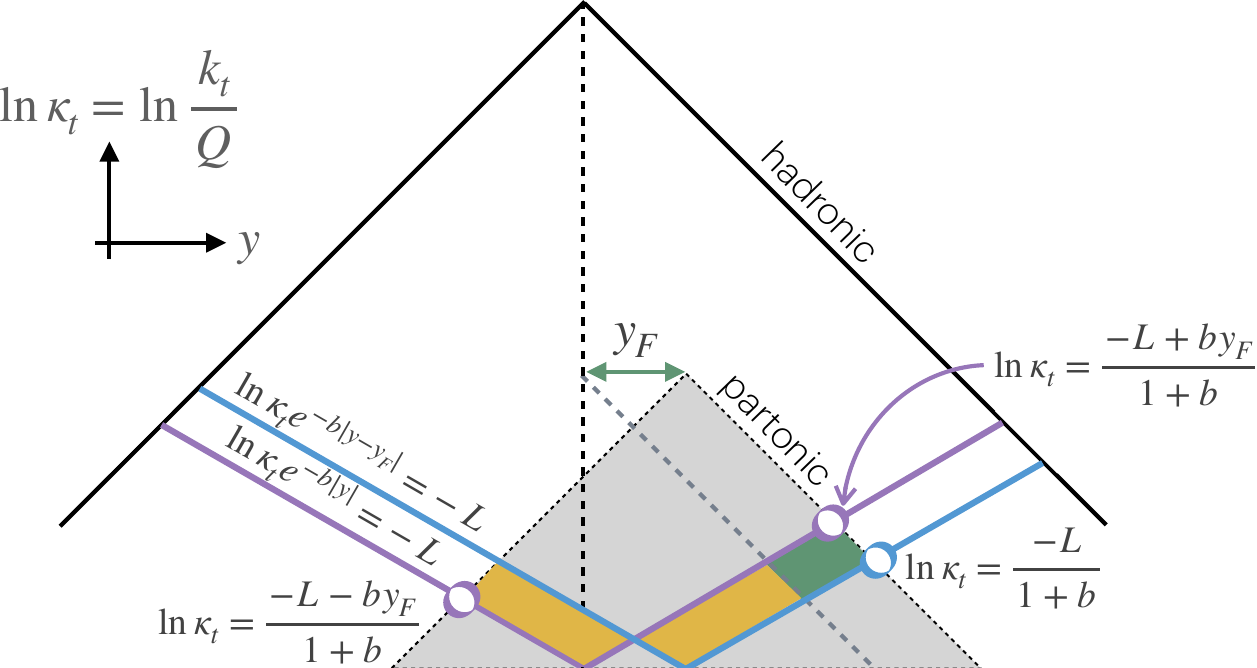}
  \end{center}
  \caption{Location of NNLL corrections in the Lund plane when
    computing $\bo$-dependent \lobs for $pp$ events with
    rapidity defined in the event frame (purple) or in the frame
    where the colour singlet has zero rapidity (blue).
  }
  \label{fig:LPframe}
\end{figure}

Notice that this term could also have been obtained from a geometric
argument, as it corresponds to the Lund plane density integrated over
the green region of Fig.~\ref{fig:LPframe}. Indeed, it can be naively
interpreted as the amount of soft-collinear radiation that is included
in the Sudakov radiator for the partonic c.o.m.~frame observable, and
needs to be removed for the hadronic c.o.m.~frame case.
One sees in Fig.~\ref{fig:LPframe} that the contributions form the two
yellow shaded areas cancel each other, so that the difference is given
by the green shaded area. This area has no logarithmic enhancement and
is therefore relevant at the NNLL accuracy.
This argument also shows the appearance of $Q e^{-\frac{L}{1+\bo}}$ as
the scale at which $\alpha_s$ has to be evaluated.

Finally, we examine the last correction, which is of hard-collinear
origin.  This arises from the fact that the collinear scale
\( \muc \), at which the PDF needs to be evaluated (represented by
empty dots in Fig.~\ref{fig:LPframe}), now receives a small
modification:
\begin{equation}
	\ln \frac{\muc}{\Q} =\frac{-L}{1+\bo} \to \frac{-L\pm b y_F}{1+\bo},
\end{equation}
where the \( + \) (\( - \)) sign applies to the incoming parton with
positive (negative) rapidity.
At NNLL accuracy, this results in the following replacement in the
scale of the parton densities in
Eqs.~\eqref{eq:sigmaS_pp},~\eqref{eq:Fdglap-Sb}
\begin{align}
	f_{\ell}\left(x,  \muf e^{-\frac{L}{1+\bo}}\right)  
    \to & f_\ell\left(x,   \muf \exp\left(\frac{-L -(-1)^\ell \bo\, y_F}{1+\bo}\right)\right)\,,
\end{align}
where we conventionally choose the beam $\ell=\ell_1$ to be in the
positive-rapidity hemisphere.
Eq.~\eqref{eq:Fdglap-Sb} is then expanded around the scale $\muf
e^{-\frac{L}{1+\bo}}$ using the DGLAP equation, neglecting
contributions beyond NNLL.

The sum of all contribution derived above could be alternatively
collected into a correction to the result in the partonic c.o.m.~frame
in Eq.~\eqref{eq:sigma_pp}, that reads
\begin{align}
	\dSframe{\had}(v,\PhiB)&= \frac{d\hat{\sigma}^{(0)}_{cd}}{d\PhiB}
  e^{-\mathcal{R}_{\rm NNLL}(v)}\FNLL(v) \frac{\bo\, y_F}{1+\bo-2\lambda} \frac{\alpha_s(\mur)}{\pi}\notag\\&\times
	\Bigg[ 2C_\ell \, y_F f_c(x_c, \muf e^{-\frac{L}{1+\bo}})\, f_d(x_d, \muf e^{-\frac{L}{1+\bo}}) \\&+ \sum_i \left( (\hat{P}_{ci}\otimes f _i)(x_{c},\muf e^{-\frac{L}{1+\bo}}) \, f_d(x_d, \muf e^{-\frac{L}{1+\bo}}) -\{ c\leftrightarrow d\}\right)\Bigg]\,. \notag
\end{align}

\subsubsection{Absence of non-global logarithms in the multi-jet case}
\label{app:NGL}

In this section, we discuss the absence of non-global
logarithms~\cite{Dasgupta:2001sh,Banfi:2002hw,Caron-Huot:2015bja,Banfi:2021xzn,Banfi:2021owj,FerrarioRavasio:2023kyg,Becher:2023vrh} 
in the resummation of \lobs, in particular when three or more coloured legs are present
already at the Born level.
For the sake of arguments, we examine one such processes, $pp\to Fj$,
where $F$ is a generic colour singlet and $j$ is a hard and resolved
final-state jet.
We then start by performing a fixed-order calculation of the leading
non-global configuration, stemming from the emission of a pair of soft
wide-angle gluons with momentum $p_{1,2}$ strongly ordered in energy.
At this order, non-global corrections arise from scenarios in which one
of the two gluons propagates inside the jet $j$ and one propagates
outside (and hence clusters with the beam in the definition of the
\lobs).
We introduce the following parameterisation for the kinematics
\begin{align}
p_n = E_n\left(1,0,0,1\right),\;\;
&p_{\bar n} = E_{\bar n}\left(1,0,0,-1\right),\;\; \\ p_\ell  =
  p_{t,\ell}(\cosh \eta_\ell,\sin \phi_\ell, \cos&\phi_\ell, \sinh
  \eta_\ell)\,,\quad \ell\in\{j,1,2\}\,,\notag
\end{align}
where $p_{n,\bar{n}}$ denote the momenta of the two incoming partons and
$p_j$ that of the final-state jet, and all quantities are defined
w.r.t.~the beam axis.
To simplify the calculation, we work in the planar, large-$N_c$ limit
($C_F=C_A/2=N_c/2$) that is customary in the study of non-global
observables. The Born will be then a collection of dipoles.
To single out the leading non-global configuration, we consider the
situation in which the two emissions are strongly ordered and we write
the corresponding emission probability as
\begin{align}\label{eq:emsn-prob}
[d p_1][dp_2]W_2 &= [d p_1][dp_2](4\pi\alpha_s)^2 N_c^2\!\!\! \sum_{\{ik\}\in {\rm
    dipoles}}\frac{(ik)}{(i1)(k1)}\left(\frac{(i1)}{(12)(i2)}+\frac{(k1)}{(12)(k2)}-\frac{(ik)}{(i2)(k2)}\right)\notag\\&\equiv
  4\,(4\pi\alpha_s)^2
    N_c^2 \frac{[d p_1][dp_2]}{p_{t,1}^2 p_{t,2}^2} \,w_2\,,
\end{align}
where
\begin{equation}
  [d p_i] = \frac{d\eta_i}{2} \frac{p_{t,i}\,dp_{t,i}\,d\phi_{i}}{(2\pi)^3}\,,
\end{equation}
and we adopted the notation $(ij) \equiv p_i\cdot p_j$. In
Eq.~\eqref{eq:emsn-prob}, we have subtracted the primary contribution
to the squared amplitude, which is a source of double logarithmic
corrections and has been already accounted for in our resummation.
To parameterise the phase space, we perform the change of variables
\begin{align}
v_i = p_{t,i} e^{-\bo|\eta_i|}\,, \quad \eta_i - \eta_j = R r_i \cos\theta_i\,, \quad \phi_i - \phi_j = R r_i \sin\theta_i\;\;\mbox{ with }i=1,2
\end{align}
for which the phase-space factor reads
\begin{equation}
4(4\pi\alpha_s)^2
N_c^2 \frac{[d p_1][dp_2]}{p_{t,1}^2 p_{t,2}^2}
=\left(\frac{\alpha_{s}}{\pi}\right)^{2} N_c^2 \frac{R^{4}}{(4\pi)^2}\prod_{i=1,2}\frac{dv_{i}}{v_{i}}dr^{2}_{i}d\theta_{i}\,.
\end{equation}
With these definitions, the clustering condition for a parton $p_i$ to
be recombined with the jet $p_j$ is
\begin{align}
  d_{j,i}^2 = r_i^2 R^2 < R^2\,.
\end{align}
Requiring that only one of the two radiated particles clusters then
leads to the following two possible angular constraints
\begin{subequations}
\begin{align}
\Theta_A &= \Theta(r_1 -1)\Theta(1-r_2)\Theta(r_1 - 2  r_2
           \cos(\theta_1 - \theta_2))\,, \quad \text{$p_1$ out, $p_2$ in}\\
\Theta_B &= \Theta(1-r_1)\Theta(r_2-1)\Theta(r_2 - 2 r_1
           \cos(\theta_1 - \theta_2))\,, \quad \text{$p_1$ in, $p_2$ out}
\end{align}
\end{subequations}
where the last $\Theta$ function enforces that the two particles
$p_{1,2}$ are not clustered first.
In the strongly ordered limit $v_1\gg v_2$, non-global logarithms
arise~\cite{Dasgupta:2001sh} if the following phase-space integral
\begin{align}~\label{eq:ngl-master}
\left(\frac{\alpha_{s}}{\pi}\right)^{2} N_c^2 \frac{R^{4}}{(4\pi)^2}\,\int &\prod_{i=1}^2 d r_i^2 d\theta_i \, w_2(\{r_i,\theta_i\})\\&\times
 \int_0^{\infty} \frac{dv_1}{v_1}  \int_0^{\infty} \frac{dv_2}{v_2}  \left(\Theta(v - V(p_1,p_2)) - \Theta(v - V(p_1))\right) \Theta_{A/B} \Theta(v_1-v_2)\notag
\end{align}
develops logarithmic terms ${\cal O}(\ln^n(v))$. 
The second term in Eq.~\eqref{eq:ngl-master} corresponds to the
virtual correction, which in the strongly ordered limit can be
approximated by simply imposing unitarity. Moreover,
$\Theta(v_1 - v_2)$ is the (strong) ordering condition.
Due to the way we parameterise the phase space, we can focus only on the $v_{1,2}$ integration to assess the presence of a logarithmically-enhanced term.

It is now straightforward to see that, when $v_1 \gg v_2$, we have
$V(p_1,p_2) = V(p_1)\neq 0$, so that no non-global logarithm can arise
from the above integral.
This is a direct consequence of the fact that the observable resolves
the substructure of jets, hence imposing a commensurate constraint
over all emissions regardless of whether they cluster with the beam or
with the jet.
To be explicit, let us consider case $A$, where only $p_2$ clusters
with the final-state jet. In the case of the \lobs in this
process (cf. Section~\ref{sec:pp}), the first emission contributes
with $v_1$, whereas the second emission contributes with
\begin{align}
p_{t,2}e^{-\bo|\eta_j|}r_2^{1+b} = v_2 r_2^{1+b} e^{-\bo(|\eta_j| - |\eta_2|)} \equiv v_2 g(r_2,\theta_2)\,.
\end{align}
For $\Sb{1}$ and $\Mb{1}$, we thus have $V(p_1)=v_1$ and
\begin{equation}
  V(p_1,p_2) = \begin{cases}
   v_1\max\left\{1,\frac{v_2}{v_1}g(r_2,\theta_2)\right\} \,\quad &\text{for $\Mb{1}$}\\
     v_1\left(1+\frac{v_2}{v_1}g(r_2,\theta_2)\right)\,\quad &\text{for $\Sb{1}$}
  \end{cases}
  \,.
\end{equation}
We therefore find that the integral is given by
\begin{align}
& \int_0^{v} \frac{dv_1}{v_1} \int_0^{v_1}
  \frac{dv_2}{v_2}\left[\Theta\left(v-V(p_1,p_2)\right) -
  \Theta(v-V(p_1))\right] = \notag \\
& \hspace{5cm}  =
  \begin{cases}
    \displaystyle-\frac{1}{2}\ln^2(\max\left\{1,g(r_2,\theta_2)\right\})\quad&\mbox{for }\Mb{1}\\
     {\rm Li}_2\left(-g(r_2,\theta_2)\right)\quad&\mbox{for }\Sb{1}
\end{cases}\,,
\end{align}
which is free of logarithmic terms.
An analogous conclusion holds for case $B$, in which $p_1$ clusters
with the jet and $p_2$ clusters with the beam.
The remaining angular integral is of ${\cal O}(1)$ as long as $R\sim
{\cal O}(1)$, which proves the absence of non-global logarithms at
this perturbative order.

Our conclusions readily extend to all perturbative orders (both in
$\alpha_s$ and in the resummation). The mechanism that led to the
cancellation of non-global logarithms in the above example relies on
the recursive nature of the observable that resolves the structure of
radiation inside the jet. As a result, for any number of emissions
$\{p_{i}\}$, the observable scales uniformly across different regions
of the solid angle, that is $V(\{p_{i}\})\sim V(p_{j}) \sim v$ where
$p_{j}$ is any one of the $\{p_{i}\}$ emissions, as long as
$R\sim {\cal O}(1)$. This guarantees that all phase-space
configurations that would typically lead to non-global logarithms now
give at most a non-logarithmic contribution to the observable whose
coupling is evaluated at the soft scale $Q e^{-L}$.
In practice, this means that beyond NNLL, the extension of the wide-angle
function $\dFwa$ will receive contributions involving the
non-trivial clustering of emissions near the jet boundary, all at
commensurate scales of order $Q e^{-L}$.

\subsubsection{Comment on possible coherence-violating effects at higher
  orders}
\label{sec:sll}
Although our observables are free of non-global logarithms at all
logarithmic orders, their logarithmic structure in the $pp$ case may
be affected by coherence-violating corrections at higher-loop
orders. Such corrections appear beyond the leading-$N_c$ level and
originate from Glauber gluons exchanges. They are known to be present
in global observables that are not sufficiently inclusive over
final-state
radiation~\cite{Banfi:2010xy,Gaunt:2014ska,Zeng:2015iba,Forshaw:2021fxs,Banfi:2025mra}.\footnote{Corresponding
  studies of coherence-violating corrections in non-global observables
  can be found, e.g., in
  Refs.~\cite{Forshaw:2006fk,Forshaw:2008cq,Forshaw:2009fz,DuranDelgado:2011tp,Becher:2021zkk,Becher:2023mtx,Boer:2023jsy,Boer:2023ljq,Boer:2024hzh,Becher:2024nqc,Banerjee:2025kkq,Dasgupta:2025cgl}.}
These studies indicate that such coherence-violating corrections might
as well appear in the case of \lobs at sufficiently high perturbative
orders. We do not consider these corrections here and leave their
study to future work.

\subsection{$N=1$ case in DIS}
\label{sec:sigmaDIS}
We now consider the \lobs at lepton-hadron colliders, $\Sigma^{ep}(v)$.
The resummation of the simple observable $\Sigma^{ep}_S(v)$ can be obtained from
the ingredients of the lepton- ($\Sigma^{\lep}_S(v)$) and hadron-collider
($\Sigma^{\had}_S(v)$) cases given in Eq.~\eqref{eq:sigmaS_ee} and
Eq.~\eqref{eq:sigmaS_pp}, respectively. We label the flavour of the incoming parton by
$c$ and that of the outgoing parton by $d$. The resummation for
$\Sigma^{\dis}_S(v,\PhiB) $ reads
\begin{align}\label{eq:sigmaS_ep}
   \Sigma^{\dis}_S(v,\PhiB) &=
                              \frac{d\hat{\sigma}^{(0)}_{c d}}{d\Phi_B}
                              \exp\{-\mathcal{R}_{\rm NNLL}(v)\}\,\bigg[f_{c}(x_1,\muc) \left(1+\frac{\alpha_s(\mur)}{2\pi} H^{(1)}_{cd}(\Phi_B)\right)\,\notag\\
                            &\!\!\! + \frac{\alpha_s(\mur)}{2\pi}\frac{1+b}{1+b-2\lambda} \bigg(\sum_{i}\big(C^{(1)}_{c i}\otimes f_i\big)(x_1,\muc) + f_{c}(x_1,\muc) \,C^{(1)}_{{\rm hc},d}\bigg) \,\bigg]\,,
\end{align}
where the hard-collinear constant $C^{(1)}_{{\rm hc},d}$ depends on
whether the flavour $d$ is a gluon or a quark.
The resummation for the cumulative cross section for the $\dis$ case
follows from the $\lep$ and $\had$ cases, and takes the form
  \begin{equation}\label{eq:sigma_dis}
    \Sigma^{\dis}(v,\PhiB) = \Sigma^{\dis}_{S}(v,\Phi_{\rm
      B}){\cal F}^{\dis}(v) +\dSdglap{\dis}(v,\PhiB)\,,
  \end{equation}
  with
  \begin{align}\label{eq:F-dis}
{\cal F}^{\dis}(v) &\,= {\cal F}^{\dis}_{\rm NLL}(v) +
                     \frac{\alpha_s(\mur)}{\pi}\delta {\cal F}^{\dis}_{\rm NNLL}(v)\,,
\end{align}
and
  \begin{align}
    \FNLL^{\dis}(v)&=\FNLL^{\had}(v)=\FNLL^{\lep}(v)\,,\\
    \FNNLL^{\dis}(v) &= \dFsc^{\dis}(v) + \dFwa^{\dis} (v)+ \dFcorrel^{\dis}(v) + \dFclust^{\dis} (v)+\dFhc^{\dis} (v)+ \dFrec^{\dis}(v)\,.
  \end{align}
The expressions for all the ingredients are given in Appendix~\ref{app:ingredients}.
As in the $\had$ case, the DGLAP correction $\dSdglap{\dis}(v,\PhiB)$
is absent for $\Mb{1}$ and for $\Sb{1}$ it reads
\begin{align}\label{eq:Fdglap-Sb-dis}
\dSdglap{\dis}(v) =&-\frac{d\hat{\sigma}^{(0)}_{c d}}{d\Phi_B}
                     \exp\{-\mathcal{R}_{\rm NNLL}(v)\}\FNLL^{\dis}(v)\frac{\alpha_s(\mur)}{\pi}\left(\psi^{(0)}(1+\RpNLL(v))+\gamma_E\right) \nonumber\\&\times \frac{1}{1+\bo-2\lambda} 
\sum_i (\hat{P}_{ci}\otimes f _i)(x_1,\muc) \,.
\end{align}
Notice that, unlike for the $\had$ case, now both the DGLAP and the
recoil corrections contribute, due to the presence of emitting partons
both in the initial and final state.

\subsection{Fixed-order validation up to second perturbative order for
$\had$}
\label{sec:fixed-order-2-legs}
In this section we discuss the validation of our resummation for the
\lobs.
As a first check, we compare the expansion of our resummation in powers
of $\alpha_s$ to an exact fixed-order calculation.
We recall that NNLL resummation implies control, at the cumulative
level, over the terms $\alpha_s L^2, \alpha_s L, \alpha_s$ (i.e.\ all
terms at NLO in the small-$v$ limit), as well as
$\alpha_s^2 L^4\,, \dots, \alpha^2_s L$ (i.e.\ all but the
${\cal O}(\alpha_s^2L^0)$ term at NNLO).
Therefore, we consider the difference between the fixed-order (fo) and resummed-expanded (res) predictions at the differential level, namely,
\begin{align}
	\label{eq:deltaFFO}
\Delta f^{(i)}(v) \equiv \frac{1}{\sigma^{(0)}}\left[\frac{d\Sigma^{(i)}_{\rm fo}}{d \ln v} - \frac{d\Sigma^{(i)}_{\rm res}}{d \ln v} \right]\,,
\end{align}
where
\begin{equation}\label{eq:expansion}
  \Sigma_{\rm res}= \sum_{i=0}^2 \Sigma^{(i)}_{\rm res} + {\cal
    O}(\alpha_s^3 \sigma^{(0)})\,,\quad \Sigma_{\rm fo}= \sum_{i=0}^2 \Sigma^{(i)}_{\rm fo} \,.
\end{equation}
As a function of $L=-\ln(v)$, we expect that $\Delta f^{(i)}(v) \to 0$
for small $v$ and $i=1,2$.
Note that the correctness of the NLO term ${\cal O}(\alpha_s L^0)$ in
our resummation is tested in the $\mathcal{O}(\alpha_s^2)$ check for
$\Delta f^{(2)}(v)$.

The fixed-order reference results are obtained from
\texttt{Event2}~\cite{Catani:1996vz} (for $\lep$),
\texttt{disorder}~\cite{Karlberg:2024hnl,Salam:2008qg,Catani:1996vz} (for $\dis$)
and \texttt{Matrix}~\cite{Grazzini:2017mhc} (for $pp$).
Here we will only show our validations for $pp \to Z$ and $pp \to H$,
while those for $\lep$ and $\dis$ are collected in
Appendix~\ref{app:fo-validation}.
The phenomenological setup for obtaining the fixed-order and resummed-expanded
result is the same as the one used to obtain our final results, and will be detailed
in Sec.~\ref{sec:matched-lhc}.
For our test we consider central scales, $\muf = \mur = Q$, where
$Q$ is the mass of the $Z$ or Higgs boson, with $M_Z = 91.1976$ GeV
and $M_H = 125$ GeV.

\begin{figure}[tb]
	\centering
    \includegraphics[width=\textwidth]{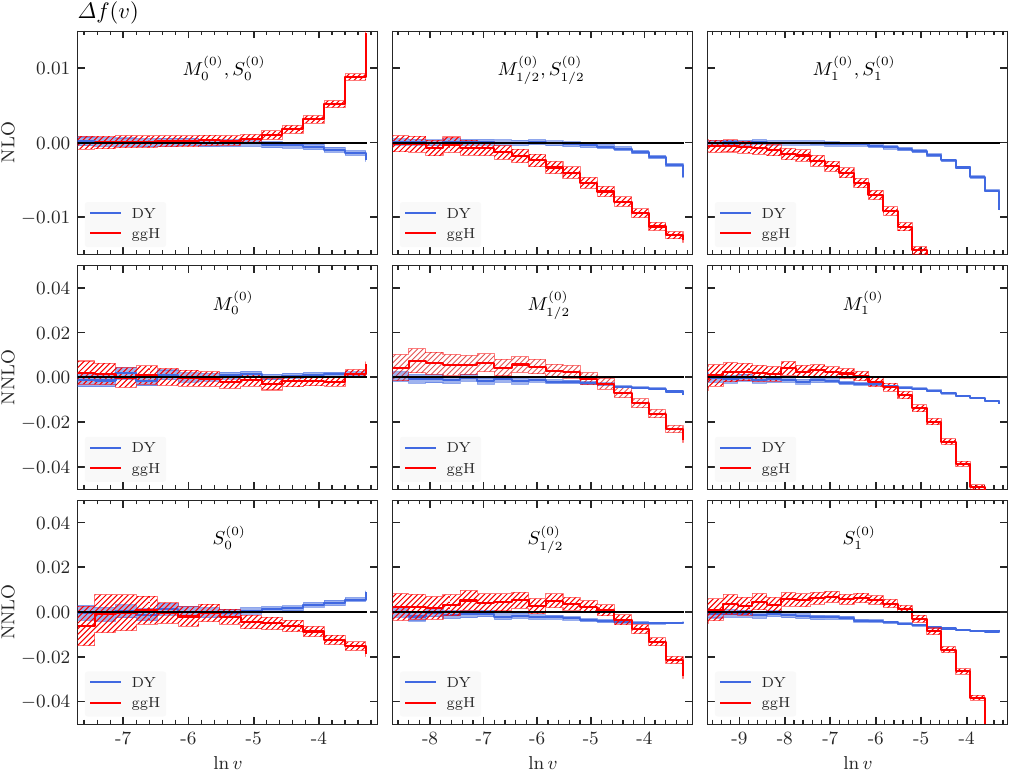}
	\caption{Difference between the fixed-order and resummed-expanded differential cross sections, normalised to the Born cross section, c.f.~Eq.~\eqref{eq:deltaFFO}, for DY (blue) and gluon fusion Higgs (red) production. We consider three values of $b$, $b=0$ (left), $b=1/2$ (middle) $b=1$ (right).
	The top panel shows the NLO comparison for $\Sb{0}$/$\Mb{0}$.
	The middle (bottom) panel shows the NNLO comparison for $\Mb{0}$ ($\Sb{0}$).
	The bands show the statistical
        uncertainty.}
	\label{fig:validation}
\end{figure}

The results are shown in Fig.~\ref{fig:validation}. In the top panel we show the
NLO comparison for $\Sb{0}$ and $\Mb{0}$, which are the same at this order
since there can be at most one emission.
We observe, as expected, that $\Delta f^{(1)}(v)$ vanishes in the limit $v\to0$ for
the values of $b$ considered ($b=0, 1/2$ and $b=1$).
The remaining panels show the NNLO results. We see that the NNLO
convergence is slightly slower, as shown in the middle ($\Mb{0}$) and
bottom ($\Sb{0}$) panels of Fig.~\ref{fig:validation}, as the power
corrections will be enhanced by additional logarithms.
For large values of $-\ln v$, we observe the expected agreement between the
fixed-order and the expanded resummed predictions.
In Fig.~\ref{fig:validation}, we observe that the $b=0$ results
converge to their asymptotic values for larger values of $v$ than for
the $b=1/2$ and $b=1$ cases, which is directly related to the
structure of their power corrections as well as the different
dependence on the radiation rapidity in the three cases. Note,
however, that the computational cost to reach the asymptotic region of
convergence is comparable for different $b$ values, as it is also
reflected on the size of the statistical errors at lower values of
$v$. Interestingly, we note that, within the same $b$ and at order
$\mathcal{O}(\alpha_{s}^{2})$, there is a mildly faster convergence
for $M^{(0)}_b$ than for $S^{(0)}_b$, which deserves further
investigations on the size of their power corrections.

We performed additional tests to validate the dependence on the
renormalisation ($\mur$), factorisation ($\muf$) and resummation
($\mul$, cf.~Eq.~\eqref{eq:res-scale-def}) scales of our predictions.
Firstly, we have verified the correctness of the scale dependence up to relative
order $\mathcal{O}(\alpha_{s}^{2})$ by repeating the comparison of the expansion
of the resummation and the fixed order results for individual scale variations.
In all cases, we find results similar to those reported in
Fig.~\ref{fig:validation} for central scales.
Secondly, we checked that the $\mur,\muf$ dependence of our NNLL result
exactly cancels at relative order $\mathcal{O}(\alpha_s^2)$, while their
dependence amounts to an $\mathcal{O}(\alpha_s^3L)$ effect.
This is done numerically through the evaluation of our NNLL
expressions and their expansion at a small values of $\alpha_s$.
In addition, we checked that the $\mul$ dependence starts at
$\mathcal{O}(\alpha_s^2L^0)$, which is expected since we have not
included $\mathcal{O}(\alpha_s^2)$ constants in our resummed result.

\section{NNLL+NNLO predictions and Monte Carlo studies for the LHC}
\label{sec:non-pert}
In this section, we discuss some phenomenological considerations for
the \lobs, including the matching of our resummed predictions to NNLO
fixed-order results at the cumulative cross-section level, as well as
the impact of non-perturbative corrections (i.e.\ hadronisation) and
underlying event.

\subsection{Matching to fixed order}
\label{sec:matching}
We now discuss the matching of our resummed predictions to a fixed
order, NNLO calculation. We will consider the $\had$ case, although an
analogous procedure can be used for the $\lep$ and $\dis$
cases. Similarly, the extension to higher orders is straightforward.

To perform the matching, we first need to ensure that at large $v$,
the resummation is turned off. Furthermore, we want to have a handle
on the size of subleading logarithmic corrections in the resummation,
to be included in the estimate of theoretical uncertainties.
We proceed in two steps. Firstly, we split the resummed logarithm as
\begin{equation}\label{eq:res-scale-def}
\ln\frac{1}{v} = \ln\frac{\mul}{Q v} + \ln\frac{Q}{\mul} = L + \ln\frac{Q}{\mul} \,,
\end{equation}
and then we expand the dependence in the resummed prediction on the
second term ($\ln\frac{Q}{\mul} $) by truncating the expansion at
NNLL, hence inducing a dependence on $\mul$ that is strictly
N$^3$LL. %
As a consequence all the ingredients of the resummation will depend on
$\mul$, and their corresponding expressions are provided in
Appendix~\ref{app:ingredients}.
The scale $\mul$ is the \textit{resummation scale}, which we will vary
around $Q$ to estimate the size of subleading logarithms in the
prediction.
As a second step, we now modify $\mul$ and $L$ as follows~\footnote{In
  the past, this was achieved by modifying the definition of the
  logarithm
  $L\to \tilde{L} = \frac{1}{p}\ln\left(1+\left(\frac{\mul}{Q
        v}\right)^p\right)$, see
  e.g.~Refs.~\cite{Banfi:2015pju,Banfi:2012jm,Bizon:2017rah,Bizon:2018foh}. Instead,
  here we adopt a dynamical resummation scale, similar in spirit to
  what was done in Ref.~\cite{Mazzitelli:2021mmm}.}
\begin{align}\label{eq:muLtilde}
\mul \to \mulbar(v) = 
\begin{cases}
  \mathfrak{r}(v)\quad &\mbox{ for } v <\vm, \\
    \mathfrak{m}(v)\quad &\mbox{ for } \vm \leq v < 1, \\
	\displaystyle Q \quad &\mbox{ for } v \geq 1\,,
\end{cases} 
\end{align}
and
\begin{equation}~\label{eq:Lmod}
L \to L \,\Theta\left(\frac{\mum}{Q}-v\right)\,.
\end{equation}
The functions $\mathfrak{r}(v)$ and $\mathfrak{m}(v)$ are polynomials
in the (dimensionless) observable $v$, whose expressions are given in
Appendix~\ref{app:matching}, where more information is provided.
They guarantee continuity of both the cumulative and differential
distributions for $v=v_M$ and $v=1$. 
The logarithm $L$ is now switched off at $v=\vm=\mum/Q\leq 1$, where
$\mum$ is a \textit{matching scale}, to be taken of order $Q$.

To match the NNLO fixed-order and NNLL resummed results, we will use a
multiplicative matching scheme inspired to those introduced in
Refs.~\cite{Banfi:2012yh,Banfi:2012jm,Banfi:2015pju,Bizon:2018foh}.
We introduce the notation for the fixed-order expansions of the
ingredients up to relative order ${\cal O}(\alpha_s^2)$ w.r.t.~the
Born, working at the level of the integrated, cumulative cross section
with any phase space cuts on the Born kinematics $\PhiB$ and a veto
$v$ on the resolution variables studied here.
We denote the fiducial cross section with any phase space cuts on the
Born phase space $\PhiB$ by $\sigma_{\rm fid}$
($\sigma_{\rm fid}=\sigma_{\rm total}$ if no cuts are applied) and
define, in addition to Eq.~\eqref{eq:expansion},
\begin{subequations}
\begin{align}
\sigma_{\rm fid} = \sum_{i=0}^2 \sigma_{\rm fid}^{(i)}\,, \quad \Sigma_{\rm fo}(v) = \sum_{i=0}^{2}\Sigma_{\rm fo}^{(i)}(v) &= \sigma_{\rm fid} - \sum_{i=1}^{2}\bar\Sigma_{\rm fo}^{(i)}(v)\,, \\
\Sigma_{L\to 0}= \left.\Sigma_{\rm res}(v)\right|_{L\to 0}  &= \sigma_{\rm fid}^{(0)}+\sum_{i=1}^{2}\Sigma_{L\to 0}^{(i)}\,,
\end{align}
\end{subequations}
where $\Sigma_{\rm fo}(v)$ ($\bar\Sigma_{\rm fo}(v)$) is the
cumulative fixed-order cross section below (above) the cut $v$, and
$\Sigma_{\rm res}$ is its resummed analogue, that is
\begin{equation}
\Sigma_{\rm res}(v) = \int_{\rm cuts} \!\! d\PhiB \Sigma^{\had}(v,\PhiB) \,,
\end{equation}
where the integral is performed taking into account any cuts on the
Born kinematics.
The quantity $\Sigma_{L\to 0}$ is obtained by setting $L=0$ in the
resummed prediction. We stress that, in general, $\ln Q/\mul\neq0$
(cf. Eq.~\eqref{eq:res-scale-def} and discussion thereof) in
$\Sigma_{L\to 0}$.
The matched  cross section can be written as
\begin{align}
		\label{eq:master-matching}
		\Sigma_{\rm mat}(v) & = \frac{\Sigma_{\rm
                                          res}(v)}{\Sigma_{L\to 0}} \Bigg[
		\Sigma_{L\to 0}+ \Sigma^{(1)}_{\rm fo}(v) -  \Sigma^{(1)}_{\rm res}(v)\\
		&\notag \, + \Sigma^{(2)}_{\rm fo}(v) -  \Sigma^{(2)}_{\rm res}(v) + \left( \frac{\Sigma_{L\to 0}^{(1)}}{\sigma_{\rm fid}^{(0)}}-
           \frac{\Sigma^{(1)}_{\rm res}(v)}{\sigma_{\rm fid}^{(0)}}\right)
           \left( \Sigma^{(1)}_{\rm fo}(v) - \Sigma^{(1)}_{\rm res}(v)\right) \notag 
		\Bigg].
\end{align}
The advantage of a multiplicative matching scheme like the one in
Eq.~\eqref{eq:master-matching} is that terms
${\cal O}(\alpha_s^2 L^0)$, formally of N$^3$LL order, are recovered
correctly from the matching, effectively promoting the accuracy of the
prediction to what is commonly labelled as NNLL$^\prime$+NNLO.

\subsection{Results at the LHC}
\label{sec:matched-lhc}
In this section we report our predictions for $\Mb{0}$ and $\Sb{0}$ in
neutral-current DY production.
We consider $pp$ collisions at a centre-of-mass energy of 13~TeV and
use the \texttt{NNPDF40MC\_nnlo\_as\_01180}
\cite{Cruz-Martinez:2024cbz} PDF set.
The required convolutions with the PDFs are handled with the
\texttt{Hoppet} package~\cite{Salam:2008qg,Karlberg:2025hxk}, whereas
the parton densities themselves are evaluated via
\texttt{LHAPDF}~\cite{Buckley:2014ana}.
Note that NNLO DGLAP evolution of the PDF set is used for our results,
even though this level of evolution is of N$^3$LL order.
We require that the $Z$ boson is exactly on-shell with a mass of
$Q=M_Z = 91.1976$ GeV.
We use the C/A algorithm to cluster the event, and take $R=0.8$
as our jet radius. The choice of the large jet radius is motivated by
the fact that it increases the sensitivity to underlying events, which
we wish to examine below. Smaller values of $R$ are on the other hand
preferable for phenomenological applications.
We assess the missing higher-order uncertainties by performing a
$7-$point scale variation of $\mur$ and $\muf$ by a factor of two
around their central values $\mur=\muf=Q$.
In addition, we vary the resummation scale $\mul$ by a factor of $2$
around $\mul=Q$, for the central choice of $\mur,\muf$.
The resulting uncertainty is taken as the envelope of these $9$
different scale variations. In our dynamical resummation scale
(cf. Appendix~\ref{app:matching}) we use $q=p$ and set the default
value to $p=2$. To assess the matching uncertainties we also consider
the predictions with $p=4$ and, separately, with a variation of the
matching scale from $\mum=Q$ to $\mum=0.8 \,Q$. The fixed-order
results are obtained from \texttt{Matrix}~\cite{Grazzini:2017mhc}.

\begin{figure}[tb]
	\centering
	\includegraphics[page=1, width=\textwidth]{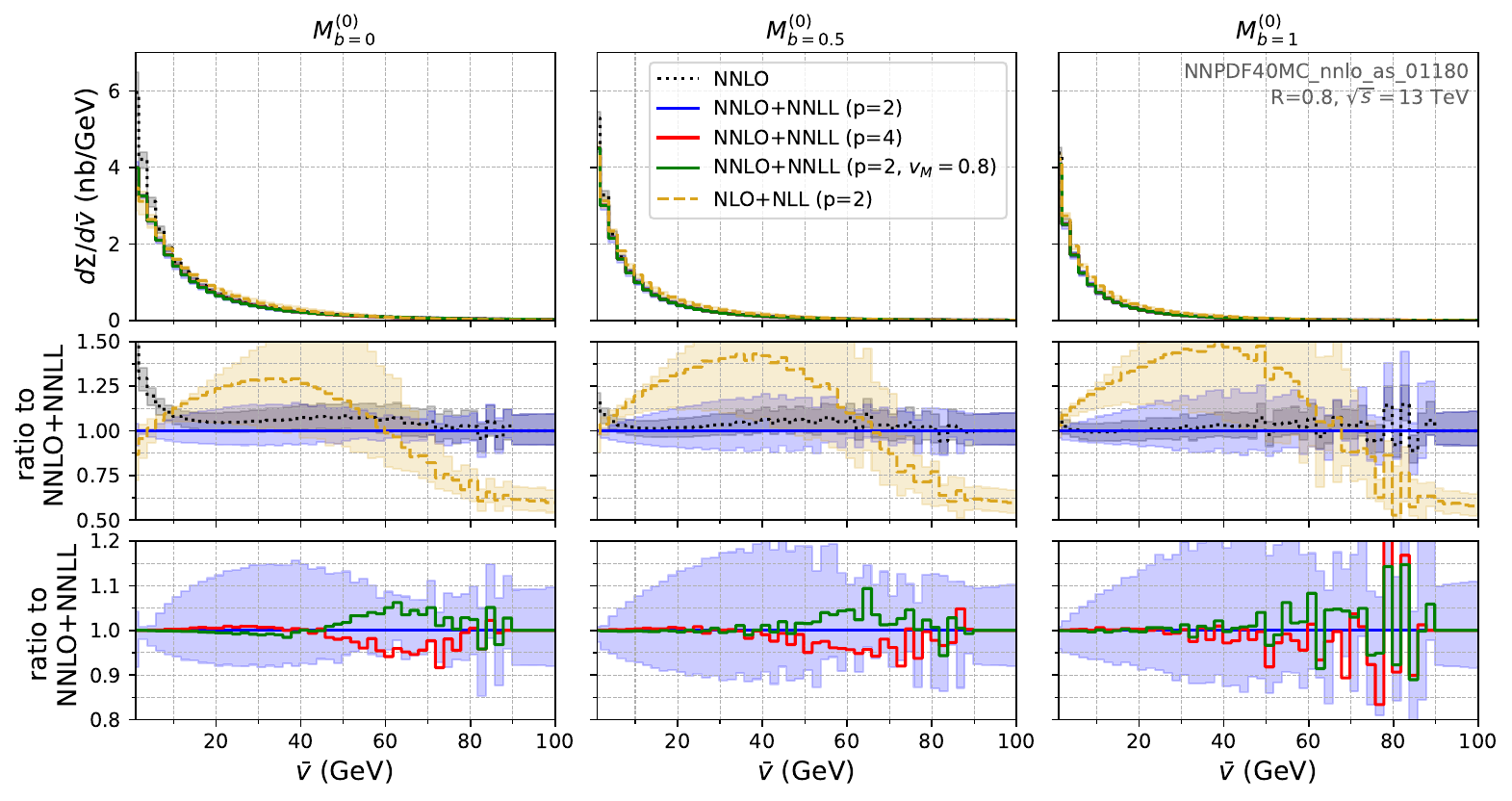}
	\caption{Differential distribution for  $\Mb{0}$ with $b=0$
		(left), $b=1/2$ (middle) and $b=1$ (right) as a function of $\bar{v} = v\,Q$. 
		The black dotted curve represents the NNLO result, the yellow dashed curve the NLO+NLL result, the blue (red) solid curve the NNLL+NNLO result with $p=2$ ($p=4$) in Eq.~\eqref{eq:muLtilde}.
		The green curve shows the NNLL+NNLO result with $p=2$ and $v_M = 0.8$. 
		Scale uncertainties (computed as explained in the text) are indicated via the opaque bands. 
		The middle panel shows the ratio of the NLO+NLL and NNLO results with respect to the
                NNLL+NNLO with $p=2$ ($v_M = 1$) result.
         	The bottom panel shows the ratio of the NNLO+NNLO result with $p=4$ and that with  $p=2$ and $v_M = 0.8$ to the $p=2$, $v_M = 1$ result. }
	\label{fig:matching-mbeta}
\end{figure}

\begin{figure}[tb]
	\centering
	\includegraphics[page=2, width=\textwidth]{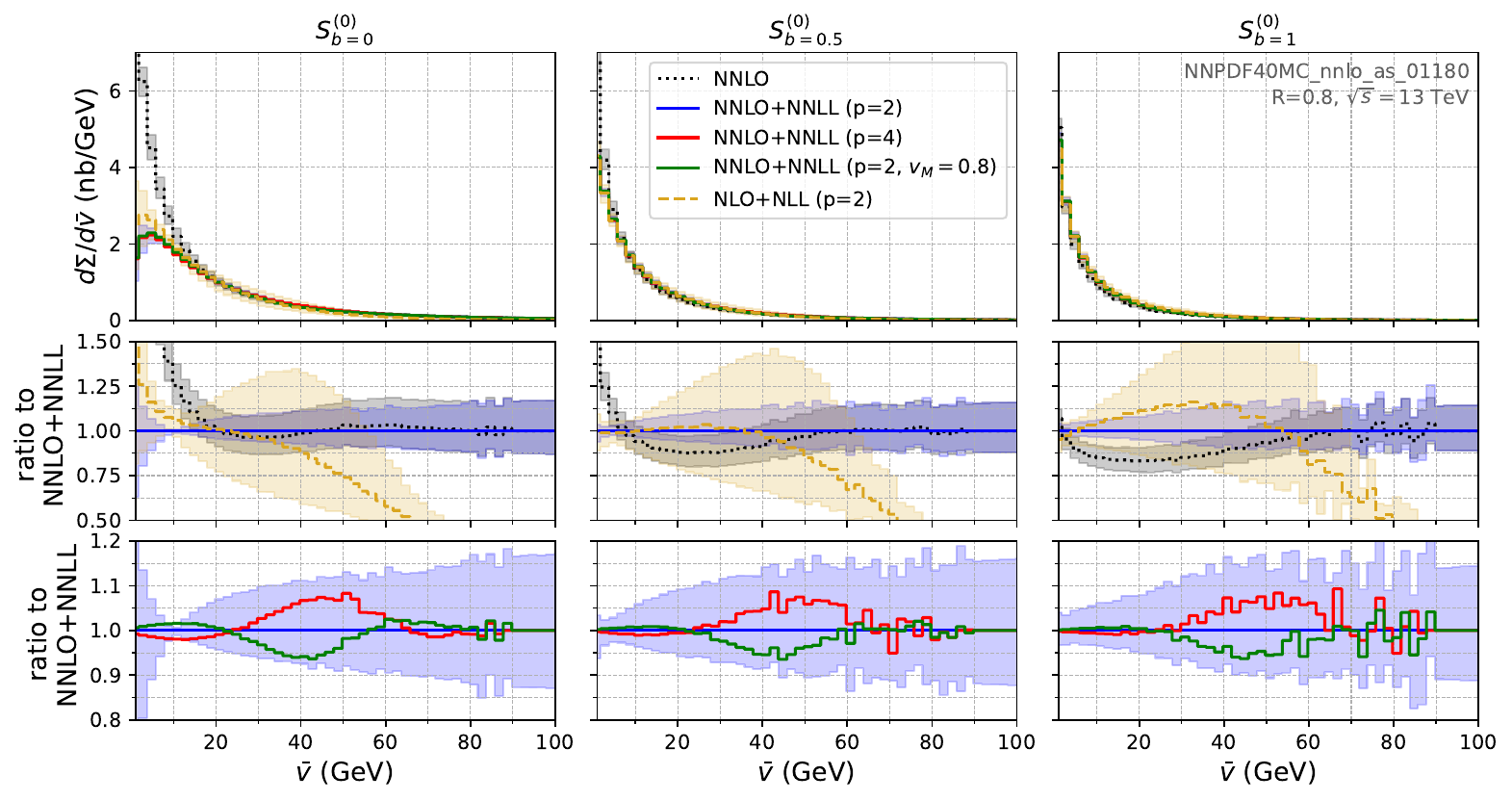}
	\caption{Same as Fig.~\ref{fig:matching-mbeta} but for $\Sb{0}$. }
	\label{fig:matching-sbeta}
\end{figure}

\begin{figure}[tb]
	\centering
	\includegraphics[page=1, width=\textwidth]{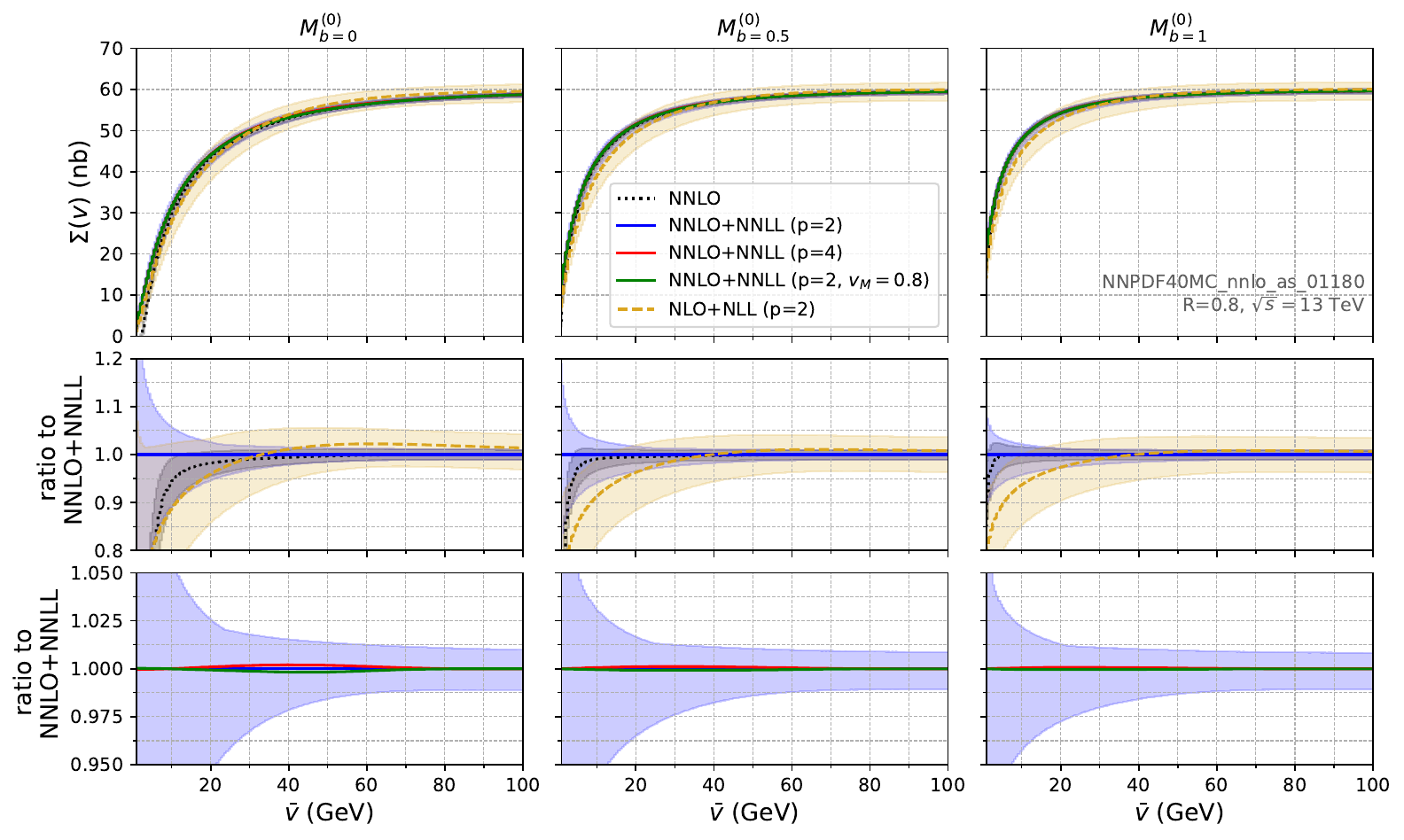}
	\caption{Same as  Fig.~\ref{fig:matching-mbeta} but for the
          cumulative distribution.
	}
	\label{fig:matching-mbeta-cumul}
\end{figure}

\begin{figure}[tb]
	\centering
	\includegraphics[page=2, width=\textwidth]{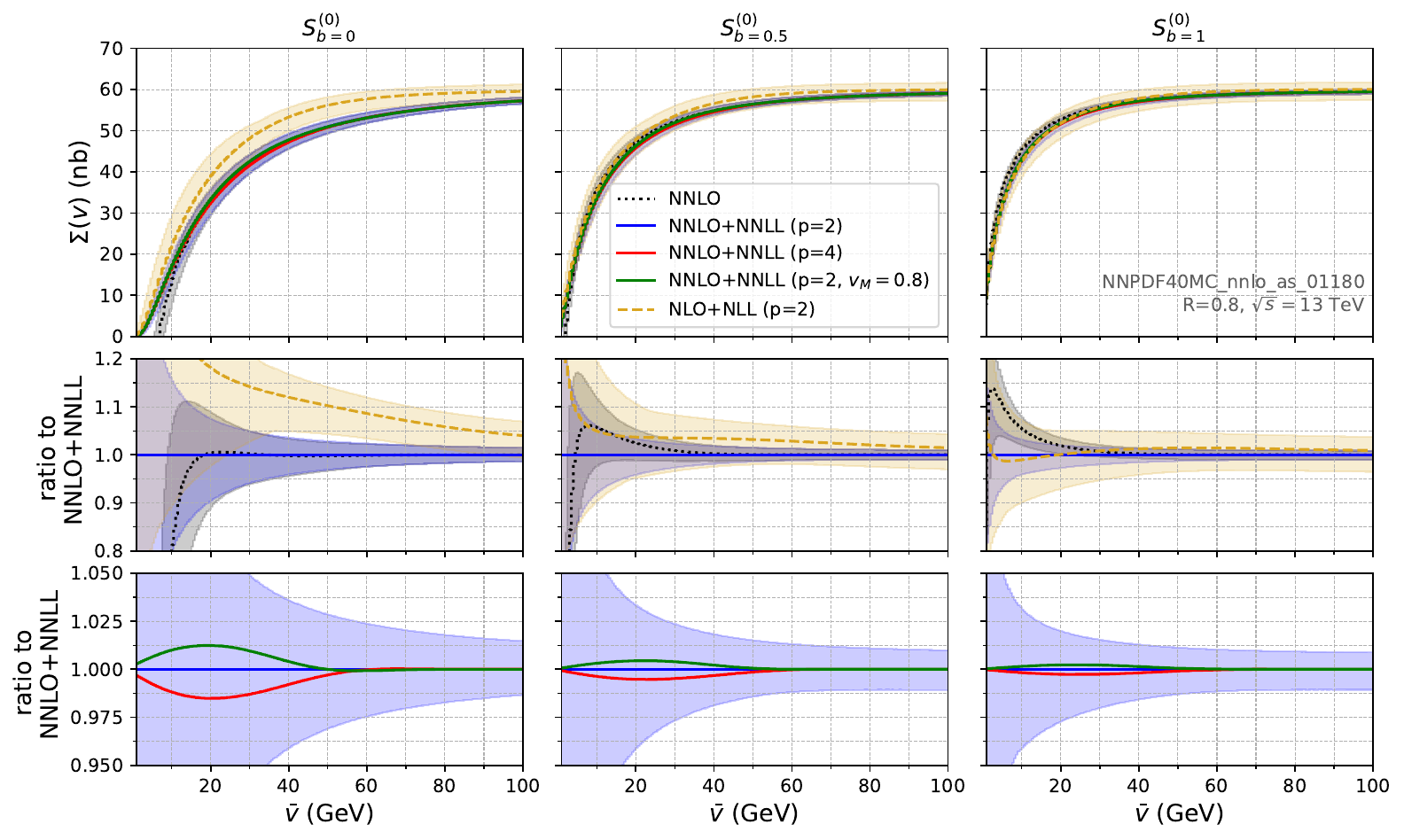}
	\caption{Same as Fig.~\ref{fig:matching-mbeta-cumul}  but for $\Sb{0}$. }
	\label{fig:matching-sbeta-cumul}
\end{figure}

In Fig.~\ref{fig:matching-mbeta} we show the differential
distributions for $\Mb{0}$ with $b=0, 1/2$ and $1$ at various levels
of accuracy, with the band representing the perturbative uncertainty
as outlined above.
To avoid confusion, we stress that the fixed-order accuracy for the
differential distributions is actually one order down w.r.t.~that of
the cumulative counterpart, i.e.~a NNLO cumulative cross section
corresponds to a NLO differential distribution.
To show the scales probed by the observables in the plots, we display a
dimensionful version $\bar{v} \equiv v\,Q$ on the $x$ axis.
One clearly sees the effect of the resummation becoming relevant at low
values of $\bar{v}$, while the matched curves tend to the fixed-order result
around $\bar{v} \sim \mum$, as guaranteed by the form of the resummation
scale $\mul$ given in Eq.~\eqref{eq:muLtilde}. 
The perturbative uncertainties are significantly reduced by going from
NLL+NLO to NNLL+NNLO, and the two uncertainty bands of the two
predictions show a good degree of overlap at small $\bar{v}$,
indicating the robustness of the error estimates. At larger values of
$\bar{v}$, where the prediction is given by the fixed order, the large
difference between the NLO and NNLO predictions is representative of
the fact that the former is only LO accurate for the $\bar{v}$
spectrum, and hence subject to large radiative corrections captured by
the higher-order predictions.
As a further check of the stability of our matching procedure, we also
show the prediction with a different value for $p$ in
Eq.~\eqref{eq:muLtilde} ($p=4$), as well as the prediction with
$\mum=0.8\,Q$. This does not affect, by design, the distributions for
$\bar{v} \to 0$ and has the effect of shifting mildly the central
value in the matching region $\bar{v}\sim \mum$, above which the
matched curves reproduce the fixed-order result. The predictions
obtained with these variations are contained within the error band,
suggesting that the matching uncertainties are under control.
Quantitatively similar features can be seen for $\Sb{0}$ shown in
Fig.~\ref{fig:matching-sbeta}.
Furthermore,
Figs.~\ref{fig:matching-mbeta-cumul},~\ref{fig:matching-sbeta-cumul}
show analogous predictions at the cumulative cross-section level.

\subsection{Impact of multi-parton scattering and hadronisation}
We now study the impact of hadronisation and underlying event on the
\lobs.
We do so by performing \pythiaeight~\cite{Bierlich:2022pfr} simulations
for $pp\to Z$ at 13 TeV. We keep the width of the $Z$-boson fixed to a
small value so as to mimic the on-shell condition used in our
perturbative calculations.
All other parameters, including the jet clustering settings and the
chosen PDF set are the same as in the previous section.
We perform the study at the level of the cumulative distributions
since they are more relevant to phenomenological applications of the
\lobs as jet vetoes. All distributions are now normalised
by the total cross section, which coincides with the Born cross
section in the \pythiaeight simulations.

\begin{figure}[tb]
	\centering
   	 \includegraphics[page=1,width=\textwidth]{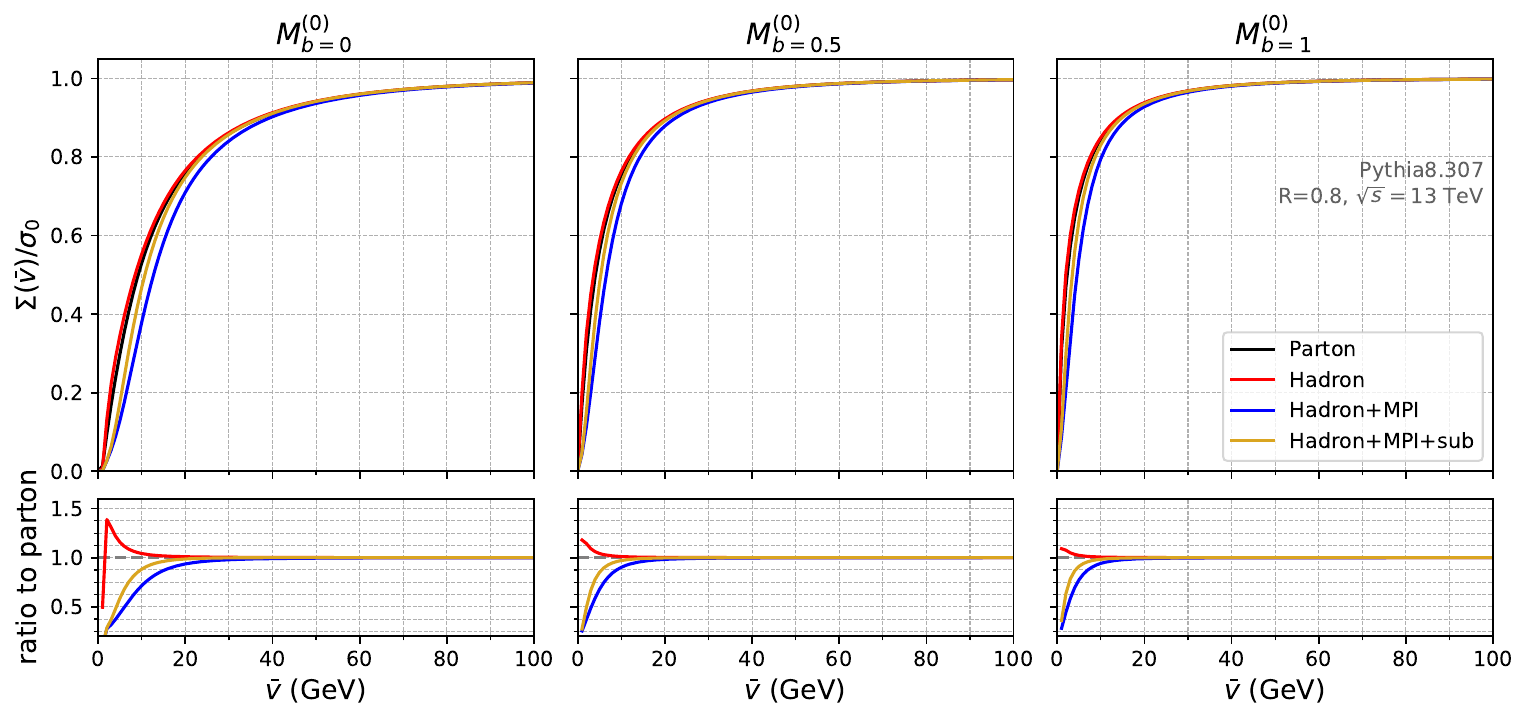}
	\caption{Cumulative distribution for  $\Mb{0}$ with $b=0$
          (left), $b=1/2$ (middle) and $b=1$ (right) as a function of $\bar{v} = v\,Q$.
   Shown are four different setups: parton-level (black), hadron-level (red), hadron-level with multi-parton interactions (blue) and
	hadron-level with multi-parton interactions and area subtraction (yellow). 
	The bottom panel displays the ratio to the parton-level
        result.}
	\label{fig:pheno-mbeta}
\end{figure}

The results for $\Mb{0}$ are shown in Fig.~\ref{fig:pheno-mbeta}. 
We observe that hadronisation corrections remain below 5\% across the
region of $\Mb{0}$ relevant for experimental analyses,
i.e.~$Q\,\Mb{0}>10$ GeV.
In turn, multi-parton interactions (MPI) lead to a sizeable
correction, up to $50\%$ at $Q\,\Mb{0}=10$ GeV for all values of $b$,
but they are reduced swiftly as the value of $\Mb{0}$ is increased,
becoming very small at scales of order $Q\,\Mb{0}\sim 25-30$ GeV.
This can be compared to the size of the NNLL+NNLO uncertainty band in
Fig.~\ref{fig:matching-mbeta-cumul} which is below 10\% at 10 GeV and
below 5\% at 20 GeV.
Moreover, we notice that the impact of the MPI activity depends on the
value of $b$, and it gets reduced for larger $b$ values in line with
the fact that radiation at larger absolute rapidities contributes less
to the observables.
An experimental measurement of these observables for several values
of the parameter $b$ could therefore inform about the rapidity
distribution of MPIs, which is an aspect that is largely unconstrained
so far.
To modulate the impact of MPI, one can use the area subtraction
algorithm~\cite{Cacciari:2007fd,Cacciari:2008gn} and compute $\Mb{0}$
using subtracted jets.\footnote{For \lobs involving Lund
  declusterings, one could also compute the Lund coordinates using
  subtracted subjets. Alternatives would include, for example, using
  direct shape subtraction~\cite{Soyez:2012hv}, or the {\tt
    ConstituentSubtractor} method~\cite{Berta:2014eza} or an
  event-wide subtraction techniques such as {\tt
    SoftKiller}~\cite{Cacciari:2014gra} or {\tt
    PUPPI}~\cite{Bertolini:2014bba} prior to the clustering.}
After area subtraction, we find less than 10\% deviations with respect
to parton-level for $Q\,\Mb{0}>10$ GeV.
Therefore, we conclude that $\Mb{0}$ is not severely affected by
hadronisation and MPI corrections, thus enabling meaningful
theory-to-data comparisons.
 
\begin{figure}[tb]
	\centering
   	 \includegraphics[page=2,width=\textwidth]{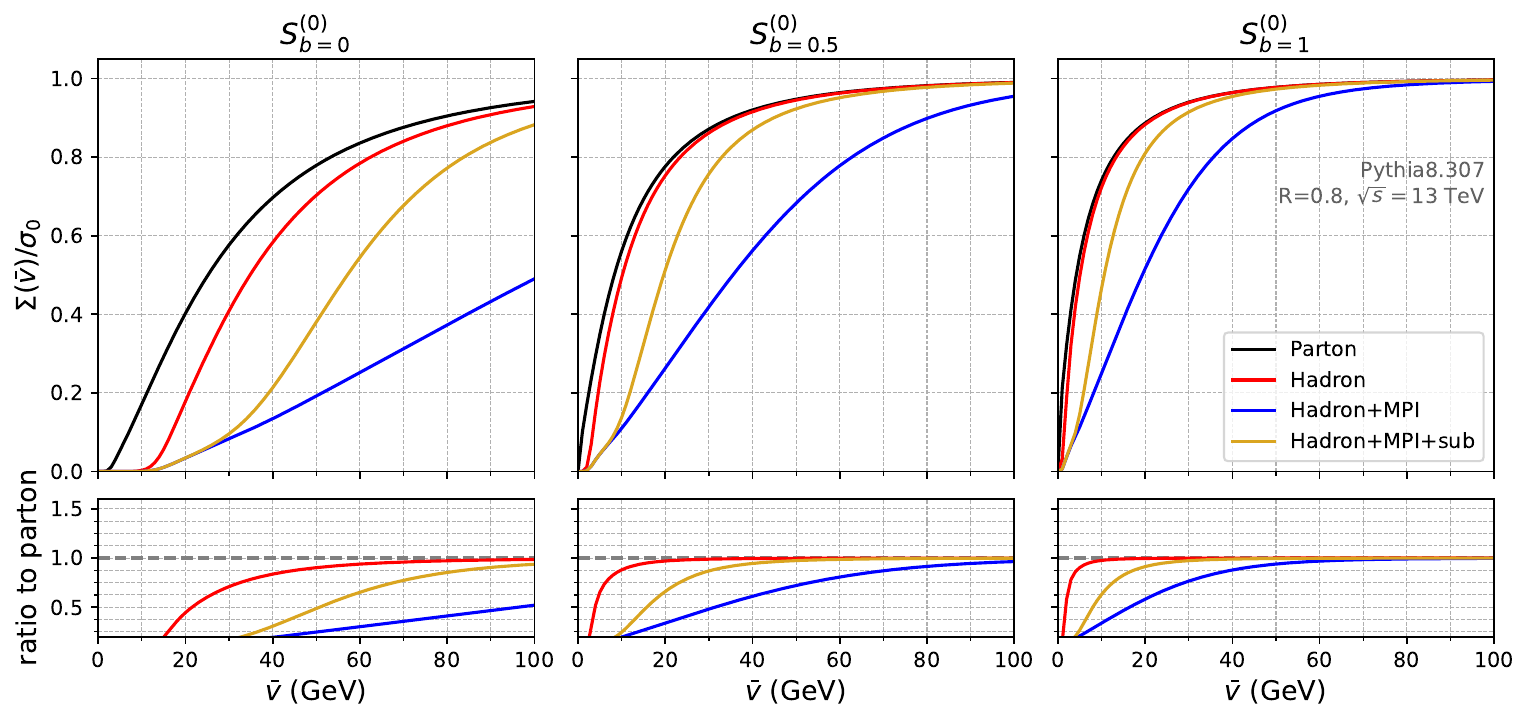}
	\caption{Same as Fig.~\ref{fig:pheno-mbeta} but for $\Sb{0}$.}
	\label{fig:pheno-sbeta}
\end{figure}

The results for $\Sb{0}$ are shown in Fig.~\ref{fig:pheno-sbeta}. 
In this case, both hadronisation and MPI corrections are much more
sizeable than for $\Mb{0}$, in particular when $b\neq 1$.
Here, hadronisation corrections lead to a $5-10$ GeV shift in the peak
of the distribution, while MPI completely distort the shape.
This is natural given the fact that the observable does not impose any
cut on the jets that enter the sum, see Eq.~\eqref{eq:LES-pp-0}, and
thus all semi-soft jets arising from MPI have a non-vanishing
contribution to $\Sb{0}$. This issue affects all additive observables
defined without a jet cut.
Area subtraction manages to bring the effect of MPI down to a factor
of $3$, which is still quite large for theory-to-data comparisons.
The situation improves slightly when choosing $b=1$.
In this case, hadronisation corrections are of the same size as for
$M^{(0)}_{b=1}$, i.e.\ below 5\%, and MPI contamination is also reduced
once including area subtraction to less than 10\% for
$Q\,S^{(0)}_{b=1}>40$ GeV.
We thus expect large deviations between experimental measurements of
$\Sb{0}$ and our NNLL+NNLO predictions. To mitigate this discrepancy,
one could think about measuring $\Sb{0}$ inside groomed jets
(cf.~Sec.~\ref{sec:groom}) instead of at event-wide level. Changing
the definition of the observable of course modifies the resummation
structure and we thus leave the study of this possibility for future
work.

\section{Conclusions}
\label{sec:conclusions}
In this work, we have introduced Lund-Tree Shapes (\lobs), a new
class of collider observables defined from the Lund-plane
representation of QCD radiation in multi-jet scattering processes.
For processes with two emitting legs, these observables were first
introduced, at NLL, for the purpose of testing the logarithmic
accuracy of parton showers in $\lep$~\cite{Dasgupta:2020fwr},
$\had$~\cite{vanBeekveld:2022ukn} and
$\dis$~\cite{vanBeekveld:2023chs} collisions. In this article, we
presented a generalisation to processes with any number of resolved
jets. We also provide a variant of \lobs that can be measured on
groomed jets, offering an additional tool for the study of jet
structure at different collider environments.
These observables provide a versatile tool for probing multi-jet final
states. At the differential level, \lobs behave as continuous, global
variables, analogous to classical event shapes and $n\to n+1$
jet-resolution parameters, allowing detailed studies of the
hierarchical and geometric structure of QCD final states. At the
cumulative level, they naturally define jet rates, useful for
resolution-based characterisation of multi-jet events or as jet
vetoes.

The generality of \lobs makes them applicable to scattering processes
with any number of resolved jets, across $\lep$, $\had$ and $\dis$
colliders, as well as within groomed jets. From a theoretical
perspective, they serve as versatile resolution variables for
phase-space slicing in higher-order calculations, for matching
fixed-order predictions to parton showers, and for testing the
logarithmic accuracy of shower algorithms. They also feature good
calculability properties, as they are free of non-global logarithmic
corrections regardless of the number of jets.

As an initial application, we have presented new NNLL predictions for
two-leg processes at $\lep$, $\had$ and $\dis$ colliders, and
NNLL+NNLO predictions for the LHC, discussing also the practical
utility of \lobs in collider phenomenology. Beyond these examples,
Lund-Tree Shapes offer a promising avenue for systematic studies of
multi-jet processes, enabling both precision QCD tests and novel analyses
of jet events at present and future collider experiments.
The \lobs can be easily calculated from the Lund jet plane
declusterings obtained from the public version of the
\texttt{LundPlane}~\cite{Dreyer:2018nbf} contrib of
\texttt{FastJet}~\cite{Cacciari:2011ma}.\footnote{Available at
  \url{https://fastjet.fr/contrib/}. The implementation of the plugin for DIS is in preparation~\cite{DISLundFastjet}.}


\section*{Acknowledgements}
We are grateful to Basem El-Menoufi for his contributions in the early
stages of this project, and Gavin Salam for discussions and comments
on the article.  This research was supported by the Italian Ministry
of Universities and Research (MUR) under the FIS grant (CUP:
D53C24005480001, FLAME) (SFR), the Ramón y Cajal program under grant
RYC2022-037846-I and from ERDF (grant PID2024-161668NB-100) (ASO), by
the Dutch Research Council (NWO) under project number VI.Veni.232.190
(MvB) and by the European Research Council (ERC) under the European
Union's Horizon 2020 research and innovation programme (grant
agreement No.\ 788223, ASO, GS). The work of LB and PM is funded by
the European Union (ERC, grant agreement No. 101044599).  Views and
opinions expressed are however those of the authors only and do not
necessarily reflect those of the European Union or the European
Research Council Executive Agency. Neither the European Union nor the
granting authority can be held responsible for them.

\appendix

\section{Perturbative expansion for resummation ingredients}
\label{app:ingredients}
In this appendix, we will provide the reader with the expressions for
the resummation ingredients not provided in the main text, needed to
reproduce the results of the paper. Following the main text, we will
use the following notation throughout the appendix
\begin{equation}\label{eq:defs}
\as \equiv \as(\mur),  \qquad L \equiv \ln\frac{\mul}{Q v}, \qquad
\lambda \equiv \as \bqcd{0} L,
\end{equation}
where $\bqcd{0}$ is the one-loop QCD $\beta$-function, and its value
is given in Appendix~\ref{app:constants}. The resummation scale $\mul$
is meant to be replaced with the dynamical definition of
Eq.~\eqref{eq:muLtilde} in all occurrences.

\subsection{Splitting functions and coefficient functions}
The $\bo$-dependent coefficient functions $C_{ki}^{(1)}(z)$ are a new result of this work. Before reporting their expression, let us introduce the regularised Altarelli-Parisi splitting functions $\hat{P}_{k,i}(z;\epsilon)$ in $d=4-2\epsilon$ dimensions 
\begin{align}
 \hat{P}_{qq}(z;\epsilon) &= C_F \left[\frac{1+z^2}{(1-z)_+} -\epsilon (1-z)+\frac{3}{2}\delta(1-z) \right]\,, \nonumber\\
 \hat{P}_{gq}(z; \epsilon) &=  C_F \left[\frac{1+(1-z)^2}{z} -\epsilon z \right]\,,\nonumber\\
 \hat{P}_{qg}(z; \epsilon) &= T_R \left[ 1-\frac{2z(1-z)}{1-\epsilon}\right] = T_R[1-2z(1-z) -2 \epsilon z(1-z) ] +\mathcal{O}(\epsilon^2)\,,\nonumber\\
  \hat{P}_{gg}(z; \epsilon) & = \frac{2 C_A \left(1+z^2-z\right)^2}{z(1-z)_+}+2 \pi \beta_0 \delta (1-z).
\label{eq:splittingfuncs}
\end{align}
We will denote by $ \hat{P}_{ki}^{(0)}(z)$ their expression for
$\epsilon\to 0$, while $ \hat{P}_{ki}^{(\epsilon)}(z)$ is the
coefficient of the ${\cal O}(\epsilon)$ term, i.e.
\begin{equation}
 \hat{P}_{ki}(z; \epsilon) = \hat{P}_{ki}^{(0)}(z) + \epsilon \hat{P}_{ki}^{(\epsilon)}(z).
\end{equation}
We also introduce the unregularised splitting function
$ P_{ki}^{(0)}(z)$, which can be obtained from $\hat{P}_{ki}^{(0)}(z)$ omitting the terms proportional to $\delta(1-z)$ and replacing $(1-z)_+ \to (1-z)$.
In term of these functions, the coefficient functions can be written
as
\begin{align}  \label{eq:coefffun}
 C_{ki}^{(1)}(z) =& - \hat{P}_{ki}^{(\epsilon)}(z)+ \frac{2 \bo}{1+\bo}(1-z)  P_{ki}^{(0)}(z)\left[ \left(\frac{\ln(1-z)}{1-z}\right)_+-\frac{\ln(z)}{1-z} \right]
 \\
 &\hspace{2cm} +2\left(\ln \frac{\Q}{\muf}+\frac{1}{1+\bo}\ln \frac{\mul}{\Q}\right)P_{ki}^{(0)}(z).
 \nonumber
\end{align}

\subsection{The Sudakov radiator}
\label{sec:sudakovrad}
The Sudakov radiator $\mathcal{R}$ is separated into a soft (s) and hard-collinear (hc) terms, i.e.
\begin{equation}
\mathcal{R}_{\rm NNLL}(v) = \sum_{\ell=1}^2 \left[ R^{\ell}_{\text{s}}(v) +  R^{\ell}_{\text{hc}}(v) \right],
\end{equation}
where we also separate contributions associated with radiation from
each leg $\ell$.  At NNLL accuracy, the soft radiator $R_s^{\ell}$
reads
\begin{align}
  R_{\rm s}^{\ell}(v)=& -\frac{\lambda}{\as \beta_0} g_1^{\ell}(\lambda) -  g_2^{\ell}(\lambda) -\frac{\as}{\pi} g_3^{\ell}(\lambda) -\frac{\as}{\pi} \delta g_3^{\ell}(\lambda)
  \end{align}
where
{\allowdisplaybreaks
\begin{align}
\label{eq:g1l}
  g_1^{\ell} (\lambda) = \frac{C_\ell}{2} &\frac{ (1+\bo-2\lambda) \ln \left(1 - \frac{2\lambda}{1+\bo}\right) - (1 - 2 \lambda ) \ln \left(1 -  2 \lambda \right)}{\pi  \bo \beta _0 \lambda}\,, \\\nonumber
  g_2^{\ell} (\lambda) 
  = \frac{C_\ell}{2}  
                        &\Bigg[ \frac{ K^{(1)} \left( \ln \left( 1 - 2 \lambda
                           \right)- (1+\bo) \ln \left( 1 - \frac{2 \lambda
                           }{1+\bo}\right)\right)}{2 \pi ^2 \bo \beta _0^2}  \\\nonumber
                         &+  \frac{\beta _1 (1+\bo) \ln ^2\left( 1 -\frac{2\lambda}{1+\bo}  \right)}{ 2 \pi \bo \bqcd{0}^3} + \frac{\bqcd1 (1+\bo)
                           \ln \left(1 - \frac{2 \lambda }{1+\bo}\right)}{\pi \bo \bqcd{0}^3} \\
                         & - \bqcd1\frac{ \ln \left( 1 - 2 \lambda \right) \left(\ln \left(1- 2 \lambda
                          \right)+2\right)}{2 \pi \bo \bqcd{0}^3}
+\frac{2}{\pi\bqcd{0}}\lnxM \ln \left(1-\frac{2 \lambda}{1+\bo} \right) \\
&    +\frac{2}{\bo\pi\bqcd{0}}\lnxmuR\left[\ln (1-2 \lambda )-(1+\bo)\ln \left(1-\frac{2 \lambda}{1+\bo} \right)\right]
                           \Bigg]\,, \nonumber\\
\nonumber
  g_3^{\ell} (\lambda) = \frac{C_\ell}{2} &\Bigg[ K^{(1)}  \frac{
                           \beta _1 \left( (1+\bo+2 \lambda )
                           \ln \left( 1 -2\lambda\right)
                           - (1+\bo)^2 (1-2\lambda) \ln \left( 1 -
                           \frac{2 \lambda }{1+\bo}\right) + 6
                           \bo \lambda^2 \right)}{2 \pi \bo
                           \bqcd{0}^3 (1-2\lambda) (1+\bo-2\lambda)}
  \\\nonumber
&+\frac{\left(\text{$\bqcd1 $}^2 (1+\bo)^2 (1-2 \lambda ) \ln ^2\left(1-\frac{2 \lambda }{1+\bo}\right)-4 \bo \lambda ^2 \left(\text{$\bqcd{0} $}
   \text{$\bqcd2 $}+\text{$\bqcd1 $}^2\right)\right)}{2 \bo
\text{$\bqcd{0} $}^4 (1-2 \lambda ) (1+\bo-2 \lambda )}\\\nonumber
&-\frac{ \ln \left(1-2\lambda\right) \left(2 \text{$\bqcd{0} $} \text{$\bqcd2 $} (1-2 \lambda )+\text{$\bqcd1 $}^2 \ln
   \left(1-2\lambda\right)+4 \text{$\bqcd1 $}^2 \lambda
 \right)}{2 \bo \text{$\bqcd{0} $}^4 (1-2 \lambda )}\\\nonumber
&+\frac{(1+\bo) \ln \left(1-\frac{2 \lambda }{1+\bo}\right) \left(\text{$\bqcd{0} $} \text{$\bqcd2 $} (1+\bo-2 \lambda )+2 \text{$\bqcd1 $}^2 \lambda
   \right)}{\bo \text{$\bqcd{0} $}^4 (1+\bo-2 \lambda )}\\
                         &- K^{(2)} \frac{  \lambda^2 }{2\pi^2 (1-2\lambda) (1+ \bo - 2\lambda) \bqcd{0}^2} \nonumber \\
& 
-\lnxmuRsq \frac{8\lambda^2}{(1-2\lambda)(1+b-2\lambda)}
\nonumber \\
 &+\lnxmuR \frac{2\bqcd{1}\left(4\bo{}\lambda^2-(1+\bo)^2(1-2\lambda)\ln\left(1-\frac{2\lambda}{1+b}\right)+(1+\bo-2\lambda)\ln(1-2\lambda)\right)
 }{\bo{}\bqcd{0}^2(1-2\lambda)(1+b-2\lambda)}  \nonumber \\
 &- \lnxmuR \frac{4 K^{(1)}\lambda^2}{\bo{}\bqcd{0}^2(1-2\lambda)(1+b-2\lambda)} \nonumber  \\
 &
 + 4 \lnxMsq 
\frac{ \bo{} \,\lambda }{ (1+\bo) (1+\bo-2\lambda) }
-  4 \lnxM\lnxmuR \cdot \frac{ 2 \lambda }{ 1+\bo-2\lambda } \nonumber \\
&
+ 4 \lnxM 
\left[
   \frac{ \bqcd{1} }{ 2 \bqcd{0}^2 }
   \frac{ 2\lambda + (1+\bo) \cdot \ln\bigl(1 - \tfrac{2\lambda}{1+\bo}\bigr) }{ 1+\bo-2\lambda }
 - \frac{ K^{(1)} \,\lambda }{ 2 \pi \bqcd{0} \bigl(1+\bo-2\lambda\bigr) }
\right]
                         \Bigg]\,,
\label{eq:g3l}                  
                         \\
  \delta g_3^{\ell}(\lambda) &= -C_\ell\frac{\pi^2}{6} \frac{\lambda }{(1+\bo-2\lambda)}\,.
  \label{eq:dg3l}   
\end{align}
}%
The constants $K^{(i)}$ and the QCD beta function coefficients $\bqcd{i}$ are given in appendix~\ref{app:constants}.
 We also introduce various derivatives of the NLL soft radiator, as they will appear in the evaluation of the transfer function.
{\allowdisplaybreaks
 \begin{subequations}
\label{eq:sudradder}
 \begin{align}
\mathcal{R}^\prime_{\text{NLL},\ell} (v)\equiv&-\frac{\partial \left[g_1(\as \bqcd0 L) L\right] }{\partial L} =  \frac{C_\ell}{\bo \pi \bqcd0} \left( \ln \left( 1- \frac{2\lambda}{1+\bo} \right) -  \ln \left( 1- 2\lambda \right) \right)\,, \label{eq:RpNLL} \\ 
\mathcal{R}^\prime_{\text{NNLL},\ell}(v) \equiv& -\frac{\partial \left[g_2(\as \bqcd0 L) \right] }{\partial L} = \frac{C_\ell \alpha_s}{\bo \pi^2 \bqcd0^2 (1-2\lambda) (1+\bo -2\lambda)} \bigg[ \bo \bqcd0 \lambda K^{(1)} - 2 \pi \bo \bqcd1 \lambda \nonumber \\
&- \pi  (1+\bo -2 \lambda) \bqcd1 \ln \left( 1- 2\lambda \right) + \pi (1+\bo) (1-2\lambda) \bqcd1 \ln \left( 1- \frac{2\lambda}{1+\bo} \right)\bigg]\nonumber\\
&
+\frac{2C_{\ell}\alpha_s}{\pi}
\left[\lnxM(1-2\lambda)+\lnxmuR\frac{2\lambda }{(1-2\lambda)(1+\bo-2\lambda)}\right]
\,, \label{eq:RpNNLL} \\
\mathcal{R}^{\prime\prime}_{\ell} (v)\equiv& \alpha_s \bqcd0 \frac{d \mathcal{R}^\prime_{\text{NLL},\ell}}{d \lambda} =  2 C_\ell \frac{\alpha_s}{\pi} \frac{1}{(1-2\lambda)(1+\bo-2\lambda)}\,.\label{eq:RppNLL} 
\end{align}
\end{subequations}
}
 Notice that by convention the subscript $s$ is dropped.
 It can be verified that the limit $b\to 0$ is continuous.
 We also define the quantities summed over both legs
 \begin{equation}
   \mathcal{R}^\prime_{\text{NLL}} (v)= \sum_{\ell} \mathcal{R}^\prime_{\text{NLL},\ell}(v), \quad
   \mathcal{R}^{\prime \prime} (v)= \sum_{\ell} \mathcal{R}^{\prime \prime}_{\ell}(v), \quad
   \mathcal{R}^\prime_{\text{NNLL}}(v) = \sum_{\ell} \mathcal{R}^\prime_{\text{NNLL},\ell}(v).
\end{equation}
At NNLL accuracy, the hard-collinear radiator reads
\begin{align}
  R_{\rm hc}(v) = - \sum_{\ell}\left[ h_2^{\ell}(\lambda) +\frac{\as}{\pi} h_3^{\ell}(\lambda)\right]\,,
  \end{align}
with
\begin{align}
  h_2^{\ell}(\lambda) =& \frac{\gamma^{(0)}_\ell}{2 \pi \bqcd0} \ln\left(1-\frac{2\lambda}{1+\bo}\right)\,,  \\
  h_3^{\ell}(\lambda) =& \gamma^{(0)}_\ell\frac{\bqcd1 \left((1+\bo) \left(\ln \left( 1- \frac{2 \lambda
                          }{1+\bo}\right)\right)+ 2\lambda \right)}{ 2\beta^2_0 \left(1+\bo - 2 \lambda\right)} -\gamma^{(1)}_\ell \frac{\lambda}{2\pi \bqcd0 (1+\bo-2\lambda)} \nonumber \\
                          &+\frac{2\gamma^{(0)}_\ell}{1+\bo-2\lambda}\left[\frac{\bo}{1+\bo}\lnxM-\lnxmuR\right]
\end{align}
The constants $\gamma_\ell^{(i)}$ are  given in appendix~\ref{app:constants}.

\subsection{The NNLL transfer function}
The transfer function $\cF(v)$ is decomposed as
\begin{equation}
{\cal F}(v) = {\cal F}_{\rm NLL}(v) +
                     \frac{\alpha_s(\mur)}{\pi}\delta {\cal F}_{\rm NNLL}(v)\,.\\
\end{equation}
The NNLL corrections can be parameterised as follows
\begin{align}
\delta \FNNLL(v) =&\dFsc(v)+\dFwa(v)+\dFcorrel(v)+\dFclust(v)+\dFhc(v)+\dFrec(v).
\end{align}
At NLL we obtain, for $\lep$, $\had$ and $\dis$,
\begin{equation}
  \FNLL(v) =
  \begin{cases}
     1 \qquad &\mbox{ for }\Mb{n_{\min}}, \\
     \displaystyle
     \frac{e^{-\gamma_E   \RpNLL(v)}}{\Gamma(1+\RpNLL(v))} \qquad &\mbox{ for }\Sb{n_{\min}}.
  \end{cases} 
  \label{eq:Fnll}
\end{equation}
Below we provide all NNLL corrections.

\paragraph{Soft-collinear correction}
The soft-collinear correction is $0$ for the $\Mb{n_{\min}}$ observable. For $ \Sb{n_{\min}}$ it reads
\begin{align}
         \label{eq:dFsc}
  \dFsc(v) =
      & - \frac{\pi}{\alpha_s} {\cal F}_{\rm NLL}(v)\sum_{\ell=1,2}  \left[   
       \mathcal{R}'_{{\rm NNLL},\ell}(v)
        \left(\psi^{(0)}(1+\RpNLL (v))+\gamma_E \right) \right. \\
       & \left. \quad +\frac{\mathcal{R}''_{\ell}(v)}{2} \left(\left(\psi ^{(0)}(1+\RpNLL (v))+\gamma_E\right)^2-
       \psi ^{(1)}(1+\RpNLL (v))+\frac{\pi ^2}{6}\right) \right]\,. \nonumber
\end{align}
where the derivatives of the soft radiator are given in
Eqs.~\ref{eq:sudradder}. The term in Eq.~\eqref{eq:dFsc} proportional
to $\mathcal{R}''_{\ell}(v)$ originates from the expansion of the
all-order functional (common in the resummation structure of additive
global event shapes)
\begin{equation}
  e^{-\frac{\mathcal{R}''}{2}\partial^2_{\mathcal{R}'}}{\cal F}_{\rm
    NLL}(v)\,,\quad \mathcal{R}''(v) =
  \sum_{\ell=1,2}\mathcal{R}''_{\ell},\,\quad \mathcal{R}'(v) = \sum_{\ell=1,2}\mathcal{R}'_{\ell}\,,
\end{equation}
that is expanded to second order neglecting N$^3$LL corrections. In
our numerical implementation we leave the above correction unexpanded
up to N$^3$LL corrections, that is we use
\begin{equation}
  e^{-\frac{\mathcal{R}''}{2}\partial^2_{\mathcal{R}'}}{\cal F}_{\rm
    NLL}(v) = e^{-\frac{\mathcal{R}''}{2}\partial^2_{\mathcal{R}'}\ln {\cal F}_{\rm
      NLL}(v)}{\cal F}_{\rm
    NLL}(v) + {\cal O}(\text{N$^3$LL})\,,
\end{equation}
to prevent the resummed distribution from becoming unphysical
(negative) to the left of the Sudakov peak, i.e.~at very small values
of $v$, where the $\Sb{n_{\min}}$ distribution is dominated by
non-perturbative corrections.

\paragraph{Wide angle correction}
\label{app:swa}
The wide-angle correction to the transfer function is zero for the
observables we study in this paper in the $\ee$ and $\dis$ cases. For
$\had$ collisions, the wide-angle correction is zero when considering
the observables defined in the partonic c.o.m.~frame, while it is
given in Eq.~\eqref{eq:framesoft2} for the hadronic c.o.m.~frame
definition.
Note that, for $ee$ collisions, the wide-angle correction is the main
difference between $M_0^{(2)}$ and the $y_{23}$ Cambridge resolution
scale, which is zero for the former but not for the latter.

\paragraph{Correlated correction}
The correlated correction reads
\begin{align}
\label{eq:dFcorrel}
\dFcorrel(v) = \cF_{\rm NLL} (v) \frac{\lambda\pi \mathcal{R}^{\prime \prime}(v)}{\alpha_s}(C_A \langle \ln f_{\rm correl} \rangle _{C_A}+ n_f T_R \langle \ln f_{\rm correl} \rangle _{n_f}).
\end{align}
We evaluate $\langle \ln f_{\rm correl} \rangle _{C_A, n_f}$ numerically, and they are defined as
\begin{equation}
\langle \ln f_{\rm correl} \rangle _{C_A, n_f} = \int_0^{\infty} \frac{d\zeta}{\zeta} \int_{-\pi}^{\pi} \frac{d\phi}{2\pi} \int_{-\infty}^{\infty} d\eta \, \mathcal{M}_{ C_A,n_f} \ln \frac{V_{\rm simple}^{\rm corr}(\zeta,\phi,\eta,\beta)}{V_{\rm sc}(\zeta,\phi,\eta,\beta)}
\end{equation}
where the variables $\eta$, $\phi$ and $\zeta$ are given by
\begin{equation}
\eta = \eta_2 -\eta_1, \quad \phi = \phi_2-\phi_1, \quad \zeta = \frac{k_{t,2}}{k_{t,1}},
\end{equation}
being $k_{t,i}$, $\eta_i$ and $\phi_i$ the transverse momentum, rapidity and azimuth of the soft emission labelled with the index $i$.
The matrix elements are given by 
\begin{subequations}
\begin{align}
\mathcal{M}_{n_f} =& \frac{\zeta  e^{2 \eta } \left(-\cos (\phi ) \left(\zeta ^2+2 \zeta  \cosh (\eta )+1\right)+\zeta ^2 \cosh (\eta )+2 \zeta +\cosh (\eta )\right)}{2 \left(\zeta +e^{\eta }\right)^2 \left(\zeta  e^{\eta }+1\right)^2 (\cos (\phi )-\cosh (\eta ))^2} \,,\\
\mathcal{M}_{C_A} =& \frac{-\zeta  \left(\left(\zeta ^2+1\right) \cosh (\eta ) \sin ^2(\phi )+\zeta  \cosh ^2(\eta ) \cos (2 \phi )-\zeta \right)}{2\left(\zeta ^2+2 \zeta  \cosh (\eta )+1\right)^2 (\cos (\phi )-\cosh (\eta ))^2} \\
& -\frac{\cos (\phi ) }{2 (\cos (\phi )-\cosh (\eta ))}-\frac{\sinh ^2(\eta ) \cos (\phi ) }{2 \left(\zeta ^2+2 \zeta  \cosh (\eta )+1\right)(\cos (\phi )-\cosh (\eta ))^2}\,. \nonumber
\end{align}
\end{subequations}
$V_{\rm simple}^{\rm corr}$ is the simple observable for a pair of correlated emissions and is given in terms of the transverse momentum and rapidity of the parent gluon $k=k_1+k_2$. In the limit $|\eta_1| \gg 1$, we have
\begin{equation}
V_{\rm simple}^{\rm corr} = \frac{k_{t,1}e^{-\bo \eta_1}}{Q} \left(\frac{\zeta +e^{\eta }}{\zeta  e^{\eta }+1}\right)^{\bo /2} e^{-\frac{\bo  \eta }{2}} \sqrt{\zeta ^2+2 \zeta  \cos (\phi )+1}.
\end{equation}
Instead $V_{\rm sc}$ is the value of the observable in the presence of
two soft and collinear emissions and it is decomposed into the sum of
two pieces
\begin{equation}
  V_{\rm sc}(\zeta,\phi,\eta,\bo) = V_{\rm clust}(\zeta,\phi,\eta,\bo)\Theta_{\rm clust} + V_{\rm no\, clust}(\zeta,\phi,\eta,\bo)(1-\Theta_{\rm clust}),
\end{equation}
where the first term refers to the case in which the two partons cluster together, and the second where they do not.
The clustering condition reads
\begin{equation}
  \Theta_{\rm clust} =
\begin{cases}
  \Theta(|\eta|<\ln 2 \cos \phi) \Theta(|\phi|<\pi/3) & \quad \mbox{ for DIS and }ee,\\
\Theta(R^2-\phi^2-\eta^2) &\quad \mbox{ for }pp,
 \end{cases}
\end{equation}
where $R$ is the jet radius.
When partons 1 and 2 do not cluster, the observable in the soft collinear limit becomes
\begin{equation}
 V_{\rm no clust}(\zeta,\phi,\eta,\bo)=
\frac{k_{t,1}e^{-\bo \eta_1}}{Q} \times
\begin{cases}
1+\zeta e^{-\bo \eta} \qquad &\mbox{ for }\Sb{n_{\min}}\,, \\
\max(1,\zeta e^{-\bo \eta}) \qquad &\mbox{ for }\Mb{n_{\min}}\,,
\end{cases}
\label{eq:Vnoclust}
\end{equation}  
and this expression is identical for all processes under consideration.
Instead when they cluster
\begin{equation}
 V_{\rm clust}(\zeta,\phi,\eta,\bo)=
\begin{cases}
  \frac{k_{t,1}e^{-\bo \eta_1}}{Q} \times \sqrt{\zeta ^2+2 \zeta  \cos (\phi )+1} \left(\frac{ \left(\zeta ^2 e^{2 \eta }+2 \zeta  e^{\eta }+1\right)}{\zeta ^2+2 \zeta  \cos (\phi )+1}\right)^{-\frac{\bo }{2}} \;\; &\mbox{ for DIS and }ee,\\
  V_{\rm simple}^{\rm corr}(\zeta,\phi,\eta,\bo) \quad &\mbox{ for }pp.
\end{cases}
\end{equation}
The numerical values of $\langle \ln f_{\rm correl} \rangle {C_A}$ and $\langle \ln f{\rm correl} \rangle _{n_f}$ for DIS and $\ee$ collisions, for $\bo = 0$, 1/2, 1, are reported in Tab.~\ref{tab:fcorrclust_eedis}.
For the case of $M_0^{(n_{\min})}$, we have verified that our results coincide with those for the Cambridge resolution variable, as presented in Ref.~\cite{Banfi:2016zlc}.

\begin{figure}[t!]
\includegraphics[width=\textwidth]{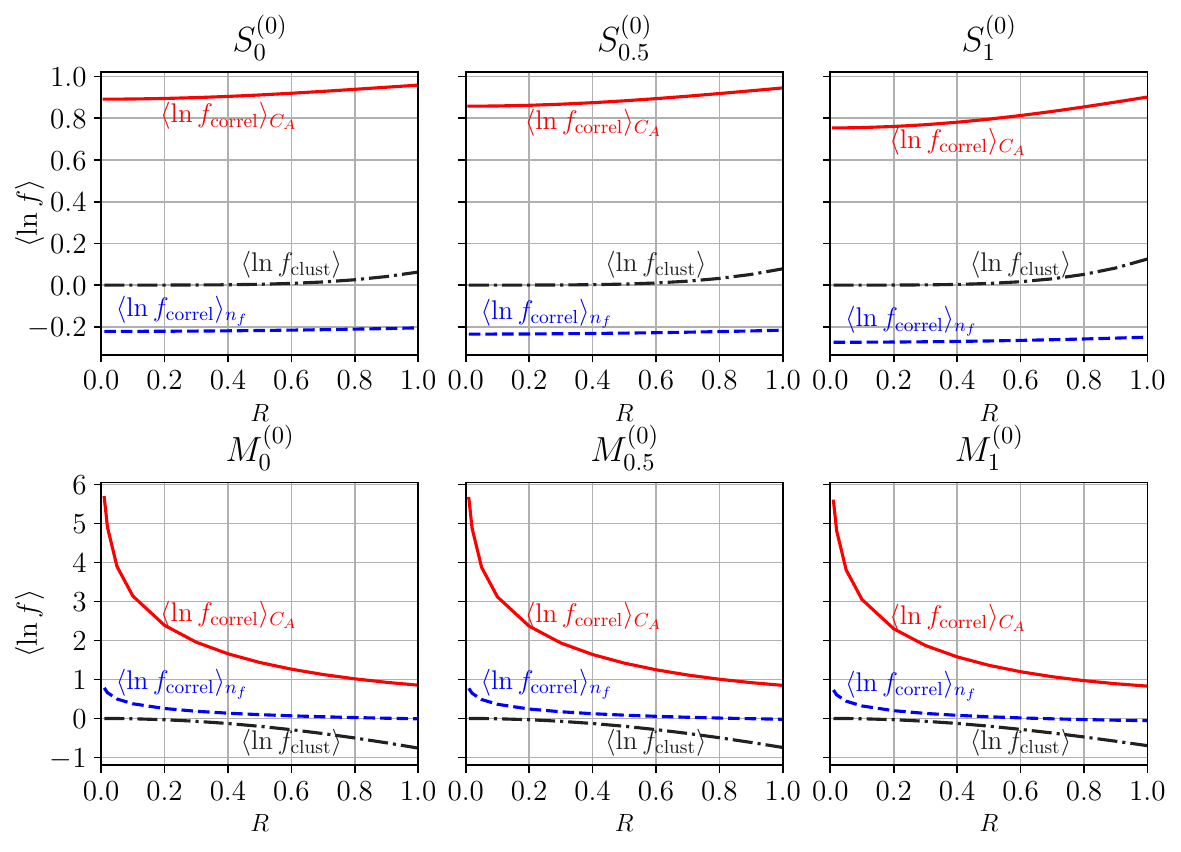}
\caption{Values for $\langle \ln f_{\rm correl}\rangle_{C_A}$ (red solid line), $\langle \ln f_{\rm correl}\rangle_{n_f}$ (blue dashed line), $\langle \ln f_{\rm clust} \rangle$ (grey dash-dotted line) for $\Sb{0}$ and $\Mb{0}$ in $\had$ collision for $\bo=0$, 1/2, 1 as a function of the jet radius $R$ for $0<R\leq 1$.
 }
  \label{fig:fcorrclust_pp}
\end{figure}

In Fig.~\ref{fig:fcorrclust_pp}, we illustrate the jet-radius ($R$)
dependence of these corrections in $pp$ collisions. Here, one can
observe that the correlated correction diverges as $\ln(1/R)$ for
$\Mb{0}$, as expected. We have also verified that, for small $R$, our
result for $M_0^{(0)}$ agrees numerically with the analytic expansion
for the jet veto given in Refs.~\cite{Banfi:2012jm,Abreu:2022sdc}.

\paragraph{Clustering correction}
The clustering correction is expressed in terms of the same quantities entering the correlated one, and reads
\begin{align}
\dFclust(v) = \cF_{\rm NLL} (v) \frac{\pi\lambda \mathcal{R}^{\prime \prime}(v)}{\alpha_s} C_{\ell} \langle \ln f_{\rm clust} \rangle,
\label{eq:dFclust}
\end{align}
with
\begin{equation}
\langle \ln f_{\rm clust} \rangle = \int_0^{\infty} \frac{d\zeta}{\zeta} \int_{-\pi}^{\pi} \frac{d\phi}{2\pi} \int_{-\infty}^{\infty} d\eta \, \ln \frac{V_{\rm simple}^{\rm indip}(\zeta,\phi,\eta,\beta)}{V_{\rm sc}(\zeta,\phi,\eta,\beta)}\Theta_{\rm clust},
\label{eq:lnfclust}
\end{equation}
where $V_{\rm simple}^{\rm indip} = V_{\rm no\, clust}$ is the simple observable for two independent emissions, and thus always equal to eq.~\eqref{eq:Vnoclust}, while
 $V_{\rm sc}$ and the clustering condition $\Theta_{\rm clust}$ are the same defined in the previous section.
 
 The numerical values of $\langle \ln f_{\rm clust} \rangle$ for DIS and $\ee$ collisions, for $\bo = 0$, 1/2, 1, are reported in Tab.~\ref{tab:fcorrclust_eedis}.
For the case of $M_0$, we have verified that our results coincide with those for the Cambridge resolution variable, as presented in Ref.~\cite{Banfi:2016zlc}.

For $\had$ collisions, Eq.~\eqref{eq:lnfclust} can be evaluated
analytically, yielding for jet radii $R \leq \pi$:
\begin{equation} \langle \ln f_{\rm clust}^{\had} \rangle = \begin{cases} \displaystyle \frac{R^4(1+\bo^2)}{16} & \mbox{ for } \Sb{0}\,,\\ \displaystyle \frac{R^4(1+\bo^2)}{16}-\frac{\pi^2 R^2}{12}& \mbox{ for } \Mb{0}\,.\\ \end{cases}
 \end{equation}
 This behaviour is also illustrated in
 Fig.\ref{fig:fcorrclust_pp}. The clustering correction for
 $M_0^{(0)}$ coincides with the jet veto result presented in
 Refs.~\cite{Banfi:2012jm,Abreu:2022sdc}.

\begin{table}[t!]
\centering
\begin{tabular}{|c|l|c|r|r|}
\hline
\multicolumn{5}{|c|}{Correlated and clustering corrections in DIS and $\ee$ collisions}\\
\hline
\multicolumn{2}{|c|}{Observable} & \multicolumn{1}{|c|}{$\langle \ln f_{\rm correl}\rangle_{C_A}$} & $\langle \ln f_{\rm correl}\rangle_{n_f}$ & \multicolumn{1}{|c|}{$\langle \ln f_{\rm clust} \rangle$}\\
\hline
\multirow{3}{*}{$\Sb{n_{\min}}$} 
                          & $\bo=0$    & $0.94426(1)$ &$-0.208545(1)$ & $0.0374791(1)$
                          \\ \cline{2-5}
                          & $\bo=0.5$ & $0.97044(2)$ &$-0.211395(1)$ & $0.0699432(1)$  \\ \cline{2-5}
                          & $\bo=1$    & $0.94845(3)$  &$-0.238064(2)$ & $0.1115572(2)$ \\ \hline
                          
\multirow{3}{*}{$\Mb{n_{\min}}$} 
                          & $\bo=0$    & $1.05074(1)$ & $0.014963(3)$ & $-0.4939429(3)$  \\ \cline{2-5}
                          & $\bo=0.5$ & $1.08702(2)$ & $0.011683(3)$ & $-0.4614788(3)$  \\ \cline{2-5}
                          & $\bo=1$    & $1.09384(3)$   &$-0.018616(2)$ & $-0.4198655(4)$ \\ \hline
\end{tabular}
\caption{Numerical values for $\langle \ln f_{\rm correl}\rangle_{C_A}$, $\langle \ln f_{\rm correl}\rangle_{n_f}$, $\langle \ln f_{\rm clust} \rangle$
for $\Sb{2}$ ($\Sb{1}$) and $\Mb{2}$  ($\Mb{1}$) in $\ee$  collisions (DIS) for $\bo=0$, 1/2, 1.
 }
 \label{tab:fcorrclust_eedis}
\end{table} 
 
\paragraph{Recoil correction}
The recoil correction is present only for final-state legs, and is
identical for $\Sb{n_{\min}}$ and $\Mb{n_{\min}}$.  For $ee$
collisions, we have
\begin{align}\label{eq:frec}
  \delta\mathcal{F}_{\rm rec}^{ee}(v) & = \frac{\mathcal{F}_{\rm NLL}(v)}{1+\bo-2\lambda}\\
  \times& 
  \begin{cases}
    \displaystyle
    C_F\frac{\left(15-4 \pi ^2\right) \bo-2 \left(\pi ^2-9+9\ln2\right)}{6} & \mbox{ for quarks,}\\[10pt]
    \displaystyle
    C_A  \frac{131-132 \ln 2-12 \pi ^2-2\left(12 \pi ^2-67\right) \bo}{36}
      + n_f \frac{24 \ln 2-23 - 26 \bo}{36}
     & \mbox{ for gluons}.
  \end{cases}\nonumber
\end{align}
For $\dis$, since we have only one final-state leg, we have
\begin{equation}
  \delta\mathcal{F}_{\rm rec}^{\dis}(v)=\frac{  \delta\mathcal{F}_{\rm rec}^{ee}(v)}{2}.
    \label{eq:dFrecDIS}
\end{equation}

\paragraph{Hard-collinear correction}
The hard collinear correction  is $0$ for $\Mb{n_{\min}}$, while for $\Sb{n_{\min}}$ it reads
\begin{equation}
  \dFhc(v) =
    - \left(\psi^{(0)}(1+\RpNLL(v))+\gamma_E\right)
          {\cal F}_{\rm \tiny NLL}(v)
  \frac{1}{1+\bo-2\lambda}\sum_{\ell} \gamma^{(0)}_\ell\,.
  \label{eq:dFhc}
\end{equation}

\subsection{Constants}
\label{app:constants}
The Casimir  factor $C_{\ell}$ for quark and gluon legs reads 
\begin{equation}
C_q \equiv  C_F = \frac{N_c^2-1}{2 N_c}, \qquad  C_g \equiv C_A = N_c,
\end{equation}
with $N_c=3$. We also consider $n_f=5$ light quarks.
The coefficients of the QCD $\beta$ function read
\begin{subequations}
	\label{eq:qcdbeta}
	\begin{align}
		\bqcd{0} =& \frac{11 C_A - 2 n_f}{12 \pi}\,,\\
		\bqcd{1} = & \frac{17 C_A^2 - 5 C_A n_f - 3 C_F n_f}{24\pi^2}\,, \\
		\bqcd{2} = & \frac{2857 C_A^3 + (54 C_F^2 - 615 C_F C_A - 1415 C_A^2) n_f + (66 C_F + 
			79 C_A) n_f^2}{3456 \pi^3}\,.
	\end{align}
\end{subequations}
The coefficients of the physical coupling
scheme~\cite{Catani:1990rr,Banfi:2018mcq,Catani:2019rvy} are given by
\begin{subequations}
	\label{eq:cuspanom}
\begin{align}
	K^{(1)} &= C_A
	\left(\frac{67}{18}-\frac{\pi^2}{6}\right)-\frac{5}{9}n_f\,,
	\\
	K^{(2)} & = C_A^2 \left( \frac{245}{24} - \frac{67}{9}\zeta_2
	+ \frac{11}{6}\zeta_3 + \frac{11}{5}\zeta_2^2\right) 
	+ C_F n_f \left(-\frac{55}{24} + 2\zeta_3\right) \\
	& \hspace{0.2cm} + C_A n_f \left(-\frac{209}{108} + \frac{10}{9}\zeta_2 - \frac{7}{3} \zeta_3\right) 
	- \frac{1}{27} n_f^2 + \frac{\pi\bqcd{0}}{2} \left(C_A\left(\frac{808}{27}-28\zeta_{3}\right)-\frac{224}{54}n_f\right)\,.\notag
\end{align}
\end{subequations}
The hard-collinear anomalous dimensions read
\begin{subequations}
	\label{eq:hardanom}
\begin{align}
	\gamma_q^{(0)} =& -\frac{3}{2} C_F\,, \qquad 
	\gamma_{g}^{(0)} =-2\pi \bqcd{0}\,, \\ 
	\gamma_q^{(1)} =& -\frac{ C_F}{2} \left(C_A \left(\frac{17}{12}+\frac{11 \pi ^2}{9}-6 \zeta_3\right)+ C_F\left(\frac{3}{4}-\pi ^2+12 \zeta_3\right)-n_f\left(\frac{1}{6}+\frac{2 \pi ^2}{9}\right) \right)\,,\\
	\gamma_{g}^{(1)} =& \frac{n_f}{2} C_F + \frac{2}{3} n_f C_A-C_A^2 \left( \frac{8}{3}+3\zeta_3\right)\,.
\end{align}
\end{subequations} 
For processes with a final-state quark/gluon leg, the collinear matching coefficients are given by
\begin{subequations}
\label{eq:coll-matching}
\begin{align}
	C_{{\rm hc}, q}^{(1)}=& \frac{C_F}{2}\left(1+\frac{7 \bo}{1+\bo}\right)\,,\\
	C_{{\rm hc}, g}^{(1)}=&\left(\frac{ \bo}{1+\bo}\left(\frac{67}{18}C_A - \frac{13}{9}T_R n_f \right)+\frac{1}{3} T_R n_f\right).
\end{align}
\end{subequations}
Finally we report the hard matching coefficients, including their scale dependence. For $e^+e^-$ and $pp$ we have for the $q\bar{q}$ and $gg$ channels respectively
\begin{align}
    H^{(1)}_q =& C_F\left(\pi^2-8\right)  -\frac{4C_F}{1+\bo}\lnxMsq-4\frac{\gamma_q^{(0)}}{1+\bo}\lnxM\,,\\
    H^{(1)}_g =& C_A\left( \pi^2+5\right) - 3 C_F-\frac{4
                 C_A}{1+\bo}\lnxMsq-4\frac{\gamma_g^{(0)}}{1+\bo}\lnxM+8\pi\bqcd0
                 \ln \frac{\mur}{\Q}\,.
\end{align}
For DIS, the constants can be obtained via crossing and read
\begin{align}
    H^{(1), \rm DIS}_{q,g} = & H^{(1)}_{q,g} - C_{F,A} \pi^2\,.
\end{align}

\section{A prescription for switching off logarithms in matching}
 \label{app:matching}
 In this section we provide additional details on the matching
 prescription introduced in Sec.~\ref{sec:matching}. The dynamical
 resummation scale given in Eqs.~\eqref{eq:muLtilde},~\eqref{eq:Lmod}
 is characterised by three regions: a \textit{resummation} region
 ($v<\vm$), a \textit{matching} region ($\vm\leq v<1$), and a
 \textit{fixed-order} region ($v\geq 1$).
 Such a simple prescription guarantees that the dependence of the
 prediction on the resummation scale $\mul$ is confined at small $v$,
 where its variation probes higher-order logarithmic corrections,
 while the matching scale $\mum$ determines the scale at which the
 matched prediction reduces to its fixed-order component.

 The function $\mathfrak{r}(v)$ is given by the polynomial
 $\mathfrak{r}(v)=r_0+\sum_{i=0}^{n}r_{i+1} v^{p+i}$, where $p$ is a
 positive power chosen such that we introduce at most quadratic power
 corrections at small $v$ ($p\geq 2$). The coefficients are determined
 by imposing the following matching conditions
\begin{align}
\mathfrak{r}(0) = r_0=\mul\,,\quad \mathfrak{r}\left(\vm\right) = \mum\,,\quad \left.\frac{d^k}{d v^k}
  \ln\frac{\mathfrak{r}(v)}{v}\right|_{v=\vm} \!\!\!= 0\,\text{ for } k = 1,\dots,n\,,
\end{align}
and solving the corresponding linear system. For $n=2$, we obtain
\begin{align}
r_0=\mul\,,\quad r_1=\frac{1}{2}\,(1+p) \vm^{-p}&\left(p \,\vm\,
                  Q-(2+p)\mul \right) \,,\\\quad
  r_2=\vm^{-1-p}\left((1-p^2)\vm\,Q+p(2+p)\mul\right)\,,\quad&
                                                               r_3=-\frac{1}{2}\,p\, \vm^{-2-p}\left((1-p)\vm\,Q+(1+p)\mul\right)\,.\notag
\end{align}
Similarly, the function $\mathfrak{m}(v)$ is given by an analogous
polynomial $\mathfrak{m}(v)=m_0+\sum_{i=0}^{2s}m_{i+1} v^{q+i}$ (where
$q$ is a positive parameter, in general different from $p$),
satisfying the constraints
\begin{align}
\mathfrak{m}\left(\frac{\mum}{Q}\right) = \mum\,,\quad  \mathfrak{m}\left(1\right) = Q\,,\quad \left.\frac{d^k}{d v^k}
  \ln\frac{\mathfrak{m}(v)}{v}\right|_{v=\vm} \!\!\!=\left.\frac{d^k}{d v^k}
  \ln\frac{\mathfrak{m}(v)}{v}\right|_{v=1} \!\!\!= 0\,\text{ for } k = 1,\dots,s\,.
\end{align}
For $s=1$ we obtain
\begin{subequations}
\begin{align}
  m_0&=\frac{Q \vm \left(-\left((\vm+1) \vm^q\right)-q
   (1-\vm) \left(\vm^q+1\right)+\vm+1\right)}{2
   \vm \left(1-\vm^q\right)-q (1-\vm)
   \left(\vm^{q+1}+1\right)}\,,\\
  m_1&=-\frac{Q \vm \left(\vm^{-q}+\vm^{q+2}-(q+1)^2
   \vm^2+2 q (q+2) \vm-(q+1)^2\right)}{(1-\vm)
   \left(2 \vm \left(1-\vm^q\right)-q (1-\vm)
   \left(\vm^{q+1}+1\right)\right)}\,,\\
  m_2&=-\frac{Q \vm^{-q} \left(-2 \vm^{2 q+2}+(\vm+1)
   \left(q (q+1) (1-\vm)^2+2 \vm\right) \vm^q-2
   \vm\right)}{(1-\vm) \left(2 \vm
   \left(1-\vm^q\right)-q (1-\vm)
   \left(\vm^{q+1}+1\right)\right)}\,,\\
  m_3&=-\frac{Q \vm^{-q} \left(-\left(\left(q^2 (1-\vm)^2+2
   \vm\right) \vm^q\right)+\vm^{2
   q+1}+\vm\right)}{(1-\vm) \left(2 \vm
   \left(1-\vm^q\right)-q (1-\vm)
   \left(\vm^{q+1}+1\right)\right)}\,.
\end{align}
\end{subequations}

 \begin{figure}[tb]
 	\centering
 	\includegraphics[width=0.48\textwidth,page=1]{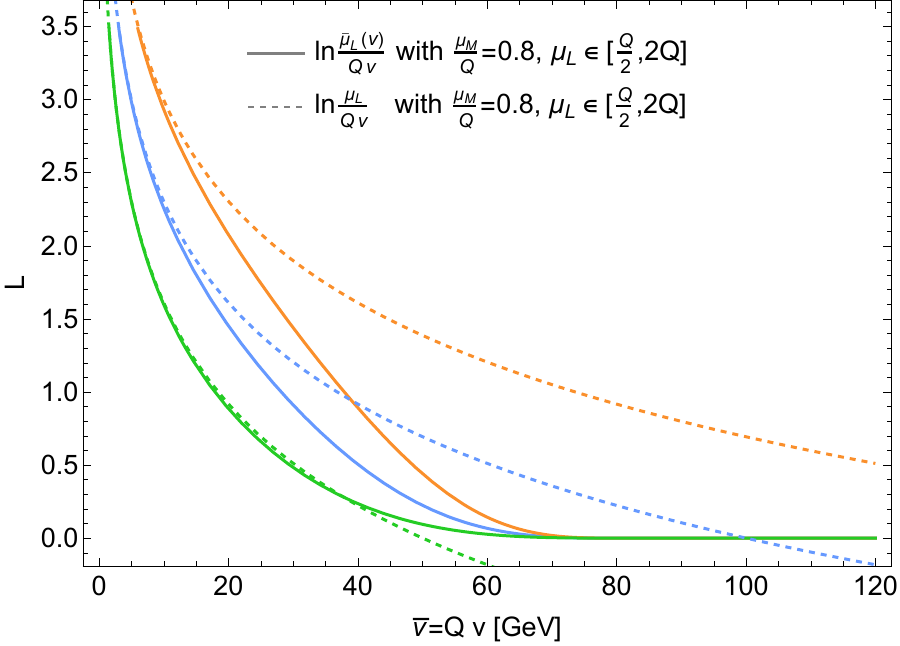}
 	\includegraphics[width=0.511\textwidth,page=1]{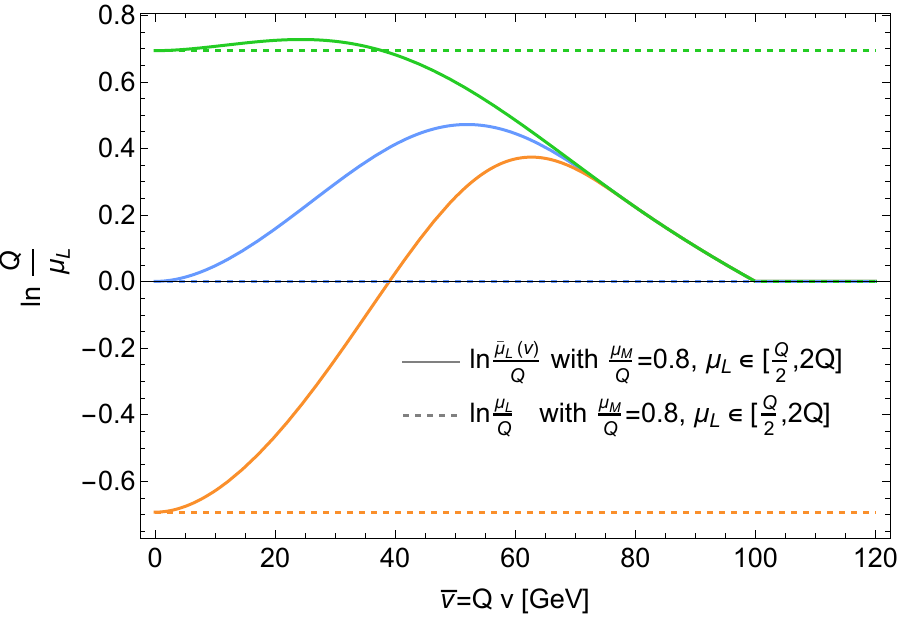}
 	\caption{Comparison between logarithms with the dynamical and
          fixed resummation scales, for a variation of the scale
          $\mul$ in Eq.~\eqref{eq:muLtilde}. The left plot shows the
          behaviour of the resummed logarithm $L$ in
          Eq.~\eqref{eq:res-scale-def}, while the right plot shows the
          term $\ln Q/\mul$ in the same equation. In all cases the
          hard scale is set to $Q=100$ GeV.}
 	\label{fig:profiled-logs}
 \end{figure}

One can verify that the limit $\vm\to 1$ is smooth.
With reference to Eq.~\eqref{eq:res-scale-def}, both $L$ and
$\ln\frac{Q}{\mul}$ are active in the resummation region, while only
the latter remains in the matching region where no large logarithms
are present. To help visualise the action of the dynamical scales, in
Fig.~\ref{fig:profiled-logs} we show an example obtained with $p=q=2$,
$Q=100$ GeV and $\mum=80$ GeV. We clearly see that both $L$ and
$\ln Q/\mul$ are affected by the prescription in the resummation
region, ensuring that at small $\bar{v}=Q v$ they agree with their
unmodified counterparts. At $\bar{v}=\mum$, the resummed logarithm $L$
is turned off according to the function $\mathfrak{r}(v)$, while only
$\ln Q/\mul$ is nonzero in the matching region $\mum \leq \bar{v} < Q$,
where it is eventually turned off at $\bar{v}=Q$ following the
function $\mathfrak{m}(v)$. It is important to stress that the
dependence of $\mathfrak{m}(v)$ completely cancels out at the level of
the matched prediction~\eqref{eq:master-matching}, so that a variation
of $\mum$ only determines the scale at which the fixed-order result is
recovered.
Finally, the fixed-order region is not affected by any of the above
scales. Together with variations of the factorisation and
renormalisation scales, this provides a flexible way to access
perturbative uncertainties in applications to phenomenology.

\section{Fixed-order validation up to second perturbative order for
  $\dis$ and $\lep$}
\label{app:fo-validation}
 \begin{figure}[tb]
 	\centering
 	\includegraphics[width=\textwidth,page=1]{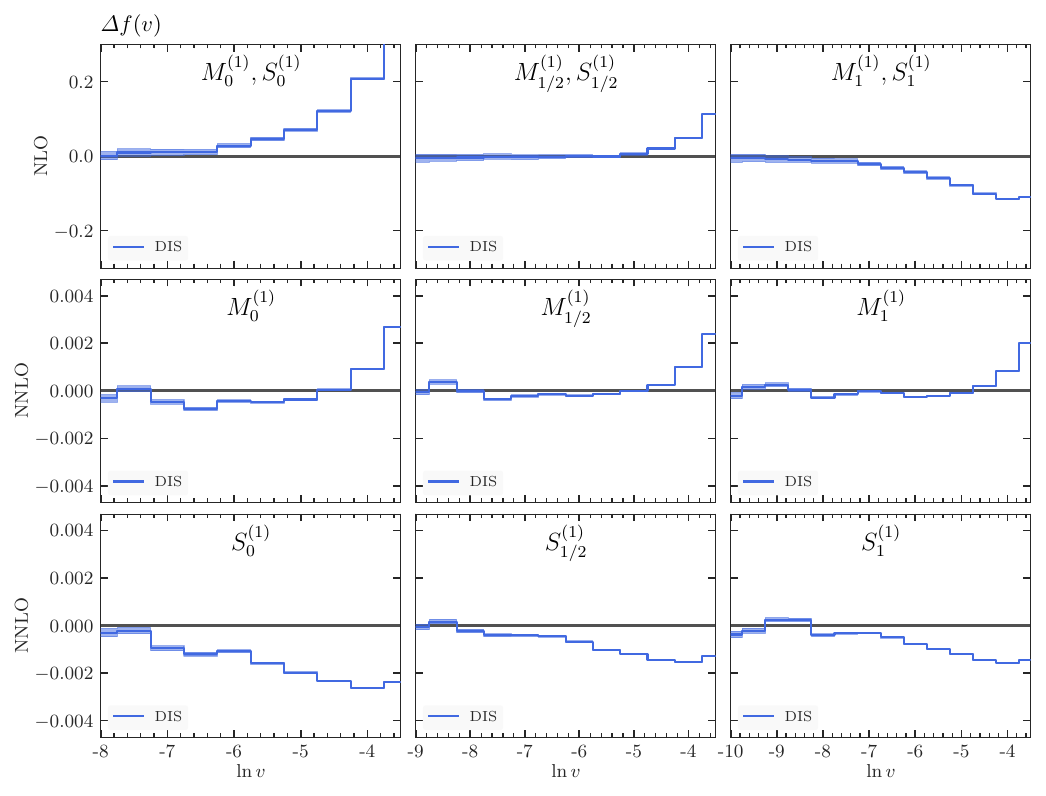}
 	\caption{Difference between the fixed-order and resummed-expanded differential cross sections, normalised to the Born cross section, c.f.~Eq.~\eqref{eq:deltaFFO}, for DIS. We consider three values of $b$, $b=0$ (left), $b=1/2$ (middle) $b=1$ (right).
 		The top panel shows the NLO comparison for $\Sb{1}$/$\Mb{1}$.
 		The middle (bottom) panel shows the NNLO comparison for $\Mb{1}$ ($\Sb{1}$).
 		The bands show the statistical
 		uncertainty.}
 	\label{fig:dis-fo-validation}
 \end{figure}
\begin{figure}[tb]
 	\centering
 	\includegraphics[width=\textwidth,page=2]{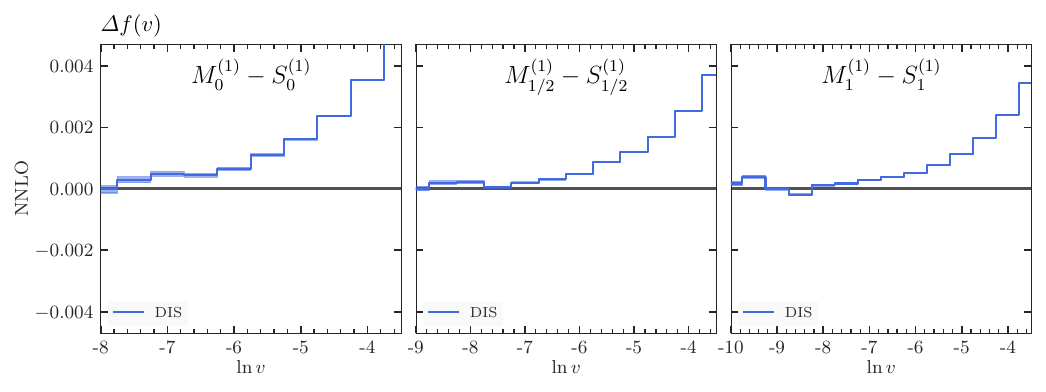}
 	\caption{Same as Fig.~\ref{fig:dis-fo-validation}, but for the difference between the $\Mb{1}$ and $\Sb{1}$ observables.}
 	\label{fig:dis-fo-m-s-validation}
 \end{figure}

This appendix contains the validation of the resummation for the \lobs for $\dis$ and $\lep$ up to second perturbative order.
The corresponding $pp$ validations were discussed in
Sec~\ref{sec:fixed-order-2-legs}.
We start by discussing the results for $\dis$.
The fixed-order reference results are obtained with
\texttt{disorder}~\cite{Karlberg:2024hnl,Salam:2008qg,Catani:1996vz} (with $\texttt{cutoff}=10^{-14}$ and
$\texttt{npow1}=6$ and $\texttt{npow2}=4$).
We perform the check at a fixed Born point given by
$x_{\rm \scriptscriptstyle DIS} = 0.3$ and $Q = 91.1876$~GeV.
We use central values for the renormalisation and factorisation
scales, and employ the central replica of the
\texttt{NNPDF40MC\_nnlo\_as\_01180} PDF set.
We then compute the quantity $\Delta f(v)$ (see
Eq.~\eqref{eq:deltaFFO}) at NLO and NNLO for $\Sb{1}$ and
$\Mb{1}$.
The results are shown in Fig.~\ref{fig:dis-fo-validation}. 
In the top panel we show the NLO comparison, and in the bottom two
panels the NNLO one.
We observe the expected convergence for $v\to 0$.
In Fig.~\ref{fig:dis-fo-m-s-validation} we also show the convergence of
the difference between the $M_b^{(0)}$ and $S_b^{(0)}$ observables,
which tends to behave better due to cancellations of the larger
logarithmic powers as well as statistical uncertainties in this
quantity.
This gives us further confidence that our prediction is
correct. 
 
\begin{figure}[tb]
 	\centering
 	\includegraphics[width=\textwidth,page=1]{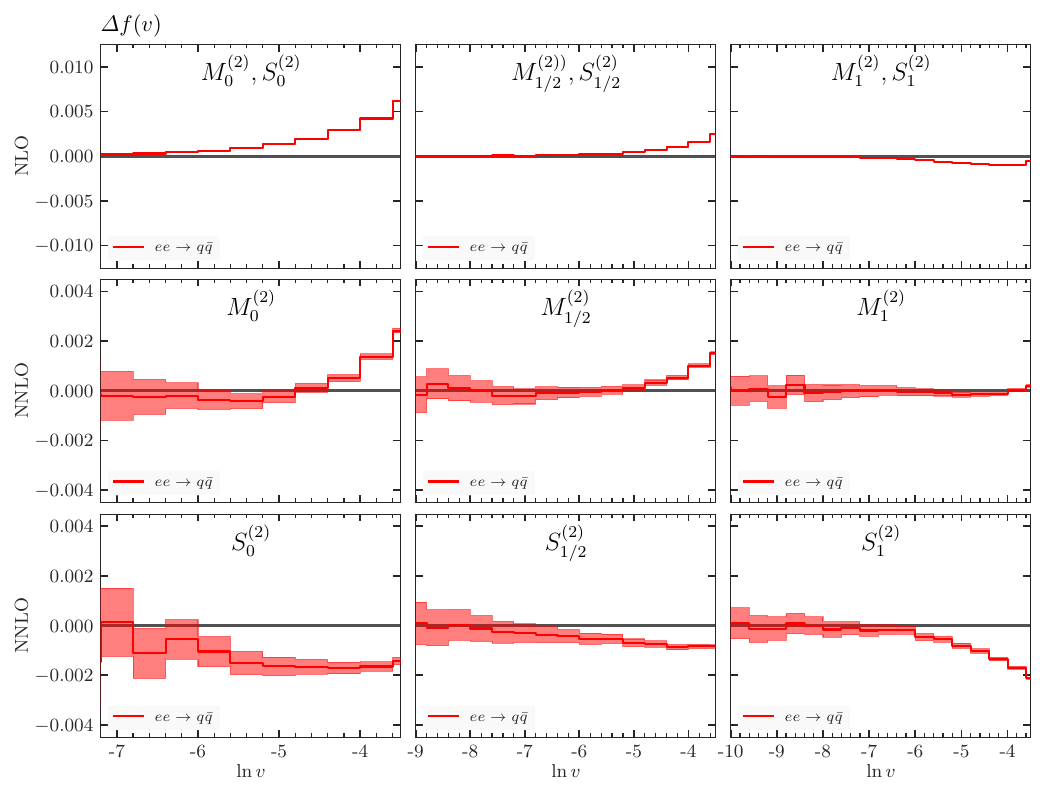}
 	\caption{Same as Fig.~\ref{fig:dis-fo-validation}, but for $e^+e^-$.}
 	\label{fig:epem-fo-validation}
 \end{figure}

We now move to the $\lep$ validation.
We ran \texttt{Event2}~\cite{Catani:1996vz} in quadruple precision
(with $\texttt{cutoff}=10^{-16}$ and
$\texttt{npow1}=\texttt{npow2}=6$) and compared the resulting
predictions with the fixed-order expansion of our resummation
prediction.
The result can be found in Fig.~\ref{fig:epem-fo-validation},
displaying a successful test for all \lobs.

\FloatBarrier 
\bibliographystyle{JHEP}
\bibliography{refs}
\end{document}